\documentclass{article}

\usepackage{arxiv}

\usepackage[utf8]{inputenc} % allow utf-8 input
\usepackage[T1]{fontenc}    % use 8-bit T1 fonts
\usepackage{hyperref}       % hyperlinks
\usepackage{url}            % simple URL typesetting
\usepackage{booktabs}       % professional-quality tables
\usepackage{amsfonts}       % blackboard math symbols
\usepackage{nicefrac}       % compact symbols for 1/2, etc.
\usepackage{microtype}      % microtypography
\usepackage{lipsum}		% Can be removed after putting your text content
\usepackage{graphicx}
\usepackage{natbib}
\usepackage{doi}
\usepackage{amsmath}
\usepackage{amssymb}
\usepackage{enumitem}
\usepackage{pdfpages}
\usepackage{blkarray}
\usepackage{mathtools}
\usepackage{natbib}
\usepackage{url}
\usepackage[lined,boxed,linesnumbered]{algorithm2e}
\usepackage{graphicx,subfigure}
\usepackage{epstopdf} 
\usepackage{array}
\usepackage{color}
\usepackage{comment}
\usepackage{ulem}
\usepackage{soul}

\title{\texttt{mvlearnR} and Shiny App for multiview learning}

\author{ 
	{\hspace{1mm}Elise F. Palzer} \\
	Division of Biostatistics and Health Data Science\\
	University of Minnesota Twin Cities\\
	Minneapolis, MN 55455 \\
	\texttt{north266@umn.edu} \\
        \And 
        \href{https://orcid.org/0000-0001-9593-4778}{\includegraphics[scale=0.06]{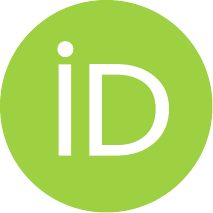}\hspace{1mm}Sandra E. Safo}\thanks{Corresponding Author: Sandra Safo, www.sandraesafo.com} \\
	Division of Biostatistics and Health Data Science\\
	University of Minnesota Twin Cities\\
	Minneapolis, MN 55455 \\
	\texttt{ssafo@umn.edu} \\
}

\renewcommand{\shorttitle}{mvlearnR}

\hypersetup{
pdftitle={A template for the arxiv style},
pdfsubject={q-bio.NC, q-bio.QM},
pdfauthor={David S.~Hippocampus, Elias D.~Striatum},
pdfkeywords={First keyword, Second keyword, More},
}

\begin{document}
\maketitle

\begin{abstract}
The package \texttt{mvlearnR} and accompanying Shiny App  is intended for integrating data from multiple sources or views or modalities (e.g. genomics, proteomics,  clinical and demographic data). Most existing software packages for multiview learning are decentralized and offer limited capabilities,  making it difficult for users to perform comprehensive integrative analysis. The new package wraps  statistical and machine learning methods and graphical tools, providing  a convenient and easy data integration workflow. For users with limited programming language, we provide a Shiny Application to facilitate data integration anywhere and on any device. The methods have potential to offer deeper insights into complex disease mechanisms.\\
%, identify biomarkers, define or redefine disease subtypes, and identify individuals at a high risk for developing a particular disease.\\ 
\textbf{Availability and Implementation:} \texttt{mvlearnR} is available from the following GitHub repository:\\
\url{https://github.com/lasandrall/mvlearnR}. The web application is hosted on shinyapps.io and available at: \url{https://multi-viewlearn.shinyapps.io/MultiView\_Modeling/} \\
\textbf{Contact: } ssafo@umn.edu \\
\end{abstract}

% keywords can be removed

\section{Introduction}
Nowadays, multiple types of data (or sometimes called views or modalities [e.g. genomics, proteomics, metabolomics]) are frequently measure on the same sets of individuals and has opened an era of research on multiview learning. It is recognized that the mechanisms that underlie complex diseases may be unraveled by approaches that go beyond analyzing each type of data separately.
However, analyzing these data types to obtain useful information and knowledge is challenging because the data are complex, heterogeneous, and high-dimensional, and requires a considerable level of analytical sophistication. Many methods have been proposed to associate multiview data (e.g. \citep{Hotelling:1936, safo2018sparse,horst1961generalized,kettenring1971canonical, lock2013joint, safo2023scalable}. 

The methods could be unsupervised or a combination of supervised and unsupervised techniques. The unsupervised methods first correlate the different data types to learn shared or view-independent low-dimensional representations \cite{Hotelling:1936, safo2018sparse,lock2013joint}. This is then followed by independent prediction analyses that use the learned low-dimensional representations. The unsupervised methods are useful for data exploration. The joint association and prediction methods combine supervised and unsupervised techniques such that assessing associations between multiple views is linked to predicting an outcome. \citep{PALZER2022107547, SIDA:2019, wang2021deep, MOMA:2022}. The goal in these methods is then to learn low-dimensional representations that have potential to predict the outcome under consideration. Since the outcome is used in deriving the low-dimensional representations, these low-dimensional representations are naturally endowed with prediction capabilities which enhances interpretability.  

Most existing software packages for multiview learning tend to be decentralized, making it difficult for users to perform comprehensive integrative analysis. The mix-omics \citep{rohart2017mixomics} package for integration offers  both supervised and unsupervised methods for multiview learning. However, the methods provided in mix-omics are limited. For instance,  the outcome types are either continuous or categorical, not allowing for other types of outcomes (e.g Poisson, time-to-event). The methods do not allow for the use of prior biological information which can enhance interpretability. Importantly, users must be well versed in the R programming langauge, which is limiting. 

We provide \texttt{mvlearnR}, an R software for multiview learning, which will serve as a comprehensive software for integrating data from multiple sources. The
new package wraps statistical and machine learning methods and graphical tools, providing a convenient and easy data integration workflow. For users with limited programming language, we provide a Shiny Application to facilitate data integration. Currently, \texttt{mvlearnR} can be used to:
\begin{itemize}
\item Prefilter each data type via differential analysis (DA). We provide both supervised and unsupervised options for DA or for filtering out noise variables prior to performing data integration. 
\item Integrate data from two sources using a variant of the popular unsupervised method for associating data from two views, i.e. canonical correlation analysis (CCA). 
\item Predict a clinical outcome using results from CCA. We provide four outcome data distribution type (i.e. gaussian, binomial, Poisson, and time-to-event data.)
\item Jointly integrate data from two or more sources and discriminate between two or more classes. We provide an additional method which allows  to incorporate prior biological structure (e.g., variable-variable relationships). These methods allow to include covariates. 
\item Visualize results from DA or  integrative analysis methods. These plots include: volcano plots, UMAP plots, variable importance plots, discriminant plots, correlation plots, relevance network plots, loadings plots, and within- and between- view biplots. These visualization tools will help unravel complex relationships in multiview data.
\item Demonstrate our integration workflow via already uploaded synthetic and real molecular and clinical data pertaining to COVID-19.
\end{itemize}

The rest of this paper presents the package and Shiny App, with more details in the Supplemetary Material. We organize the paper as follows:
First, we present the implementation details of the package and web application.
Next, we present the filtering, supervised and unsupervised integration and visualization methods used in greater detail. Then, we demonstrate how \texttt{mvlearnR} can be used on real data and discuss the proper interpretation of the results. Finally, we discuss the limitations of the package and web application and potential future directions. 

\section{Methods and Implementation} 
In this section, we give details about the package and web application and summarize the methods implemented. 
In Table S1 of the supplementary material, we provide the currently available functions in \texttt{mvlearnR} and their descriptions.

\subsection{The \texttt{mvlearnR} Web App and Package} \label{subsec:package}
\begin{figure*}
    \centering
    \includegraphics[width=\textwidth]{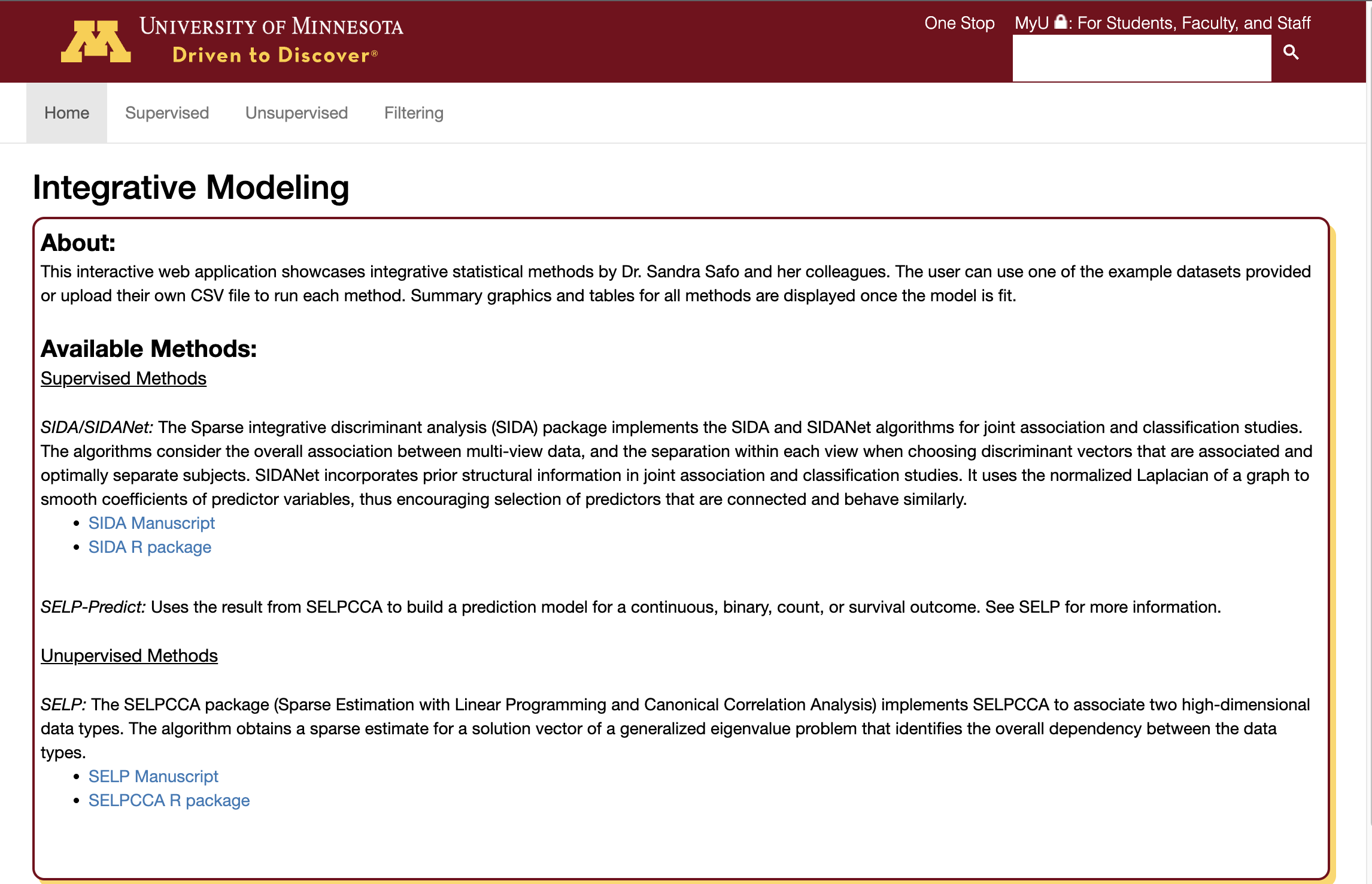}
    \caption{Multiview Shiny App Interface: Our Shiny App will allow non-users of R to seamlessly conduct integrative analysis.  The web application has four tabs. The first tab, `Home', provides a brief overview of the methods and related links. The second tab, `Supervised', is where the user will implement supervised integrative analysis methods. The third tab, `Unsupervised', is where the user will implement unsupervised integrative analysis methods.  We provide options for the user to upload their own data or use example data. These tabs produce outputs of the model including classification performance, variable importance tables and plots, and several other plots to help the user understand the results. The fourth tab, `Filtering', is where the user has the option to filter and preprocess their data to a customizable lower dimensional subset prior to data integration, using supervised and unsupervised filtering methods. }
    \label{fig:interface}
\end{figure*}

The \texttt{mvlearnR} web app consists of a user-friendly interface ( Figure \ref{fig:interface}), it is ideal for users with limited programming expertise in R, and it can be used anywhere and on any device. Leveraging state-of-the-art unsupervised\cite{SafoBiomSELP} and supervised \cite{safosida:2019} integrative analysis methods, \texttt{mvlearnR} web server and package enable researchers to  integrate molecular and clinical data, ultimately reducing the gap from raw molecular data to biological insights. The web application has four tabs. The first tab, `Home', provides a brief overview of the methods and related links (Figure \ref{fig:interface}). The second tab, `Supervised', is where the user will implement supervised integrative analysis methods. The third tab, `Unsupervised', is where the user will implement unsupervised integrative analysis methods.  We provide options for the user to upload their own data or use example data. These tabs produce outputs of the model including classification performance, variable importance tables and plots, and several other plots to help the user understand the results. The fourth tab, `Filtering', is where the user has the option to filter and preprocess their data to a customizable lower dimensional subset prior to data integration, using supervised and unsupervised filtering methods. The  web application uses the R Shiny framework and is hosted at shinyapps.io. 

In the R-package, we provide  real data pertaining to COVID-19 severity and status.  The data are from a study conducted by \cite{overmyer:2020} that collected blood samples from 102 participants with COVID-19 and 26 participants without COVID-19 and quantified for metabolomics, RNA sequencing (RNA-Seq), proteomics, and lipidomics. We provide in this package the proteomics and RNA-seq data as preprocessed in \cite{Danietal:2022}. Disease severity was measured using the World Health Organization (WHO) 0-8 disease specific scale (8 denotes death), and a score out of 45 that indicates the number of hospital free days (HFD-45) \cite{overmyer:2020}. These two outcome variables and other metadata (e.g. age, sex, comorbidities) are provided in the R-package. We refer the reader to \cite{overmyer:2020} for more details on available data. The R-package can be downloaded from GitHub at \url{https://github.com/lasandrall/mvlearnR}. We next describe the methods and functions. 

\subsection{Data Import and Filtering} \label{subsec:filtering}
We provide real data pertaining to COVID-19 severity and status and several simulated datasets to demonstrate the use of the package. Simulated data for two views and a binary outcome could be read into R as  \texttt{data(sidaData)}, and \texttt{data(selpData)}. The COVID-19 data can be imported into R as \texttt{data(COVIDData)} [Figure S3]. This is a list with 3 entries:  Proteomic, RNASeq, and Clinical data. Integrative analysis methods sometimes perform poorly on large datasets so we provide supervised and unsupervised methods to filter data to help users focus on variables that are more likely to yield meaningful findings after integration. 
In the R-package, the function \texttt{filter.supervised()} [Figure S5] can be used to filter each view when an outcome is available via the four methods: linear, logistic, t-test, and Kruskal-Wallis (KW) test. Supervised filtering allows the user to filter variables based on their association with an outcome.  P-values can be adjusted for multiple hypothesis testing. The function \texttt{filter.unsupervised()} can be used to filter each view using unsupervised methods such as variance and interquartile range (IQR) filtering. We provide an option to log2 transform variables, scale variables to have variance one, center variables to have mean zero, or normalize variables to have mean zero and variance one.  \textbf{Quality control checks (e.g. batch correction) and other form of normalizations specific to a particular omics data should be done outside \texttt{mvlearnR}.} 

The web application provides the filtering, scaling, centering, and normalization options. Regarding the data to be uploaded for filtering, the user can upload i) Train and Test Sets, for when uploaded data have already been split into training and testing sets. Filtering will be conducted only on the training data. After filtering is complete, a new test set will be constructed to contain the same variables as the filtered training data; ii) Full dataset, for when uploaded data have not been split into training and testing sets. If the user would like the app to create separate training and testing sets, we provide an option for this via the `Pct in Training set' tab.

\subsection{Unsupervised methods for associating data from two sources}
We provide the sparse canonical correlation analysis (CCA) method, SELPCCA, proposed in \cite{safo2018sparse}, and described in details in the Supplementary Material for unsupervised data integration. CCA \cite{Hotelling:1936} is a  multivariate linear dimension reduction method for maximizing association between data from two sources. CCA finds weighted combinations of variables in each data that maximize the overall dependency structure between pairs of data. In the context of data integration strategies described in \cite{picard2021integration} (e.g., early, intermediate, and late data integration), CCA falls under the intermediate strategy. The classical CCA finds linear combinations of all available variables, and since these weights are typically nonzero, it is difficult to interpret the findings.  SELPCCA \cite{safo2018sparse} is a variant of CCA that shrinks some of the weights of the low-dimensional representations to zero, thus allowing to identify relevant variables contributing to the overall dependency structure in the data. SELPCCA is thus a feature extraction and feature selection method, falling under the intermediate strategy umbrealla for data integration.  The function \texttt{cvselpscca()}[Figure S8] can be used to obtain low-dimensional linear representations that maximize associations between pairs of data, and  to identify key variables contributing to the maximum correlation between the pairs of data. The main inputs to \texttt{cvselpscca()} are the train data and the number of canonical vectors, `ncancorr', to be estimated, which defaults to 1 if not specified.

The output of \texttt{cvselpscca()} include: `hataplha' and `hatbeta', representing the loadings or canonical vectors for the two views, respectively; `optTau', the optimal tuning parameter for each data type, and `maxcorr', estimated canonical correlation coefficient, which shows the strength of the association between the data types. The canonical vectors could be visualized, for more insight (see Section on Visualizations and Supplementary Material). Since these loadings are standardized to have unit norm, a variable with larger weight contributes more to the association between the views. Please refer to the Section Use Case for a demonstration of the \texttt{cvselpscca()} function.

On the web application, the SELPCCA method is located on the `Unsupervised' tab. The user can upload their data and then use the default hyper parameters to obtain the canonical vectors. After hitting the `Run Model' tab, a notification button notifies the user that the model is running. After completion, the top 20 selected variables from each view is printed out.

\subsection{Prediction with learned low-dimensional representations from unsupervised methods}
Since SELPCCA is an unsupervised method, it can only be used to identify variables contributing to the maximal
association between two views. SELPCCA is ideal as an exploratory method to explore variables that contribute to the overall dependency structure between the views. If an outcome is available, one can associate the learned low-dimensional representation(s)
with the outcome. We provide the function \texttt{selpcca.pred()}[Figure S15] for this purpose where the results from the SELPCCA model are used to build a generalized linear model (GLM) or Cox prediction model.  The required option `family' is a string specifying the type of prediction model to build. Options are gaussian, binomial, Poisson, and survival. When family $=$ `survival', a  Cox proportional model will be fitted. Otherwise a GLM will be used. 
The function \texttt{predict()} can be used to predict out of sample data using the learned low-dimensional representations (Figure S16). The performance of these new predictions can be assessed on a training data using the function \texttt{PerformanceMetrics() } [Figure 16]. Currently two family options are provided: `binomial' and `gaussian'.

On the web application, the SELP-Predict method is located on the `Supervised' tab. The results from the SELPCCA model are used to build a generalized linear model (GLM) or Cox prediction model. To implement this function,  the user uploads their data, sets the distribution family, and can use the default hyper parameters to obtain the canonical variates. The required option `family' is a string specifying the type of prediction model to build. Options are gaussian, binomial, poisson, and survival. When family $=$ `survival', a  Cox proportional model will be fitted. Otherwise a GLM will be used. After model implementation, the user can view the model estimates, obtain some prediction estimates, and visualize the top 20 selected variables for each view.

\subsection {Supervised methods for associating data from two or more sources}
Sparse integrative discriminant analysis (SIDA) \cite{SIDA:2019} is an integrative analysis
method for jointly modeling associations between two or more views and creating separation of classes within each view.  The algorithm considers the overall association between multiview data, and the separation within each view when choosing discriminant vectors that are associated and optimally separate subjects. Thus, SIDA  combines the advantages of  linear discriminant analysis (LDA)\cite{Hotelling:1936}, a supervised learning method for maximizing separation between classes in one view, and CCA, an unsupervised learning method for maximizing correlation between two data types, and falls under the intermediate strategy for data integration described in \cite{picard2021integration}. 
SIDA allows the user to  select key variables that contribute to the maximum association of the views and separation of the classes. The function \texttt{cvSIDA()} performs n-fold cross-validation to select optimal tuning parameters for SIDA based on training data, and  predicts training or testing class membership.  The function \texttt{cvSIDANet()} incorporates prior structural information (e.g. gene-gene connectivity) in SIDA via the use of the normalized Laplacian of a graph, thus encouraging selection of predictors that are connected and behave similarly. This enhances interpretability. Covariates, if available, can be included, via the option \texttt{WithCovariates$==$TRUE}.

\section {Visualizations}
Results from the supervised filtering approach could be visualized via volcano plots using the function \texttt{volcanoPlot()}. The filtered or original data could be visualized via uniform manifold approximation projection (UMAP)\citep{mcinnes2018umap} with the function \texttt{umapPlot()}. We provide the function \texttt{VarImportancePlot()} to visualize the weights (in absolute value) of the low-dimensional loadings from \texttt{cvselpscca()}, \texttt{cvSIDA()}, and \texttt{cvSIDANet()}. Since the low-dimensional loadings are standardized to have unit norm, a variable with larger weight contributes more to the association between the views (for the unsupervised integrative analysis methods) or to the association between the views and the discrimination of classes within each view (for the supervised integrative analysis methods). We provide the function \texttt{DiscriminantPlots()} and \texttt{CorrelationPlots()} to visualize the class separation within each view, and correlations between pairs of views, respectively. We provide the function \texttt{networkPlot()} for relevance network that shows variable-variable connections between pairs of views. We provide the function \texttt{LoadingsPlot()} to visualize the loadings  for each view
and to demonstrate the relationships between pairs of variables within each view. We provide the function \texttt{WithinViewBiplot()} to show the
scores and loadings together for a specific view. The function \texttt{BetweenViewBiplot()} shows the scores and loadings from pairs of views together.

\section {Use Case}
\subsection{Demonstration of SELPCCA and SELP Predict}
In this Subsection, we demonstrate the use of SELPCCA on multiomics data pertaining to COVID-19. Our goal is to associate Proteomics and RNASeq data to identify proteins and genes driving the overall dependency structure between the two molecular data. We then associate the canonical variates with COVID-19 status  in a logistic regression model to investigate whether the canonical variates are able to discriminate between individuals with and without COVID-19.

We load the data in as \texttt{data(COVIData)} [Figure S3]. The number of cases (COVID-19) and non-cases (non-COIVD-19) is 98 and 22, respectively. There are 264 proteins and 5,800 genes. In this analysis, View 1 corresponds to the proteomis data, and View 2, the RNASeq data. We subset the data into 90\% training, and 10\% testing, keeping the proportion of cases and non-cases similar to the proportion in the whole data [Figure S4]. We filter data using the function \texttt{filter.supervised()} [Figure S5] with the options: `method'$=$ `logistic'; `padjust' $=$TRUE; `adjmethod'$=$ BH; `standardize' $=$ TRUE. Our outcome is disease status (`DiseaseStatus.Indicator').  After univariate filtering by removing proteins and genes that are not statistically significant (adjusted p-value $>$ 0.05), 87 proteins and 2,573 genes remain. We use the function \texttt{volcano()} to obtain volcano plots for proteins and genes [Figure S6]. We use the function \texttt{umapPlot()} to obtain UMAP plots of the filtered data to visualize how well the samples are separated [Figure S7]. 

To fit SELPCCA, we invoke the function \texttt{cvselpscca()} and set the number of canonical variates to 2 [Figure S8]. From running SELPCCA, we observed that 78 proteins and 32 genes have nonzero coefficients on the first CCA vector, which suggests that these proteins and genes maximize correlation between the proteomics and RNASeq data (estimated correlation is 0.636). Further, 54 proteins and 9 genes have nonzero coefficients on the second CCA vector, with estimated correlation 0.599. The top 20 proteins (shown as Uniprot IDs, UID) and genes with highest absolute loadings for the first CCA vector are shown in Figure S9. These figures are obtained with the function \texttt{VarImportancePlot()}. Some of the highly ranked proteins for the first CCA vector include: Immunoglobulin lambda variable 3-1 (UID P01715), HLA class I histocompatibility antigen, B alpha chain (UID P30491), Alpha-2-HS-glycoprotein (UID P02765). Some of the highly ranked genes on the first CCA vector include: ubiquitin conjugating enzyme E2 C (UBE2C), cell division cycle 6 (CDC6), cyclin A2 (CCNA2), and DEP domain containing 1B (DEPDC1B). We note that some of the highly-ranked proteins (e.g. HLA class I histocompatibility antigen, B alpha chain [UID P30491] and Surfactant, pulmonary-associated protein B, isoform CRA\_a [UID D6W5L6]) and genes (e.g. CDC6  and CCNA2) each had high log-odds ratios for discriminating between COVID-19 cases and non-cases.  

We observe that the first canonical variate for each view is able to separate those with and without COVID-19 in the train (Figure S10) and test (Figure S11) sets. We use the function \texttt{WithinViewBiplot()} to visualize  the sample discrimination, the canonical loadings for each view,  and assess how the top variables in each view are related to each other (Figure S12). The protein Immunoglobulin lambda 4-69 (UID A0A075B6H9) appears to be highly correlated with the protein Ferritin light chain (UID P02792). The gene UBE2C is loaded on the first CCA vector, and the genes UPK3A and GPR35 are loaded on the second CCA vector. We use between-view biplots to visualize biplots for both views (Figure S13). This plot allows us to assess how genes and proteins are related. Solid black vectors represent loading plots for the first view (proteins). Dashed red vectors represent loadings plot for the second view (genes).  We generate this plot with the function \texttt{BetweenViewBiplot()}. The protein  Immunoglobulin lambda variable 3-1 (UID P01715) appears to be positively correlated with the gene UBE2C. 

\begin{figure*}
    \centering
    \includegraphics[width=7.5in]{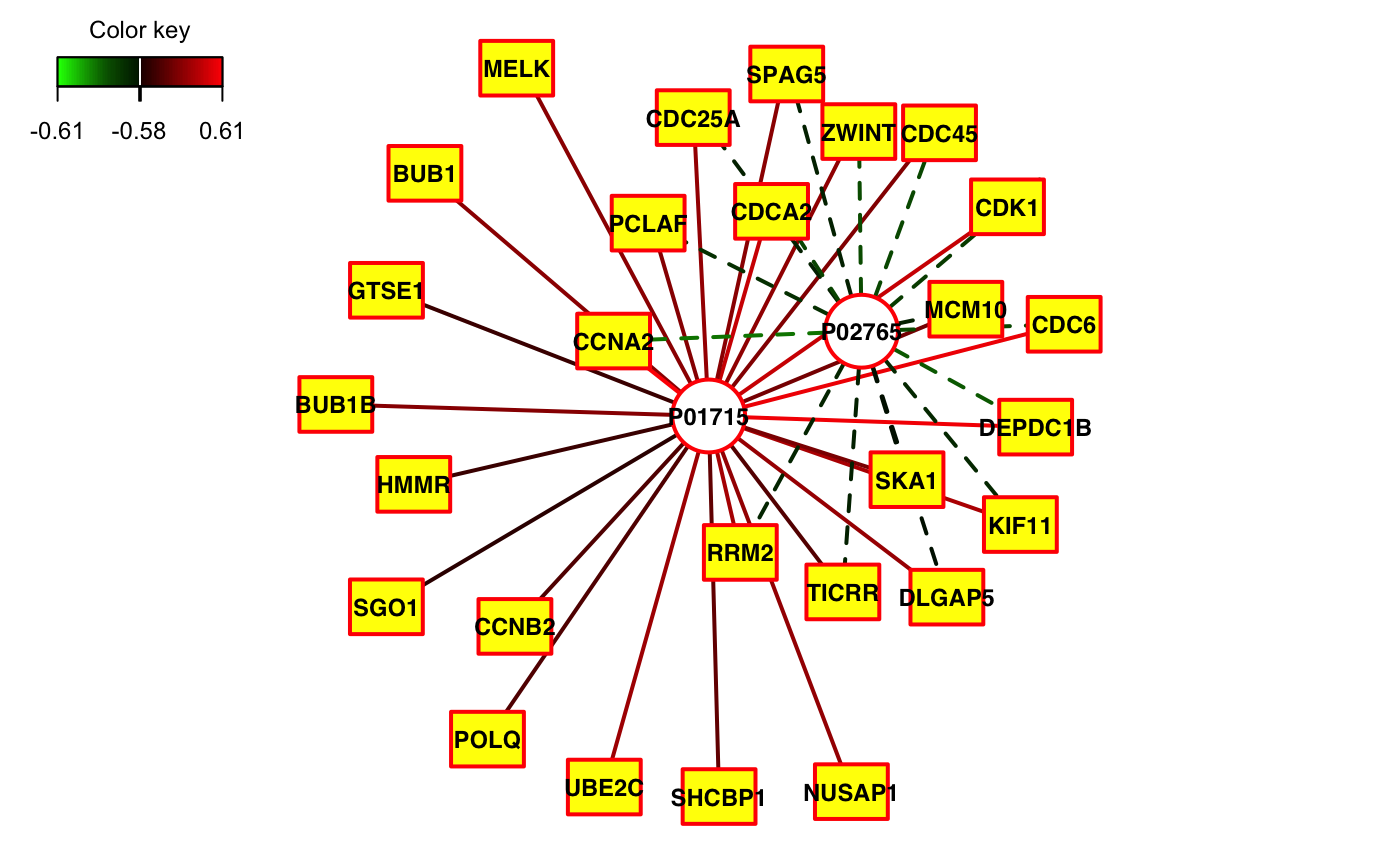} 
    \caption{Relevance network plot for SELPCCA. The nodes of the graph represent variables for the pairs of views, and edges represent the correlations between pairs of variables. Dashed and solid lines indicate negative and positive correlations, respectively. Circle nodes are View 1 variables (proteins), and rectangular nodes are View 2 variables (genes). We show edges with correlations at least 0.58.  The plot suggest that the protein P01715 is highly positively correlated with many genes, and the protein P02765 is highly negatively correlated with many genes.  }
    \label{fig:rnselp}
\end{figure*}

For more insight into the association between the genes and proteins, we invoke the relevance network plot function \texttt{networkPlot()} [Figures \ref{fig:rnselp} and S14]. The nodes of the graph represent variables for the pairs of views, and edges represent the correlations between pairs of variables. Dashed and solid lines indicate negative and positive correlations, respectively. Circle nodes are View 1 variables (proteins), and rectangular nodes are View 2 variables (genes). We show edges with correlations at least 0.58.  The plot suggests that the protein Immunoglobulin lambda variable 3-1 (UID P01715) is highly positively correlated with many genes (including CDC6, CCNA2, UBE2C), and the protein Alpha-2-HS-glycoprotein (UID P02765) is highly negatively correlated with many genes (including CCNA2 and CDC6). 

In terms of prediction, we fitted a logistic regression model on the training data with the predictors as the first two canonical variates. We used the function \texttt{selpscca.pred()} for this purpose [Figure S15]. Our results suggest that the first canonical variates for proteins and genes are significantly associated with COVID-19 status (p-value $<$ 0.05).  We predicted the test data from the learned model with the \texttt{predict()} function, and obtained train [Figure S16] and test [Figure S17] prediction estimates (e.g. accuracy, sensitivity, F1 etc) with the \texttt{PerformanceMetrics()} function. In Figures S15 and S16, we observe that both train and test accuracy and F1 score are high, suggesting that the first two canonical variates potentially discriminate those with and without COVID-19.

\subsection{Demonstration of SIDA}
Unlike SELPCCA which is an unsupervised method for integrating data from multiple sources, SIDA is a supervised data integration method.  We demonstrate the use of SIDA on multiomics data pertaining to COVID-19. Our goal is to associate the proteomics and RNASeq data and discriminate between COVID-status in a joint model. We further identify proteins and genes that maximize both association and discrimination. We apply the function \texttt{cvSIDA()} [Figure S18] to obtain estimated SIDA discriminant vectors, correlation coefficients, and variables potentially contributing to the association of the views and the discrimination between samples within each view. 

From implementing SIDA, we observed that 26 proteins and 23 genes have nonzero coefficients, which suggests that these proteins and genes maximize both correlation between the proteomics and RNASeq data (estimated correlation from train data is 0.42) as well as separation between those with and without COVID-19. The top 20 proteins (shown as Uniprot IDs, UID) and genes with highest absolute loadings are shown in Figure S19.  Some of the highly ranked proteins include:  (UID P04196),  (UID P14543), (UID E9PEK4). Some of the highly ranked genes include: (GOLGA8Q),  (ADGB), (TNFRSF6B), and  (SLC25A41). 

 \begin{figure*}
    \centering
    \includegraphics[width=7.5in]{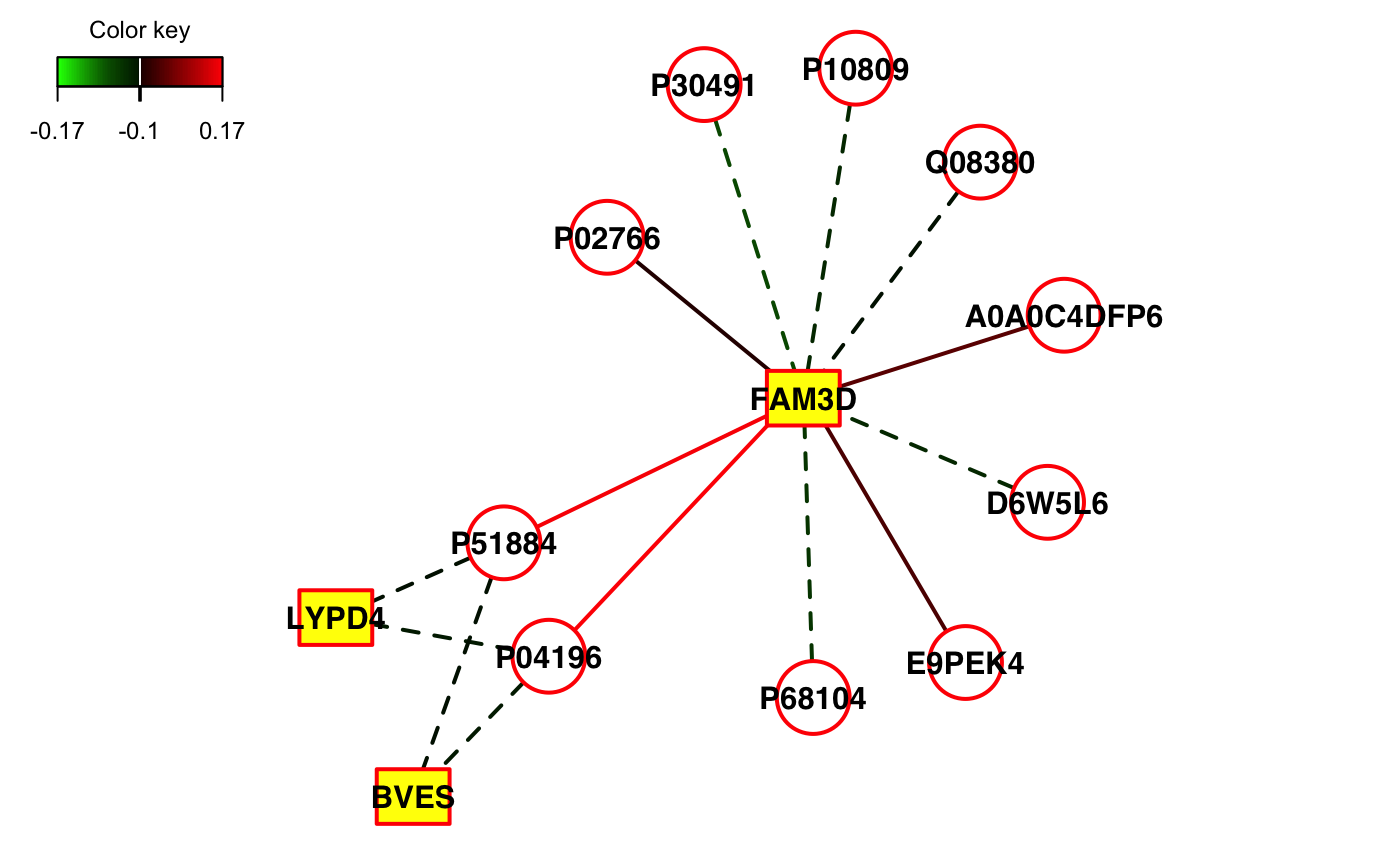} 
    \caption{Relevance network plot for SIDA. The nodes of the graph represent variables for the pairs of views, and edges represent the correlations between pairs of variables. Dashed and solid lines indicate negative and positive correlations, respectively. Circle nodes are View 1 variables (proteins), and rectangular nodes are View 2 variables (genes). We show edges with correlations at least 0.1.  The plot suggest that the gene FAM3D is negatively  correlated with many proteins (e.g. PO2766, P30491, Q08380), and positively correlated with proteins such as A0ADC4DFP6, E9PEK4, P04196, D6W5L6.  }
    \label{fig:rnsida}
\end{figure*}

We use the function \texttt{DiscriminantPlots()} [Figure S20] to visualize the separation of COVID-19 cases and non-cases. From Figures S20 and S21, the classes are well-separated in both the training (Figure S20) and testing sets (Figure S21). We use the function \texttt{CorrelationPlots()}  to visualize the strength of the association between the proteins and genes and separation as well. From Figure S22, we notice that the views are moderately correlated, and the classes are well-separated. For more insight into the association between the genes and proteins, we invoke the relevance network plot function \texttt{networkPlot()} [Figure S23]. The plot suggest that the gene FAM3D is negatively  correlated with many proteins (e.g. PO2766, P30491, Q08380), and positively correlated with proteins that include A0ADC4DFP6, E9PEK4, P04196, D6W5L6.  

In terms of prediction, we obtain the train and test error upon running \texttt{cvSIDA} [Figure S18]. We used the function \texttt{PerformanceMetrics()} to obtain other performance metrics. In Figures S24 and S25, we observe that both train and test performance metrics are high, suggesting that SIDA discriminant scores are able to discriminate those with and without COVID-19. The estimated train correlation is 0.41. Further, the performance metrics, especially test performance metrics, are better for SIDA than SELPCCA, which suggests that in this application, joint modeling of association and separation is better.

\section{Discussion and Future Work}
We have introduced an R package, \texttt{mvlearnR} for integrating data form multiple sources. The package wraps statistical and machine learning methods and graphical tools, providing a convenient and easy data integration workflow. For users with limited
programming language, we provide a Shiny Application
to facilitate data integration. Our multiview dashboard will enable easy, user-friendly comprehensive integrative analysis of molecular and clinical data from anywhere and on any device, without needing to know the R language.  We offer a friendly web user interface using the R Shiny framework where users can integrate multiple datasets, visualize and download results in easy to use format. Currently, linear multivariate methods for integrative analysis and biomarker identification are provided in \texttt{mvlearnR}, and the methods can only be used for integrating cross-sectional data.  However, we have developed integrative analysis methods for disease subtyping \citep{zhang2022robust} and for biomarker identification where we model nonlinear relationships between data from multiple sources and a clinical outcome  \cite{wang2021deep, iDeepViewLearn:2021, safo2023scalable}. These methods, and other methods we develop in the future,  will eventually be added to \texttt{mvlearnR} and  the accompanying web application.  Thus, we envision \texttt{mvlearnR} and our web application to be a one-stop place for comprehensive data integration, for both users of R (or Python) and non-users of these software.

\section{Acknowledgements and Funding}
This work was supported by the Award Number 1R35GM142695 of the National Institute of General Medical Sciences of the National Institutes of Health. The content is solely the responsibility of the authors and does not represent the official views of the National Institutes of Health.
\section*{Data Availability Statement}
The COVID-19 molecular and clinical data used to demonstrate our integration workflow were obtained from \url{https://doi.org/10.1016/j.cels.2020.10.003}. Pre-processed data follow  approach described in \url{https://doi.org/10.1371/journal.pone.0267047}.

\bibliographystyle{unsrtnat}
\bibliography{references.bib}

\end{document}

\maketitle
The package \texttt{mvlearnR} and accompanying Shiny App is intended for integrating data from multiple sources
(e.g. genomics, proteomics, metabolomics). Most existing software packages for multiview learning are decentralized,
making it difficult for users to perform comprehensive integrative analysis. The new package wraps statistical and machine
learning methods and graphical tools, providing a convenient and easy data integration workflow. For users with limited
programming language, we provide a Shiny Application to facilitate data integration. The methods have potential to offer
deeper insights into complex disease mechanisms. Currently, \texttt{mvlearnR} can be used for the following:
\begin{itemize}
\item Prefiltering of each omics data via differential analysis (DA). We provide both supervised and unsupervised options for DA or for filtering out noise variables prior to performing data integration. 
\item Integrating data from two sources using a variant of the popular unsupervised method for associating data from two views, i.e. canonical correlation analysis (CCA). 
\item Predicting a clinical outcome using results from CCA. We provide four outcome data distribution type (i.e. gaussian, binomial, Poisson, and time-to-event data.)
\item Supervised integrative analysis (one-step) for jointly associating data from two or more sources and classifying an outcome. This method allows to include covariates. 
\item Supervised integrative analysis (one-step) for jointly associating structured data from two or more sources and classifying an outcome. This method allows to include covariates. 
\item Visualizatiing results from the DA or our integrative analysis methods. These plots include: volcano plots, UMAP plots, variable importance plots, discriminant plots, correlation plots, relevance network plots, loadings plots, and within- and between- view biplots. These visualization tools will help unravel complex relationships in the data. 
\item Demonstrating our integration workflow via already uploaded synthetic and real molecular and clinical data pertaining to COVID-19.
\end{itemize}
Currently, linear multivariate methods for integrative analysis and biomarker identification are provided in \texttt{mvlearnR}. However, we have developed integrative analysis methods for disease subtyping [1] and nonlinear integrative analysis methods for biomarker identification [2-4] that will eventually be added to \texttt{mvlearnR} and  the accompanying web application. Other methods we develop in the future will be added. Thus, we envision \texttt{mvlearnR} and our web application to be a one-stop place for comprehensive data integration, for both users of R (or Python) and non-users of these software. 

\section{The Methods}
 The methods SELPCCA and SIDA in \texttt{mvlearnR} have been described in [5] and [6], respectively. For completeness sake, we describe these methods below. Since these methods  are extensions of canonical correlation analysis (CCA) [7] and linear discriminant analysis (LDA) [7], we briefly describe these two methods. 
\subsection{Linear Discriminant Analysis}
Suppose we have one set of  data $\bX$ with $n$ samples on the rows and $p$ variables on the columns.  Assume the $n$ samples are partitioned into $K$ classes, with $n_k$ being the number of samples in  class $k, k=1, \ldots,K$. Assume each sample belongs to one of the $K$ classes and let $y_i$ be the class membership for the $i$th subject.  Let ${\bX_{k}}=(\bx_{1k},\ldots,\bx_{n_{k},k})^{\smt} \in \Re^{n_k \times p}$
, $\bx_{k} \in \Re^{p}$ be the data matrix for class $k$. Given these available data, we wish to predict the class membership $y_{j}$ of a  new subject $j$  using their high-dimensional information $\bz_{j} \in \Re^{p}$. Fishers linear discriminant
analysis (LDA) may be used to predict the class membership. For a $K$ class prediction problem, LDA finds $K-1$ low-dimensional vectors (also called discriminant vectors), which are linear combinations of all available variables, such that projected data onto these discriminant vectors have maximal separation between the classes and minimal separation within the classes.  The within-class and between-class separation are defined respectively as:  $ \bSw = \sum\limits_{k=1}^{K}\sum\limits_{i=1}^{n}(\bx_{ik}-\hat{\bmu}_{k})(\bx_{ik}-\hat{\bmu}_{k})^{\smt};  ~~~~\bSb = \sum\limits_{k=1}^{K}n_{k}(\hat{\bmu}_{k}-\hat{\bmu})(\hat{\bmu}_{k}-\hat{\bmu})^{\smt}.$
Here, $\hat{\bmu}_{k} = (1/n_{k})\sum_{i=1}^{n_{k}}\bx_{ik}$ is the mean vector for class $k$, $\hat{\bmu}$ is the combined class mean vector and is defined as $\hbmu=(1/n)\sum\limits_{k=1}^{K}n_{k}\hbmu_{k}$. The solution to the LDA problem is given are the eigenvalue-eigenvector pairs $(\widehat{\lambda}_{k},\widehat{\bbeta}_{k})$, $\widehat{\lambda}_{1}>\cdots> \widehat{\lambda}_{k}$ of $\bSw^{-1}\bSb$ for $\bSw \succ 0$.  Note that LDA is only applicable to one set of data.

\subsection{Canonical Correlation Analysis}
 Canonical correlation analysis on the other hand is applicable when there are two views. Now, suppose we have two sets of data. $\bX^{1} =(\bx^1_{1},\cdots,\bx^1_{n})^{\smt} \in \Re^{n \times p}$ and  $\bX^{2} = (\bx^2_{1},\cdots,\bx^2_{n})^{\smt} \in \Re^{n \times q}$, $p,q > n$, all measured on the same set of $n$ subjects. The goal of CCA [7]is to find linear combinations of the variables in $\bX^{1}$, say $\mathbf{u}=\bX^{1} \balpha$ and in $\bX^{2}$, say $ \mathbf{v}=\bX^{2} \bbeta$, such that the correlation between these linear combinations is maximized, i.e $\rho=\max cor(\mathbf{u}, \mathbf{v})= \max cor(\bX^{1} \balpha,\bX^{2} \bbeta)$. But 
 \[ \max_{\balpha,\bbeta} cor(\bX^{1} \balpha,\bX^{2} \bbeta) = \frac{\balpha^{\smt} \bSigma_{12} \bbeta }{ \sqrt{\balpha^{\smt} \bSigma_{11} \balpha}  \sqrt{\bbeta^{\smt} \bSigma_{22} \bbeta} }, \]
 where $\bSigma_{12}$ is  the $p \times q$ population covariance between $\bX^1$ and $\bX^2$, and  $\bSigma_{11}, and \bSigma_{22}$ are the population covariances of $\bX^1$ and $\bX^2$, respectively. The population cross-covariances  $\bSigma_{12}$, $\bSigma_{11}$, and $\bSigma_{22}$ are estimated by their sample versions $\bS_{12}$, $\bS_{11}$, and $\bS_{22}$, respectively. For low-dimensional settings, these sample versions are consistent estimators of the population covariances, for fixed $p$ and $q$, and large $n$. However, for high-dimensional settings, these sample versions are regularized.   
 The vectors $\mathbf{u}$ and $\mathbf{v}$ are called the first canonical variates for $\bX^1$ and $\bX^2$, respectively.  We note that the  correlation coefficient is invariant to scaling, thus one can choose the denominator be one and then consider the equivalent problem:
 \[ \max_{\balpha,\bbeta} cor(\bX^{1} \balpha,\bX^{2} \bbeta) = \balpha^{\smt} \bSigma_{12} \bbeta~~ \mbox{s.t}~~ \balpha^{\smt} \bSigma_{11} \balpha=1,~~~  \bbeta^{\smt} \bSigma_{22} \bbeta=1.  \]
 Using Langragian multipliers, it can be shown that the CCA solution to this optimizaiton problem is given by $\widehat{\balpha}=\bS_{1}^{-1/2}\bbe_{1}, \widehat{\bbeta}=\bS_{2}^{-1/2}\bbf_{1}$, where $\bbe_{1}$ and $\bbf_{1}$ are the first left and right singular vectors of $\bS_{1}^{-1/2}\bS_{12}\bS_{2}^{-1/2}$ respectively. 
 We note that maximizing the correlation is equivalent to maximizing the square of the correlation. To obtain subsequent canonical variates, one can impose the following orthogonality constraints in the optimization problem: $\balpha^{\smt}_{i} \bSigma_{11}\balpha_{j}=\bbeta^{\smt}_{i} \bSigma_{22}\bbeta_{j} =\balpha^{\smt}_{i} \bSigma_{12}\bbeta_{j} =0$, for $i\ne j$, $i,j=\min(p,q,n)$.

\subsection{Sparse CCA via SELP (SELPCCA)}
The classical CCA method finds linear combinations of all available variables, and since these weights are typically nonzero, it is difficult to interpret the findings.  SELPCCA [5] is a variant of CCA that shrinks some of the weights of the low-dimensional representations $\balpha$ and $\bbeta$ to zero, thus allowing to identify relevant variables contributing to the overall dependency structure in the data. In particular, the authors in [5] proposed to solve the following optimization problem: 
\begin{eqnarray}
\min_{\balpha} \|\balpha\|_1~~~~ s.t~~~~~ \|\bS_{12}\tilde{\bbeta}_1 - \tilde{\rho}_1\tilde{\bS}_{11}\balpha        \|_{\infty}  \le \tau_1 
\nonumber\\
\min_{\bbeta} \|\bbeta\|_1~~~~ s.t~~~~~ \|\bS_{21}\tilde{\balpha}_1 - \tilde{\rho}_1\tilde{\bS}_{22}\bbeta        \|_{\infty}  \le \tau_2.
\nonumber\\
\end{eqnarray}
Here, $(\tilde{\balpha}_1, \tilde{\bbeta}_1)$ are the first nonsparse solution to the CCA optimization problem, and $\tilde(\rho)_{1}$ is the corresponding eigenvalue. Of note, $\tilde{\bS}_{11}$ and $\tilde{\bS}_{22}$ are regularized versions of ${\bS}_{11}$ and $\tilde{\bS}_{22}$, respectively. The authors considered two forms of regularizations: $\tilde{\bS}_{11} = \bI_p$ and $\tilde{\bS}_{22} = \bI_q$, for standardized data,  and the ridge regularization: $\tilde{\bS}_{11} = \bS_{11} + \sqrt{\log(p)/n}\bI_p$ and $\tilde{\bS}_{22} = \bS_{22} + \sqrt{\log(q)/n} \bI_q$. Also, $\tau_1$ and $\tau_2$ are tuning parameters which are chosen by cross-validation. See [5] for more details. For subsequent canonical directions $\balpha_i$, and $\bbeta_i$, $i>1$, the authors solved the above sparse optimization problem on deflated data. 

\subsection{Prediction with SELPCCA}
CCA and SELPCCA are unsupervised multivariate methods, with their main goal of finding canonical vectors that maximize the overall dependency structure between two views. In some biomedical applications, it is usually the case that an outcome variable, say $\mathbf{y}$, is available and a specific interest might be to investigate how these canonical variates are related to the outcome of interest. For this purpose, one can build regression models to associate the outcome with these canonical variates. In our package, we provide \texttt{selpPredict()} for this purpose. We provide options gaussian, binomial, poisson, and survival to model continuous, categorical, count, or time-to-event data, respectively. We also provide the function \texttt{predict()} to predict
out of sample data using the learned low-dimensional
representations or canonical variates. 

\subsection{Sparse Integrative Discriminant Analysis (SIDA) for two or more views }
Instead of considering the two-step CCA approach proposed above when there is a clinical outcome in addition to the views, the authors in [6] developed the method, SIDA, for jointly maximizing association between two or more views while also maximizing separation between classes in each view. Although CCA is applicable to only two views, SIDA is applicable to two or more views. SIDA
 combines the advantages of  LDA, a supervised learning method for maximizing separation between classes in a view, and CCA, an unsupervised learning method for maximizing correlation between two data types. 
Suppose there are $d=1,\ldots,D$ views. Let $\bX^d=[\bX^d_{1}, \bX^d_{2},\ldots,\bX^{d}_{K}]$, $\bX^{d}\in \Re^{n\times p_d}, \bX^{d}_{k} \in \Re^{n_k \times p_d}, k=1,\ldots,K,~ d=1,2,\ldots,D$ be a concatenation of the $K$ classes in the $d$-th view. Let $\bSb^{d}$ and $\bSw^d$ be the between-class and within-class covariances for the $d$-th view. Let  $\bS_{dj}, j<d$ be the cross-covariance between the $d$-th and $j$-th views. Define $\mathcal{M}^d=\bS_w^{d^{-1/2}}\bS^d_b\bS_w^{d^{-1/2}}$ and 
$\mathcal{N}_{dj}=\bS_w^{d^{-1/2}}\bS_{dj}\bS_w^{j^{-1/2}}$. The authors solved the following optimization problem for the basis discriminant vectors $\bGamma^d$: 
%\begin{small}
\begin{eqnarray*} \label{eqn:intequiv2}
	\max_{\bGamma^1,\cdots,\bGamma^D}  \rho\sum_{d=1}^D\text{tr}(\bGamma^{{d\smt}}\mathcal{M}^d \bGamma^d) 
	+ \frac{2(1-\rho)}{D(D-1)}\sum_{d=1,d\ne j }^{D}\text{tr}(\bGamma^{{d\smt}}\mathcal{N}_{dj}\bGamma^j\bGamma^{{j\smt}}\mathcal{N}_{jd}\bGamma^d) \\
	~\mbox{s.t}~\text{tr}(\bGamma^{{d\smt}}\bGamma^d)=K-1.
\end{eqnarray*} 
Of note, $\rho$ controls the influence of separation or association in the optimization problem.  The second term sums the unique pairwise squared correlations and weights them by $\frac{D(D-1)}{2}$ so that the sum of the squared correlations  is one.  The authors showed that the nonsparse basis discriminant directions for the $d$-th view, $\widetilde{\bGamma}^d$, that maximize both associations and separations are given by the eigenvectors corresponding to the eigenvalues ($\bLambda^d$) that iteratively solve the following eigensystems:
%\begin{small}
\begin{eqnarray}\label{eqn:GEVitermultiple}
\left({c}_1\mathcal{M}^{1} + {c}_1\mathcal{M}^{1^{\smt}}+ c_2\overline{\mathcal{N}}_{1j} + c_2\overline{\mathcal{N}}_{1j}^{{\smt}}\right)\bGamma^1&=&\bLambda_{1}\bGamma^1, \nonumber\\
\vdots\nonumber\\
\left({c_1}\mathcal{M}^{D} +{c}_1\mathcal{M}^{1^{\smt}} + c_2\overline{\mathcal{N}}_{Dj} + c_2\overline{\mathcal{N}}_{Dj}^{{\smt}}\right)\bGamma^D&=&\bLambda_{D}\bGamma^D,
\end{eqnarray} 
%\end{small}
where ${c_1}=\rho$ and ${c_2}=\frac{2(1-\rho)}{D(D-1)}$, and $\overline{\mathcal{N}}_{dj}=\sum_{d,j }^{D}\mathcal{N}_{dj}\bGamma^{j}\bGamma^{j^{\smt}}\mathcal{N}_{jd}$, $d,j=1,\ldots D, j\ne d$ sums all unique pairwise correlations of the $d$-th and the $j$-th views.

For sparsity or smoothness, they considered the following optimization problems:
\begin{small}
\begin{eqnarray}\label{eqn:joint2}
\min_{\bGamma^1} \mathcal{P}(\bGamma^1) \quad &\mbox{s.t}& \quad \|(c_1\mathcal{M}^1 + c_1\mathcal{M}^{1^{\smt}} + c_2\overline{\mathcal{N}}_{1j} + c_2\overline{\mathcal{N}}_{1j}^{{\smt}} )\widetilde{\bGamma}^1- \widetilde{\bLambda}_{1}\bGamma^1 \|_\infty \leq\tau_{1}\nonumber\\
\vdots\nonumber\\
\min_{\bGamma^D} \mathcal{P}(\bGamma^D) \quad &\mbox{s.t}& \quad \|(c_1\mathcal{M}^D + c_1\mathcal{M}^{D^{\smt}} + c_2\overline{\mathcal{N}}_{Dj} + c_2\overline{\mathcal{N}}_{Dj}^{{\smt}} )\widetilde{\bGamma}^D - \widetilde{\bLambda}_D\bGamma^D\|_\infty \leq\tau_{D}. 
\end{eqnarray}
\end{small}
The authors considered two forms of penalty term $\mathcal{P}(\bGamma^d)$, depending on whether sparsity only (data-driven) or sparsity and smoothness (knowledge-driven) is desired: $\mathcal{P}(\bGamma^d)= \sum_{i=1}^{p_d}\|\bgamma_{i}^d\|_{2},$ or $\mathcal{P}(\bGamma^d)=\eta\sum_{i=1}^{p_d}\|\bgamma^{\mathcal{L}_n}_{i}\|_{2} + (1-\eta) \sum_{i=1}^{p_d}\|\bgamma_{i}\|_{2}$. In the latter, 
$\bgamma^{\mathcal{L}_{n}}_{i}$ is the $i$-th row of the matrix product $\mathcal{L}_n\bGamma^d$, and $\mathcal{L}_n$, is the normalized Laplacian of a graph, which is view-dependent. Essentially, the first term in this penalty acts as a smoothing operator for the weight matrices $\bGamma^d$ so that variables that are connected within the $d$-th view are encouraged to be selected or neglected together. This was termed SIDANet (SIDA for structured or network data). SIDANet is applicable when there exists prior biological information in the form of variable-variable connections, which guides the detection of connected variables that maximize both separation and association. 

\section{Visualizations of discriminant scores and canonical correlation variates}
\begin{figure}[!h]
\begin{tabular}{ccc}
         \centering
         \includegraphics[width=0.32\textwidth]{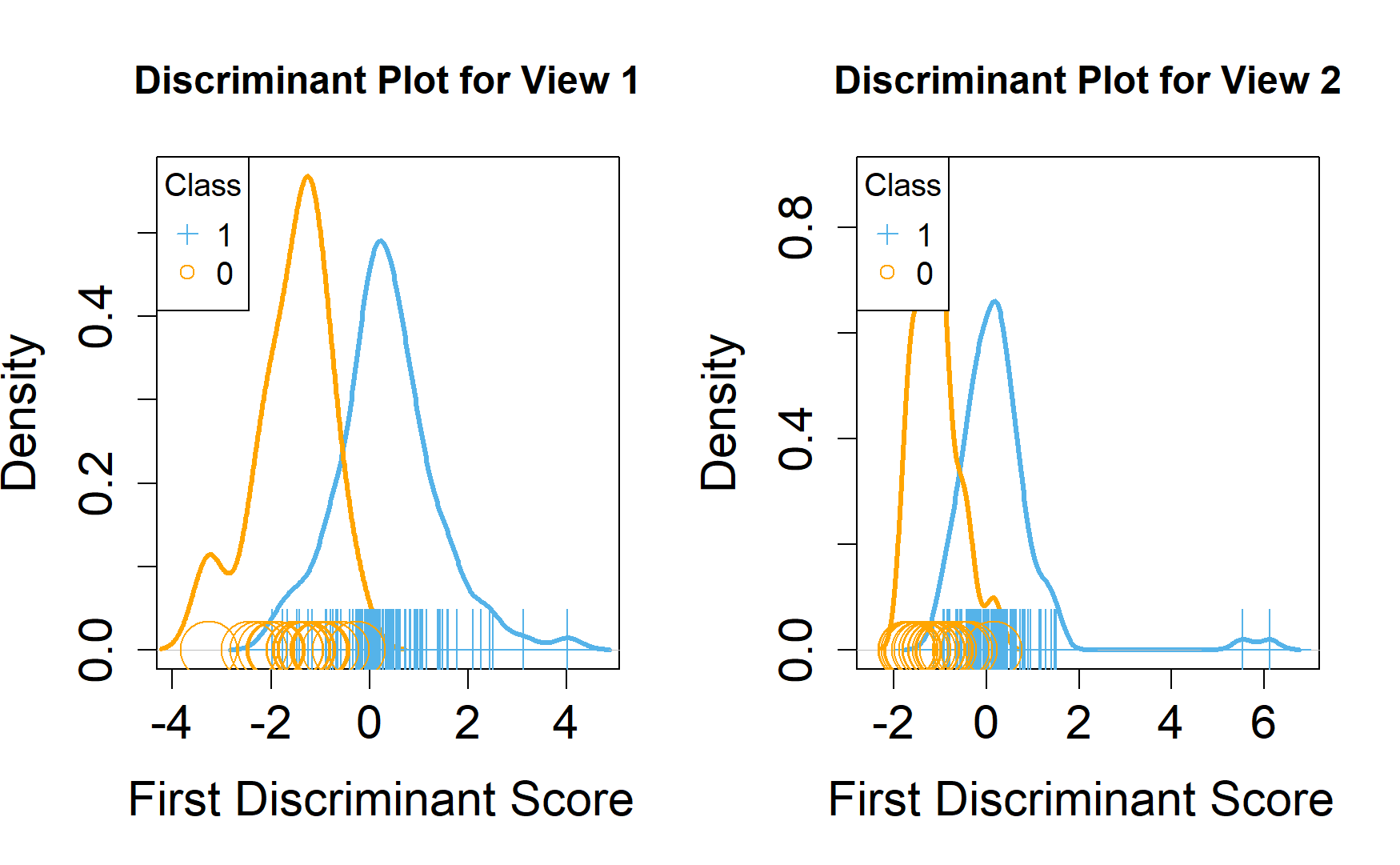}&\includegraphics[width=0.32\textwidth]{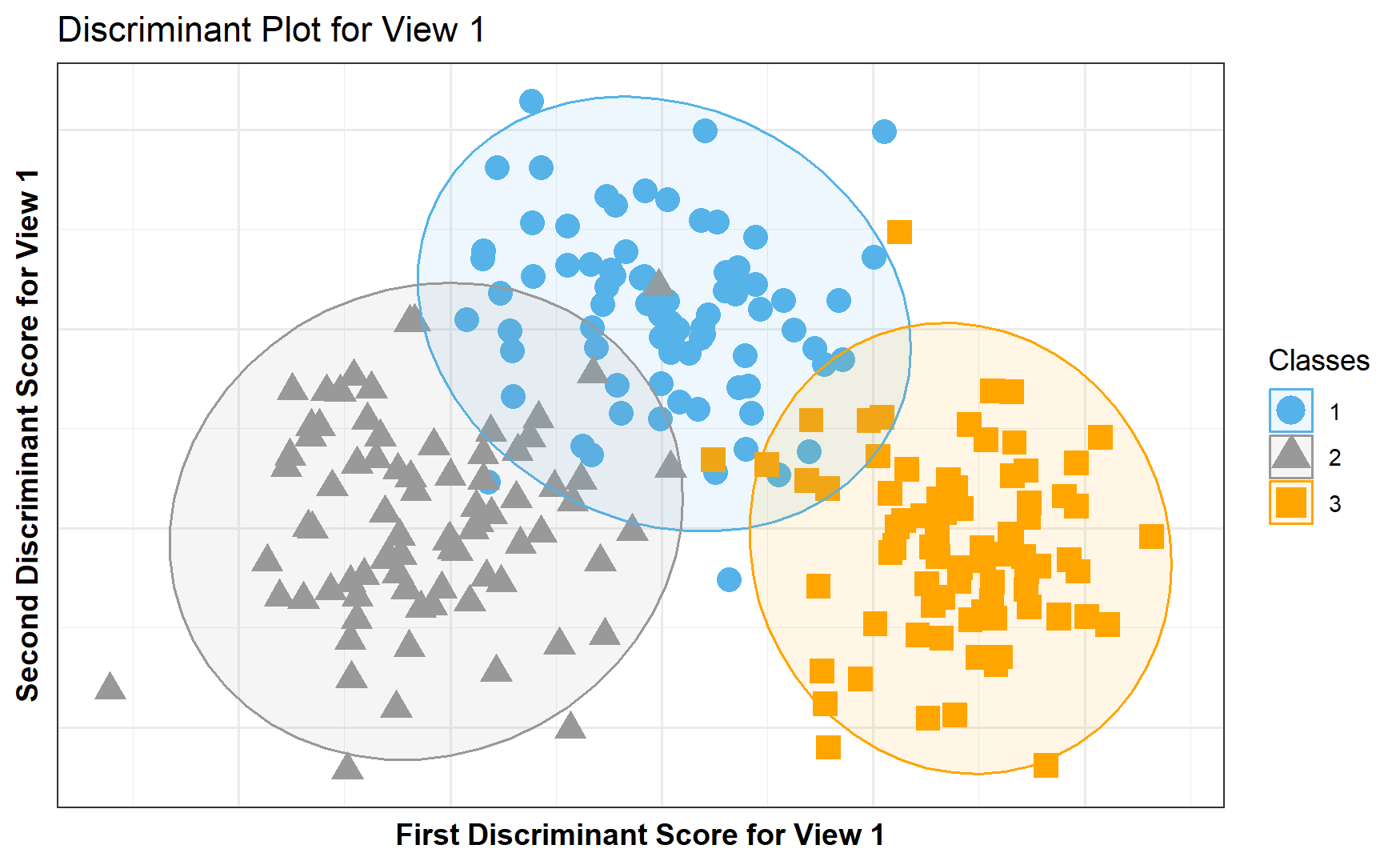} &\includegraphics[width=0.32\textwidth]{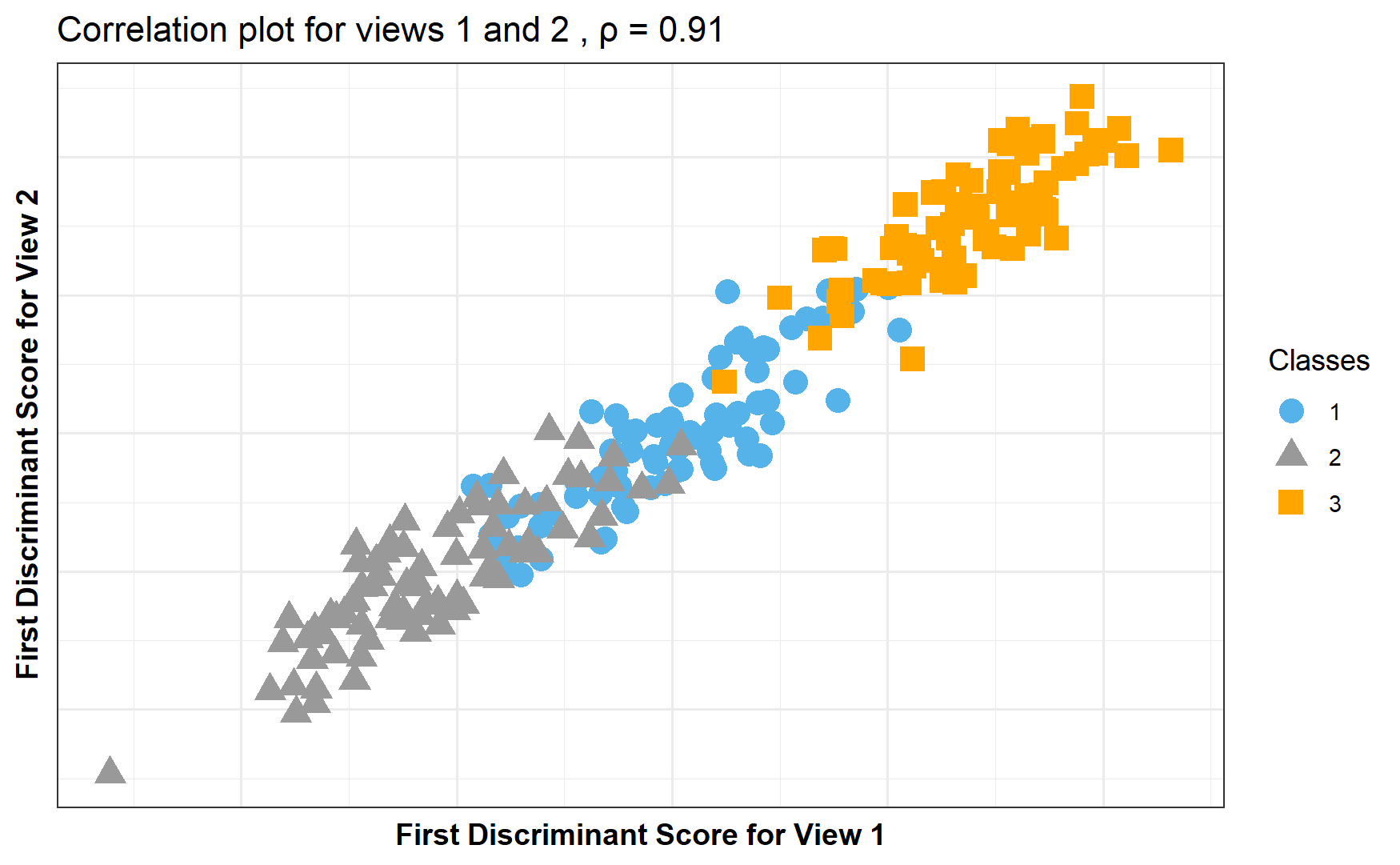}\\
\end{tabular}
    \caption{Sample discriminant plots for two classes (left panel), for three classes (middle panel) and correlation plot (right panel). Discriminant plots can demonstrate whether the discriminant vectors separate the classes. Correlation plots demonstrate the strength of association between pairs of views. It also shows how well the classes are separated. }
    \label{fig:corranddiscplots}
\end{figure}

Once the discriminant vectors or canonical correlation vectors have been estimated using SIDA/SIDANet or SELPCCA respectively, the scores for each view, representing projections of each view onto these vectors can be estimated. We provide \texttt{DiscriminantPlots()} and \texttt{CorrelationPlots()} to visualize these scores. The  \texttt{DiscriminantPlots()} function can be used to visualize the first (assuming only two classes) or the first  and second discriminant scores (assuming $>$ two classes). Each point on this plot represent a score for an individual. Discriminant plots can showcase how well the discriminant vectors separate the classes. Correlation plots on the other hand can be used to visualize the strength of association between pairs of views as well as the separation of the classes. Figure \ref{fig:corranddiscplots} demonstrates sample figures that can be generated by our \texttt{DiscriminantPlots()} and \texttt{CorrelationPlots()} in \texttt{mvlearnR}. We provide different color options for the graphs. 

\section{Visualizations of variables selected }
\subsection{Relevance Network}  
Relevance Network (RN) [8-9] have been proposed to visualize pairwise relationships of variables in integrative analysis. Here, the nodes represent variables, and edges represent variable associations. To construct RN, the correlation matrix of pairwise associations is obtained from the data. Then, for  a specific cutoff (say 0.7), if the estimated correlation coefficients is greater (in absolute value) than this cutoff, an edge is drawn between these two variables; otherwise, no edge is drawn and these two variables are deemed not associated for this threshold. Further, variables/nodes with no link are not represented in the RN. The cutoff is used to make the graph less dense, especially when the number of variables is high.  RN have potential to shed lignt into the complex associations between different types of data. We provide the function \texttt{networkPlot()} to visualize the pairwise relationships of variables selected by SELPCCA, SIDA, and SIDANet. We follow ideas [9] to construct the RN for SELPCCA, SIDA, and SIDANet. In particular, instead of computing the Pearson correlation coefficients between each pair of variables in each view separately, we construct  bipartite graph (or bigraph) where variables/nodes from view $\bX^d$ are connected to variables/nodes from view $\bX^{j}$, $d \ne j, d,j=1,\ldots,D$. Similar to [9] we construct the bipartite graphs using a pairwise similarity matrix directly obtained from the outputs of our integrative analysis methods. 

Suppose we have applied SELPCCA to two views and we have obtained the canonical loadings $\hat{\balpha}_l$ and $\hat{\bbeta}_l$, $l= 1,\ldots,L$, for views 1 and 2, respectively. For $l>2$, it is likely that within a view,  different variables will have nonzero coefficients for each canonical loading. Therefore, we deem a variable to be selected if it has a nonzero coefficient in at least one canonical loading. Let $\widehat{\bA}=(\hat{\balpha}_1^{\smt},\ldots,\hat{\balpha}_L^{\smt})^{\smt} \in \Re^{p \times L}$ be a concatenation of all $L$ canonical loadings for view 1. Let  $\widehat{\bB}=(\hat{\bbeta}_1^{\smt},\ldots,\hat{\bbeta}_L^{\smt})^{\smt} \in \Re^{q \times L}$ be defined similarly for view 2.  Given data with these selected variables, we construct the canonical variates: $\bU = \bX^1_{selected}\widehat{\bA}$ and $\bV= \bX^2_{selected}\widehat{\bB}$. Let $p'$ and $q'$ be the number of selected variables in views 1 and 2, respectively. 
To construct the $p' \times q'$ similarity matrix for views 1 and 2, we first project data for the selected variables to $\bU$, for view 1,  $\bV$ for view 2, or to a combination of $\bU$ and $\bV$. Since $\bX^1$ and $\bX^2$ are analyzed simultaneously in CCA, the projection of the data onto a combination of $\bU$ and $\bV$ seems natural. Thus, we define $\bZ=\bU + \bV$. As noted in [9], the $\bZ$ variables are closest to $\bX^{d}$ and $\bX^{j}$. Then, for the $i$th variable $X^{1}_i \in \bX^1_{selected}$ (similarly $X^{2}_i \in \bX^2_{selected}$)  and the $l$th component $\bz_l \in \bZ$, we compute the scalar inner product $x^{d}_{il} = \langle X^{d}_i, \bz_l \rangle, d=1,2; j=1\ldots p' \text{(or~ $q'$)},~l=1,\ldots,L$. Assuming that the variables $X^{d}_i$,  and $\bz_l$ are standardized to have variance 1, $x^{d}_{il} = \langle X^{d}_i, \bz_l \rangle = \text{cor}(X^{d}_i,\bz_l)$.  
We compute the similarity matrix $\bM$ between views $1$ and $2$ as $\bM=\bx^1\bx^{2'\smt} \in \Re^{p' \times q'}$,  where $\bx^1$ is a $p' \times L$ matrix with the $il$th entry $x^1_{il}$ and $\bx^2$ is a $q' \times L$ matrix with the $il$th entry $x^2_{il}$. Each entry in the similarity matrix is between $-1$ and $1$. 

We construct the similarity matrix for SIDA and SIDANet in a similar fashion. Of note, in SIDA and SIDANet, $l=K-1$. Although, the $l_{2,1}$ norm encourages the same variables to be selected across all $K-1$ components, empirically, this is not usually the case. Therefore, we obtain overall variable selection as described above. We obtain the discriminant vectors for the $d$th and $j$th views as $\bU = \bX^1_{selected}\widehat{\bGamma}^d$ and $\bV= \bX^2_{selected}\widehat{\bGamma}^j$,  respectively. Given $\bU$ and $\bV$, we compute the similarity matrix for SIDA and SIDANet in a similar fashion. 

\begin{figure}[!h]
\begin{tabular}{cc}
         \centering
         \includegraphics[width=0.5\textwidth]{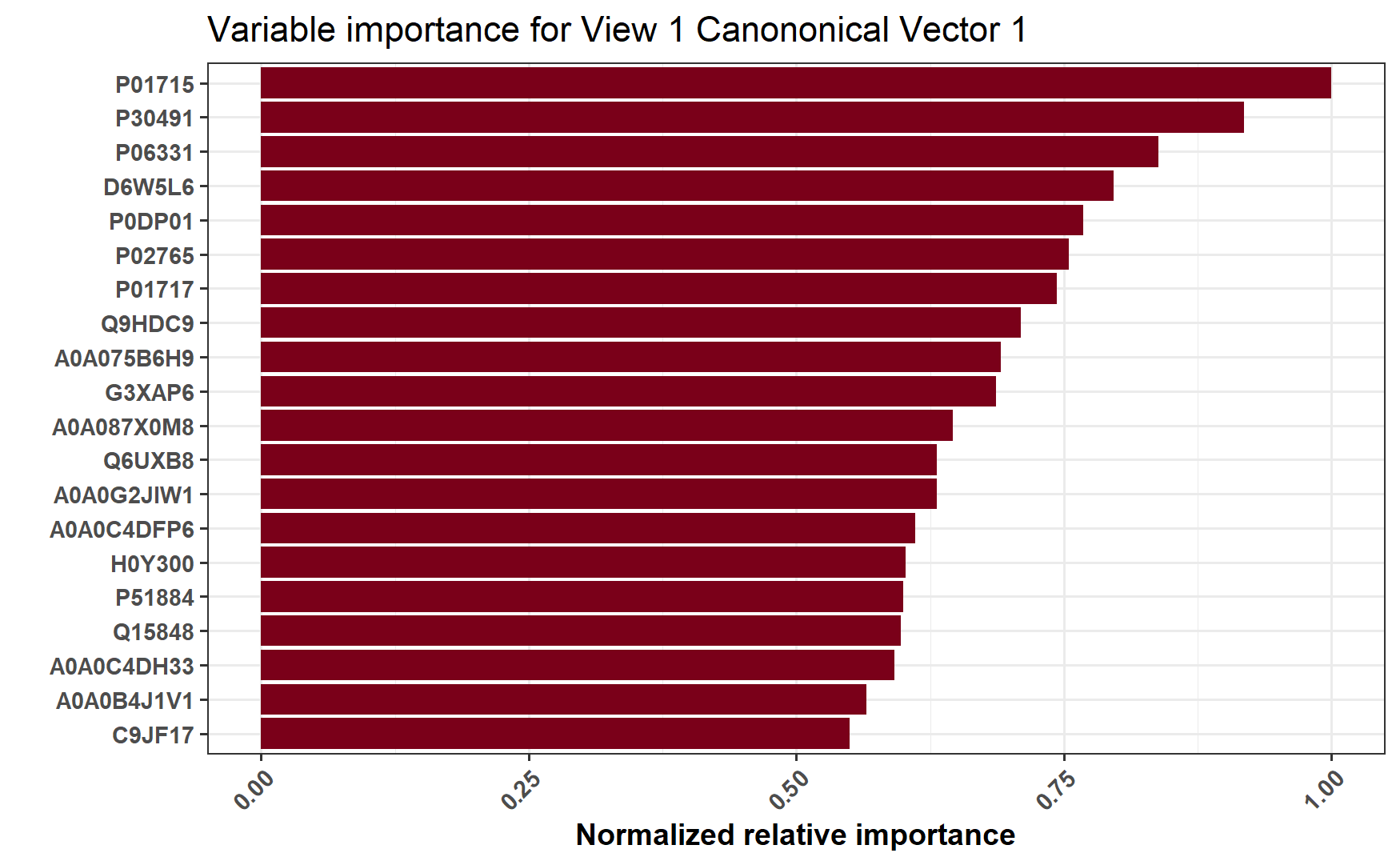}&\includegraphics[width=0.5\textwidth]{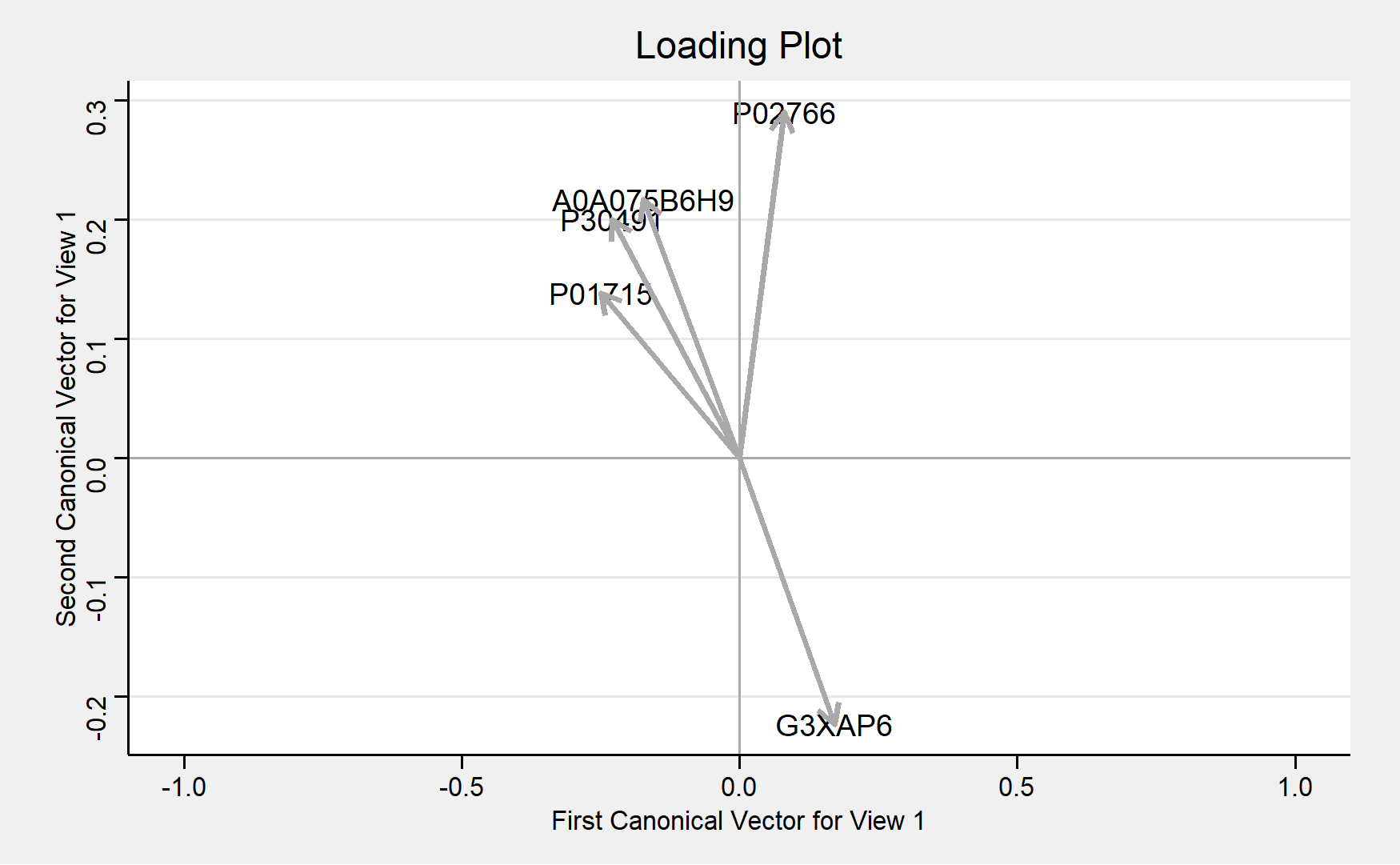} \\
         \includegraphics[width=0.5\textwidth]{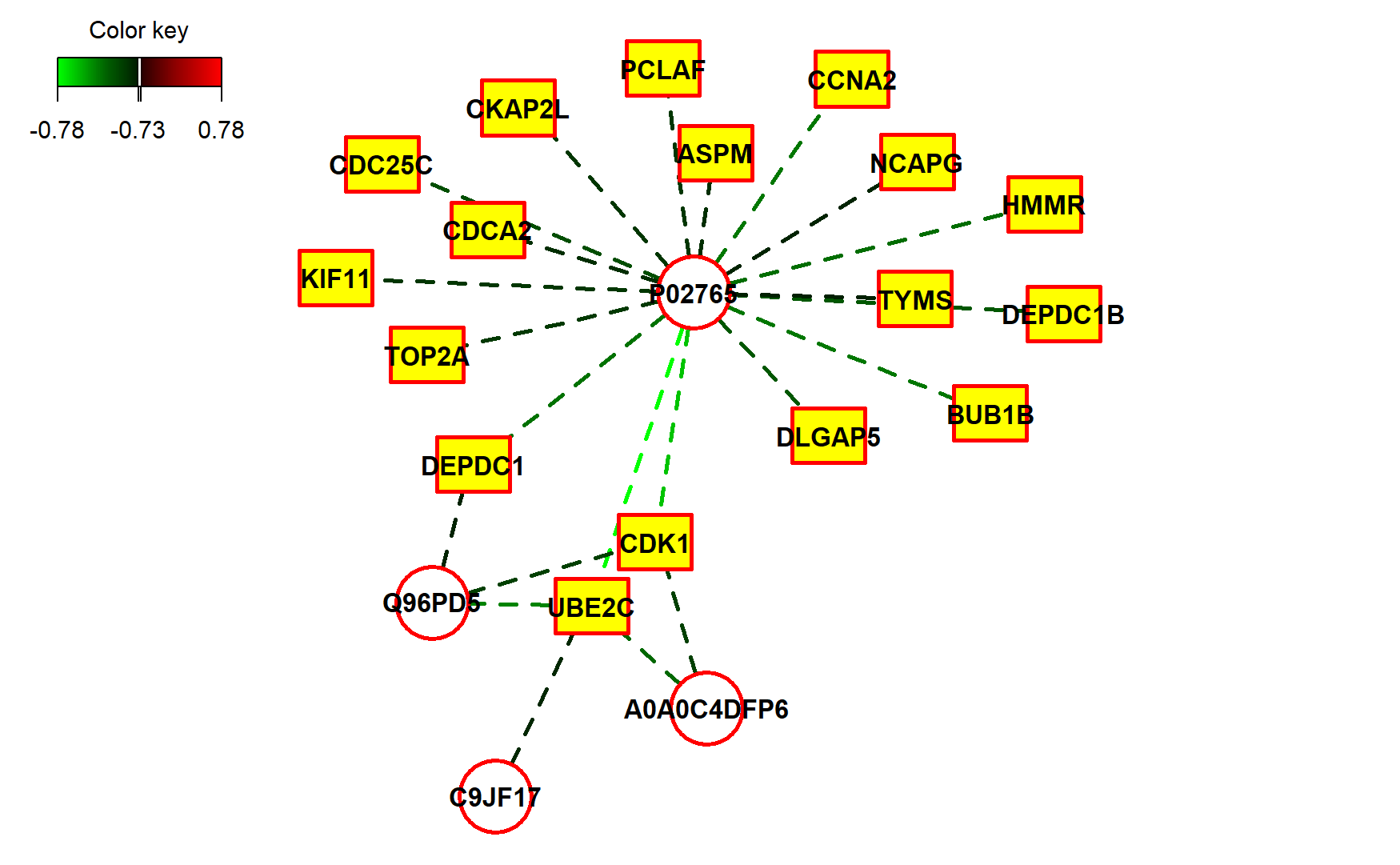}&\includegraphics[width=0.5\textwidth]{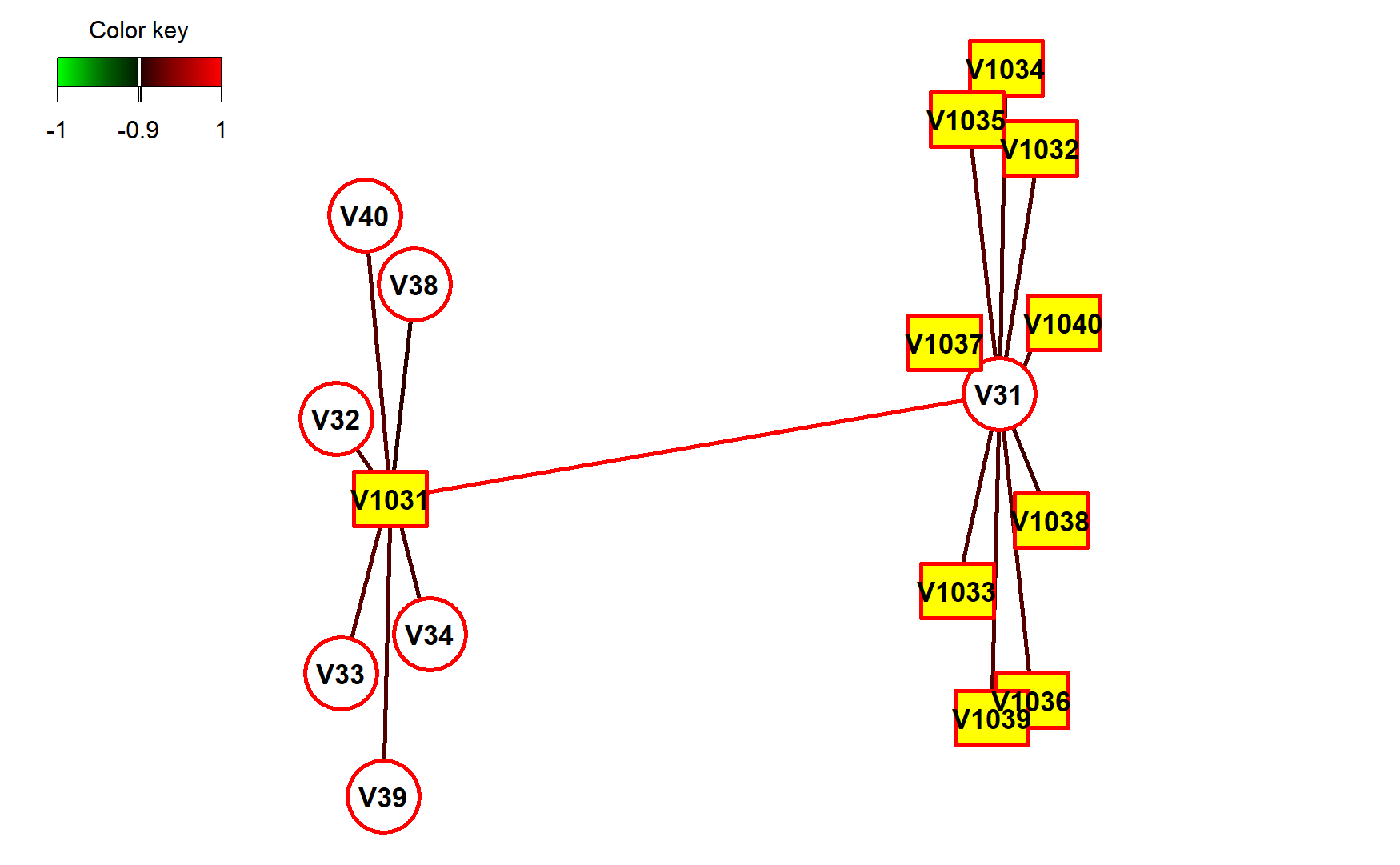}\\
         \includegraphics[width=0.5\textwidth]{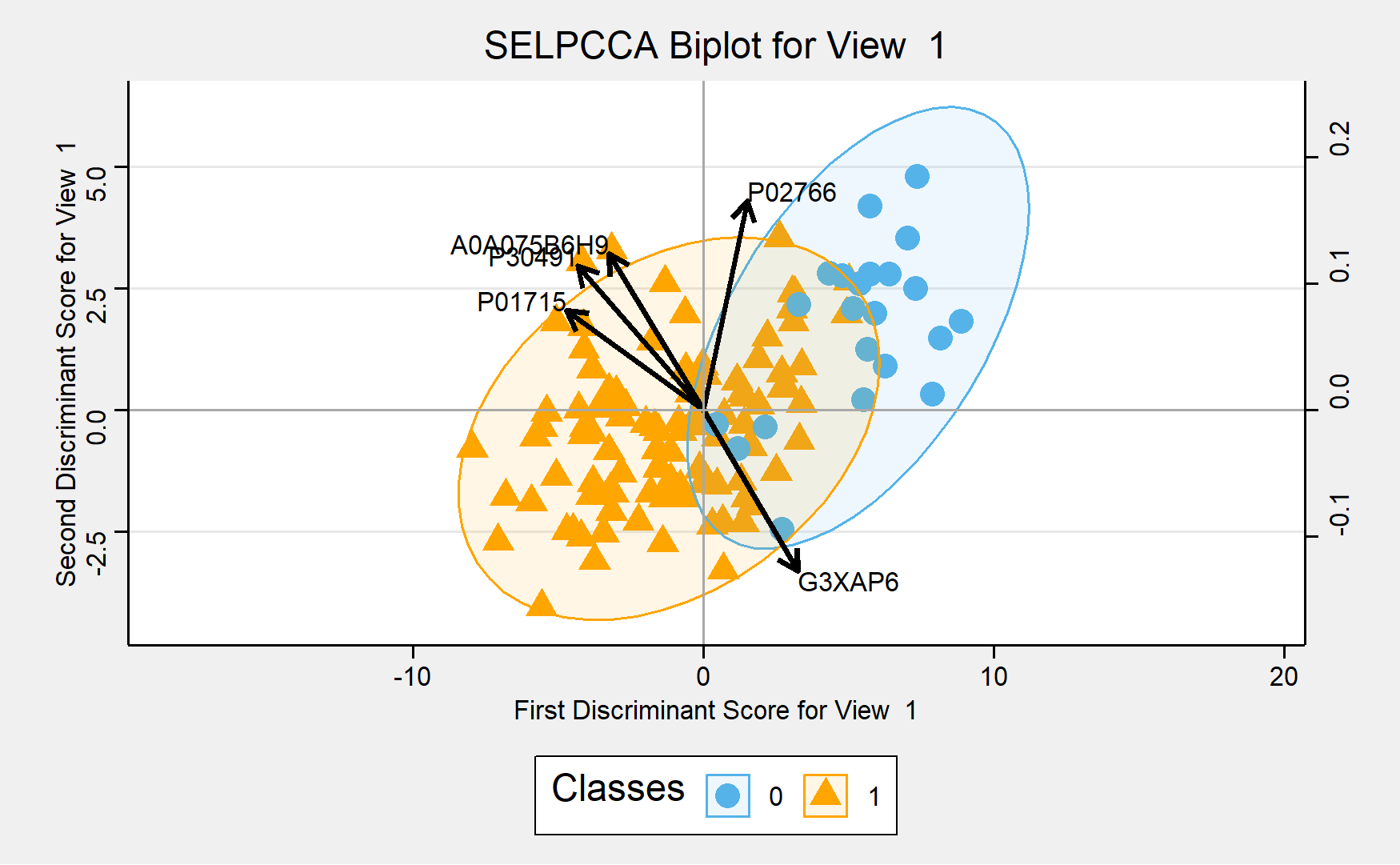}&\includegraphics[width=0.5\textwidth]{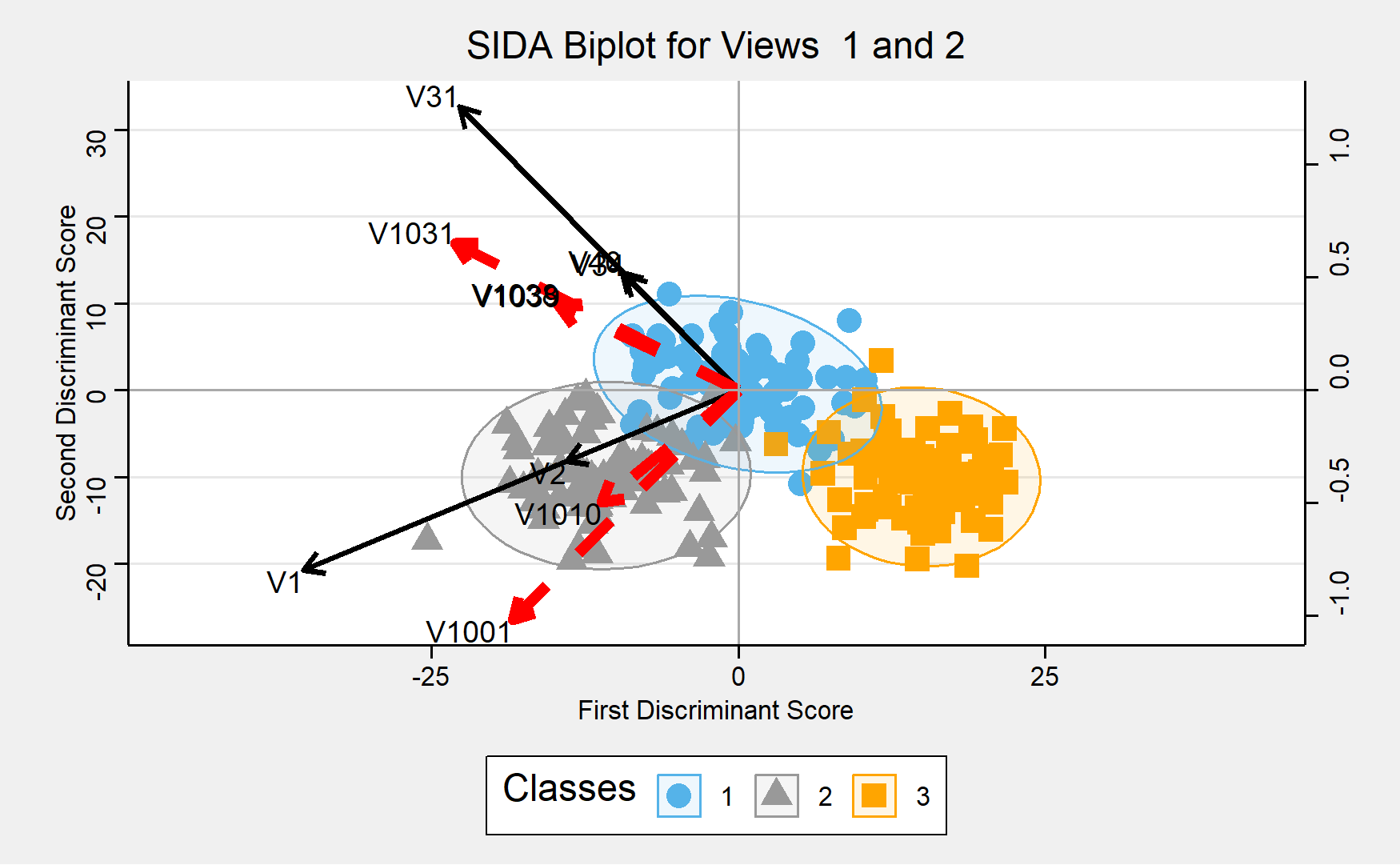}\\
\end{tabular}
    \caption{Sample plots for visualizing variables within and between views. \textit{Top left panel}: variable importance plot showing the relative importance of each variable. Variables with largest absolute loadings rank high. The weights for each variable is normalized relative to the largest weight. \textit{Top right panel}: Loading plots. We provide loading plots for each view to demonstrate the relationships between pairs of variables within each view. Variables that are farther from the origin have higher impact on the axis the vector is close to. Vectors that are close to each other suggest that the pairs of variables are correlated positively.  Vectors with an angle of about 180 degrees suggest that the variables have negative correlation. Vectors with a 90 degree angle between them suggest that the two variables are not correlated. \textit{Middle left and right  panels}: Relevance network showing variable-variable connections between pairs of views with correlations $> 0.73$ [left panel] and $> 0.9$ [right panel]. Dashed and solid lines indicate negative and positive correlations, respectively. Color key gives the direction of the associations where green indicates negative correlations and red shows positive correlations. \textit{Bottom left panel}: Within-view biplot. This plot shows the scores and loadings together for a specific view. Bottom right panel: Between-view biplot. This plot shows the  scores and loadings from pairs of views together. The scores are the sum of scores for each view. Solid and dashed lines represent vectors for Views 1 and 2, respectively. In this example, variable V31 (from View 1) has a higher correlation with variables V1031 and V1038 from View 2 compared to Variable V1001 from View 2. This finding is supported by the edge between V31 and V1038 and V1031 in the network plot but no edge between V31 and V1001 given a correlation cutoff of 0.9.     }
    \label{fig:variableplots}
\end{figure}

We generate the relevance network from the similarity matrix. The nodes of the graph represent variables for the pairs of views, and edges represent the correlations between pairs of variables. The function \texttt{networkPlot()} in \texttt{mvlearnR} can be used to generate relevance networks.  Please refer to the bottom left and right panels of Figure \ref{fig:variableplots} for sample relevance network plots. Dashed and solid lines indicate negative and positive correlations respectively. Circle nodes are View 1 variables, and rectangular nodes are View 2 variables. We provide an option to tune the network through a correlation cutoff. For instance in the middle right panel of Figure \ref{fig:variableplots}, we only show graph for variables with correlations greater than $0.9$. The plot suggest that variable V1031 from View 2 is highly correlated with variable V31 from View 1. Since there is no edge between variable V38 and variables V1037, it indicates that the correlation between these pairs of variables is smaller than $0.9$.

\subsection{Variable importance plots}
We provide the function \texttt{VarImportancePlot()} [Top left panel of Figure \ref{fig:variableplots}] to visualize the weights (in absolute value) of the loadings. Since the  loadings are standardized to have unit norm, a variable with larger weight contributes more to the association between the views (for SELPCCA) or to the association between the views and discrimination of classes within each view (SIDA and SIDANet). We only show the top 20 variables and their weights but one can view data matrix for all variables. 
\subsection{Loadings plots} \label{sec:loadings}
We provide the function \texttt{LoadingsPlot()} [Top right panel of Figure \ref{fig:variableplots}] to plot discriminant and canonical correlation vectors. These graphs are useful for visualizing how selected variables from SIDA/SIDANet and SELPCCA contribute to the first and second discriminant (for SIDA and SIDANet) or canonical correlation (for SELPCCA) vectors.  Variables farther from the origin and close to the first or second axis have higher impact on the first or second discriminant/canonical vectors, respectively. Variables farther from the origin and between both first and second axes have similar higher contributions to the first and second discriminant/canonical correlation vectors. In both situations, for SIDA and SIDANet, this suggests that these variables contribute more to the separation of classes and association of views. For SELPCCA, this suggests that these variables contribute more to the association between the two views. This plot can only be generated for classification and association problems with 3 or more classes (SIDA and SIDANet),
or for CCA problems with two or more canonical correlation vectors requested (i.e. ncancorr $> 1$ for SELPCCA). The angle between two vectors also give an indication of how the two variables are correlated. In particular, vectors that are close to each other suggests that the variables have high positive correlation. Vectors that are about 90 degrees indicate that the two variables are uncorrelated. Vectors that have an angle close to 180 degrees indicate that the two variables have negative correlation.

\section{Visualization of loadings and scores simultaneously}
Biplots are useful for representing both loadings  and discriminant scores/canonical correlation variates. We present biplots for each view and between views. In particular, 
we provide the function \texttt{WithinViewBiplot()} to visualize the scores and loadings for each view separately [Figure 1, Bottom left panel].  We also provide the function \texttt{BetweenViewBiplot()} to graph scores and loadings for pairs of views. The scores are the sum of scores for the two views (Refer to Section on relevance network for more explanation).  Please refer to Figure 1, Bottom right panel for a sample biplot between views. In this graph, dashed red vectors represent loadings plot for the second view. 
And solid black vectors represent loadings plot for the first view. Please refer to section \ref{sec:loadings} for a brief discussion on interpreting loadings plots.

\clearpage
\section{Demonstration of SELPCCA and SELPCCA Predict on COVID-19 Data}

\begin{figure*}[!hb]
    \centering
    \includegraphics[width=\textwidth]{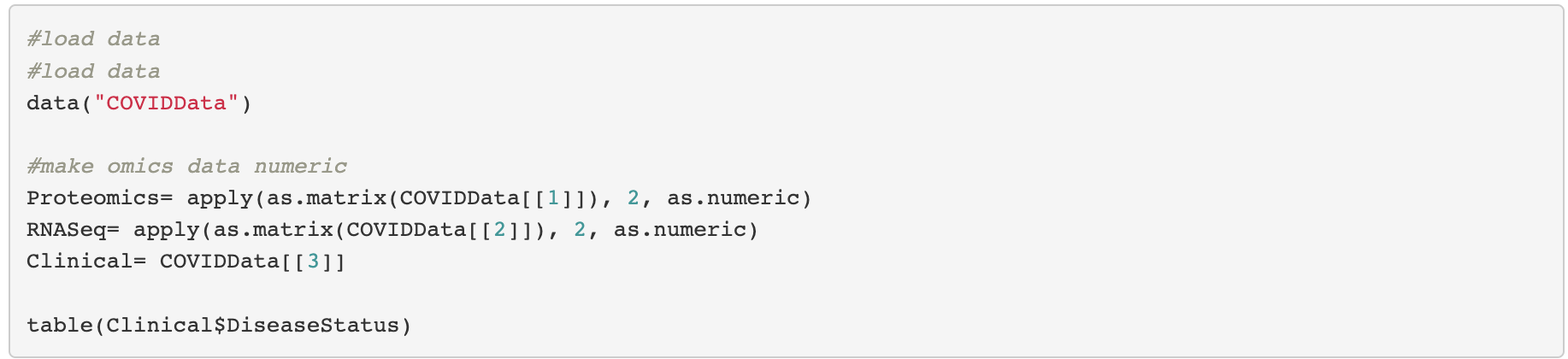} 
    \caption{Import COVID-19 data}
    \label{fig:import}
\end{figure*}

\begin{figure*}[!hb]
    \centering
     \includegraphics[width=\textwidth]{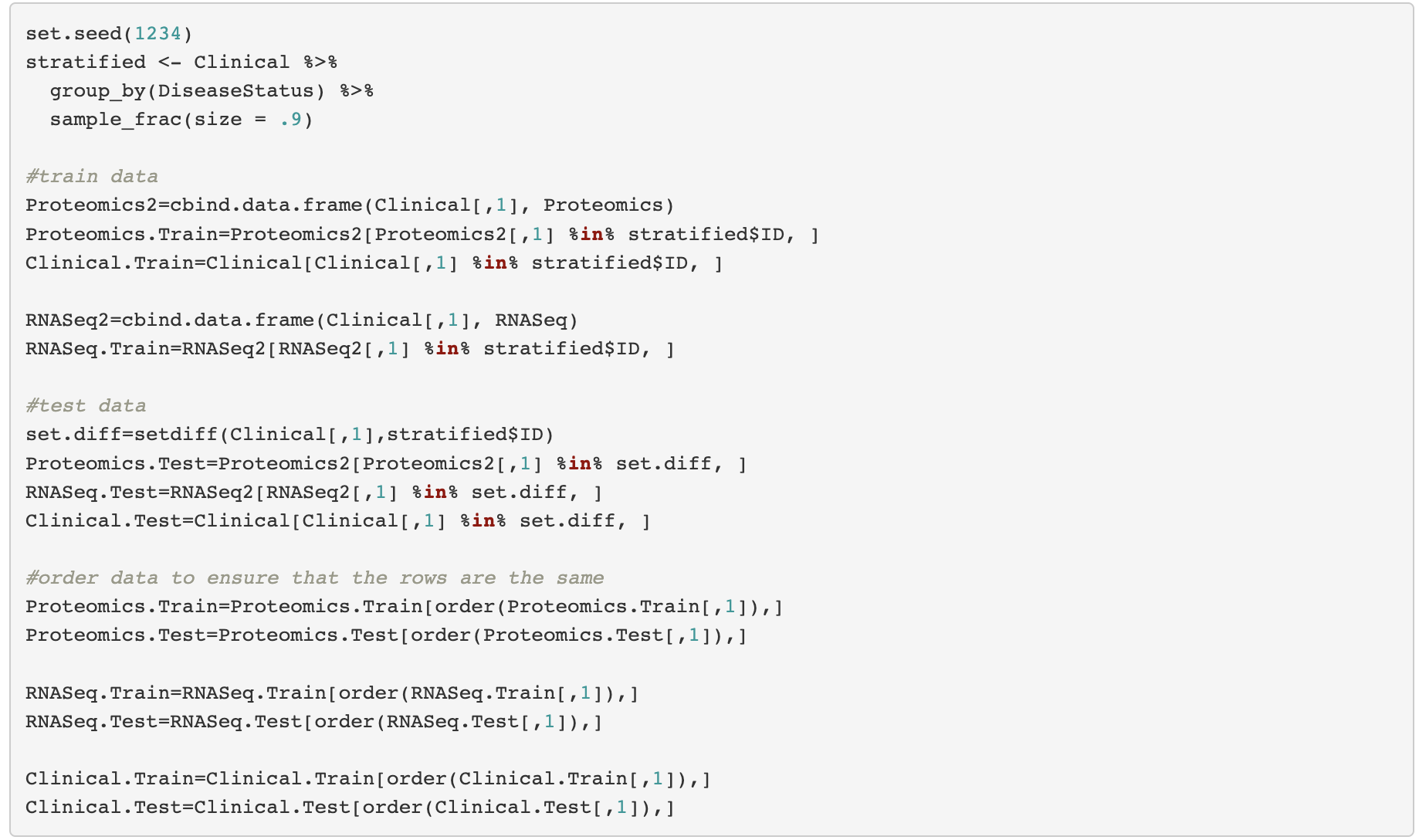}
    \caption{Train and Test splits. We split the data into 90\% train and 10\% test, keeping the proportion of COVID-19 and non-COVID-19 cases as in the original data. }
    \label{fig:traintest}
\end{figure*}

\begin{figure}[!h]
         \centering        
         \includegraphics[width=6.5in]{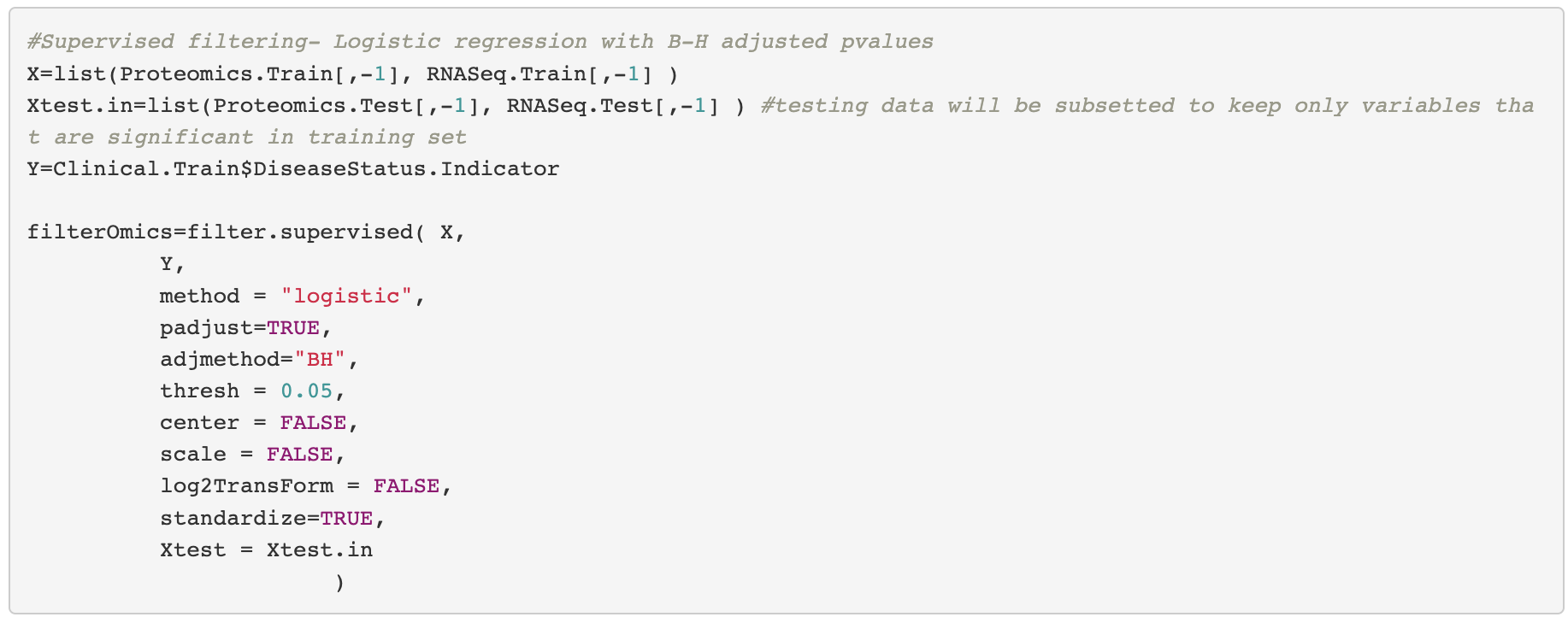}              \includegraphics[width=6in]{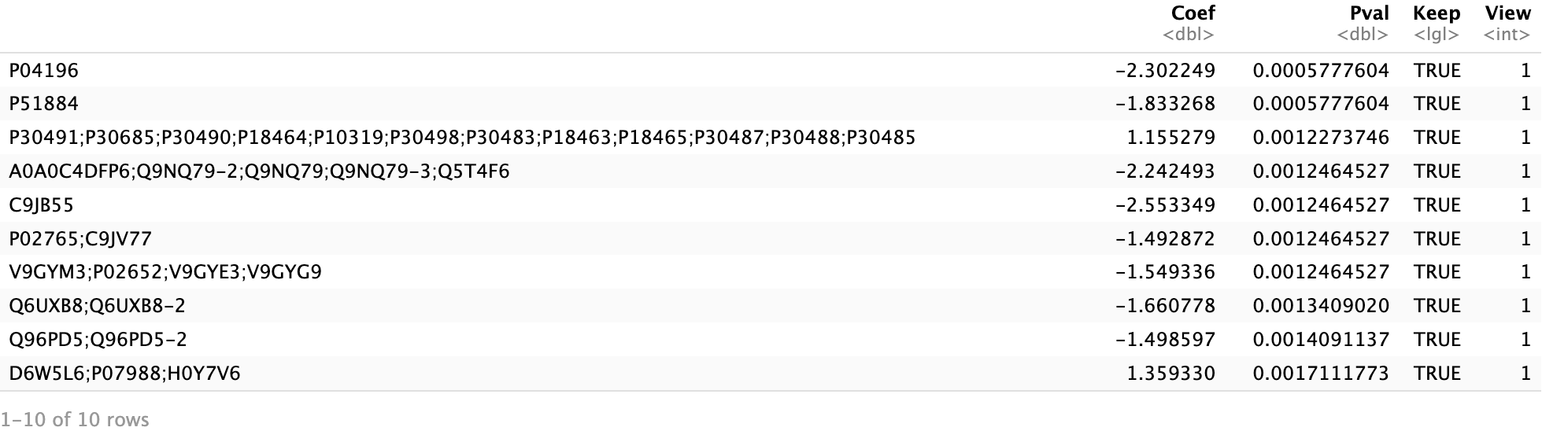}  \includegraphics[width=6in]{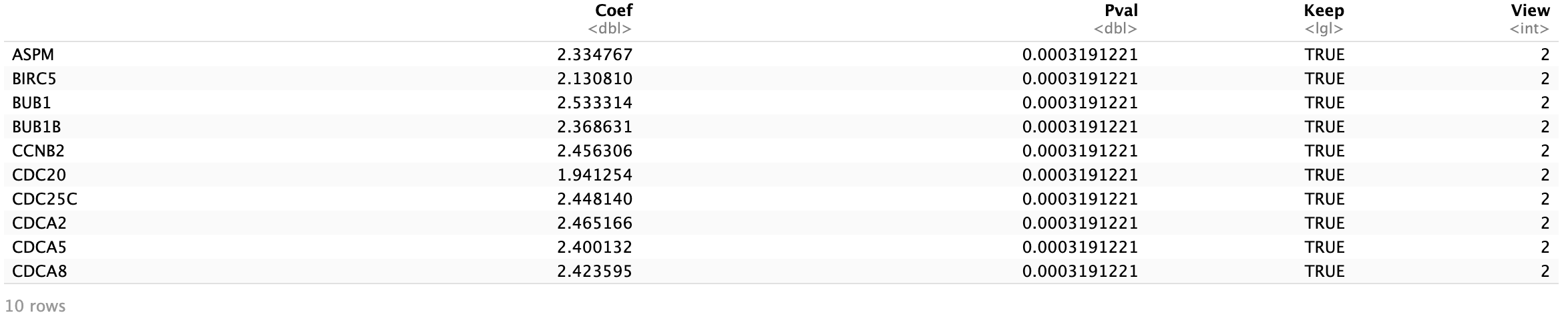}\\
    \caption{Supervised filtering with logistic regression. Example output table from supervised filtering showing Top 10 (sorted by p-values) molecules  with estimated coefficients (Coef), P-value (Pval), whether the variable is kept or filtered out (Keep), and the view (View). View 1 is for proteomics data, and View 2 is for RNASeq data.}
    \label{fig:filterout}
\end{figure}

    \begin{figure*}[!hb]
    \centering
     \includegraphics[width=\textwidth]{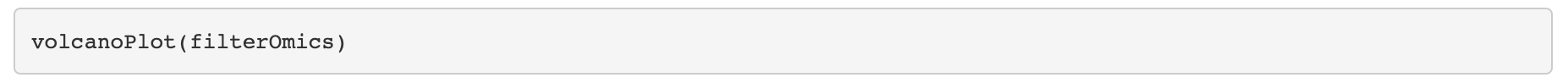} 
    \includegraphics[width=\textwidth]{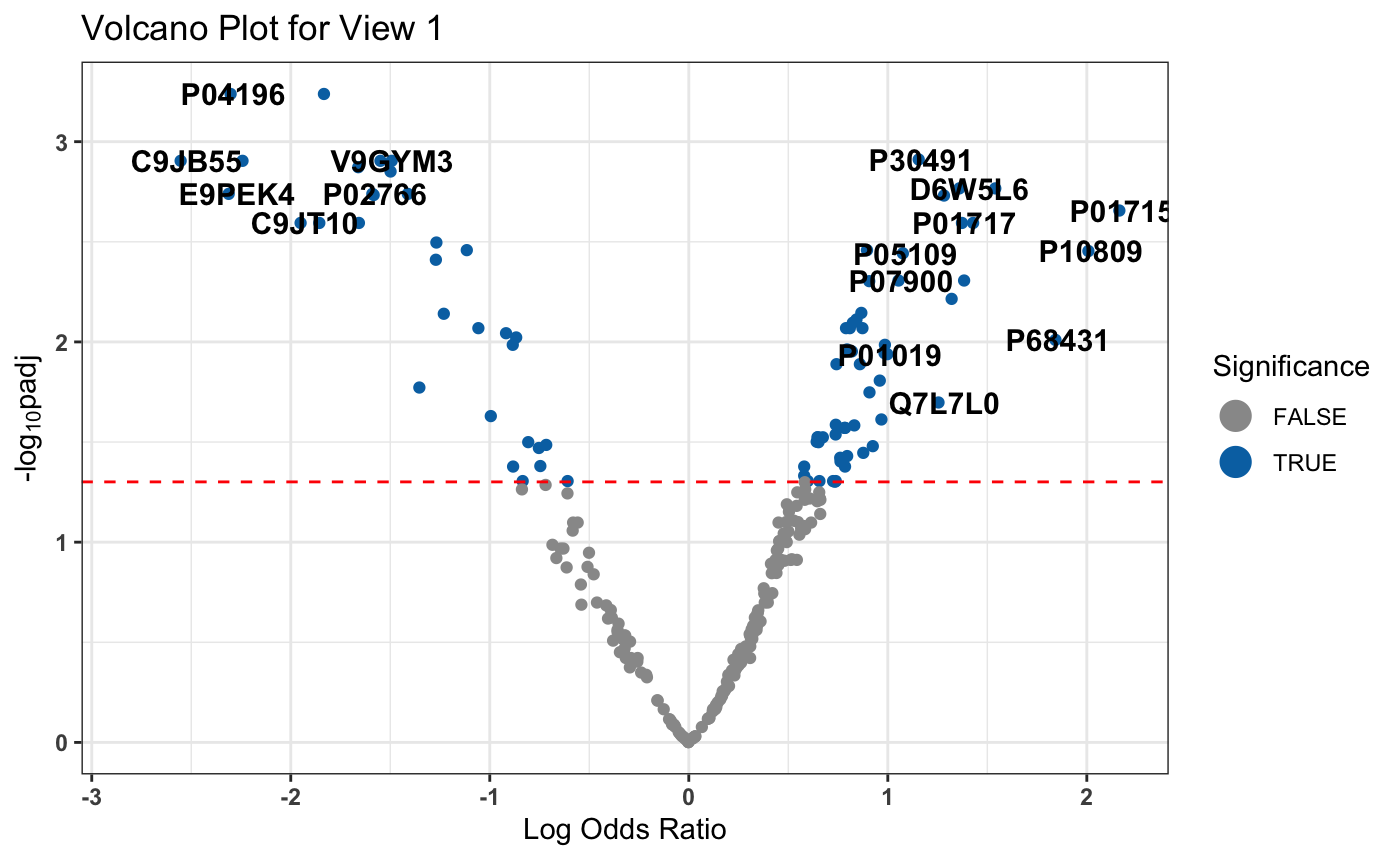}  \includegraphics[width=\textwidth]{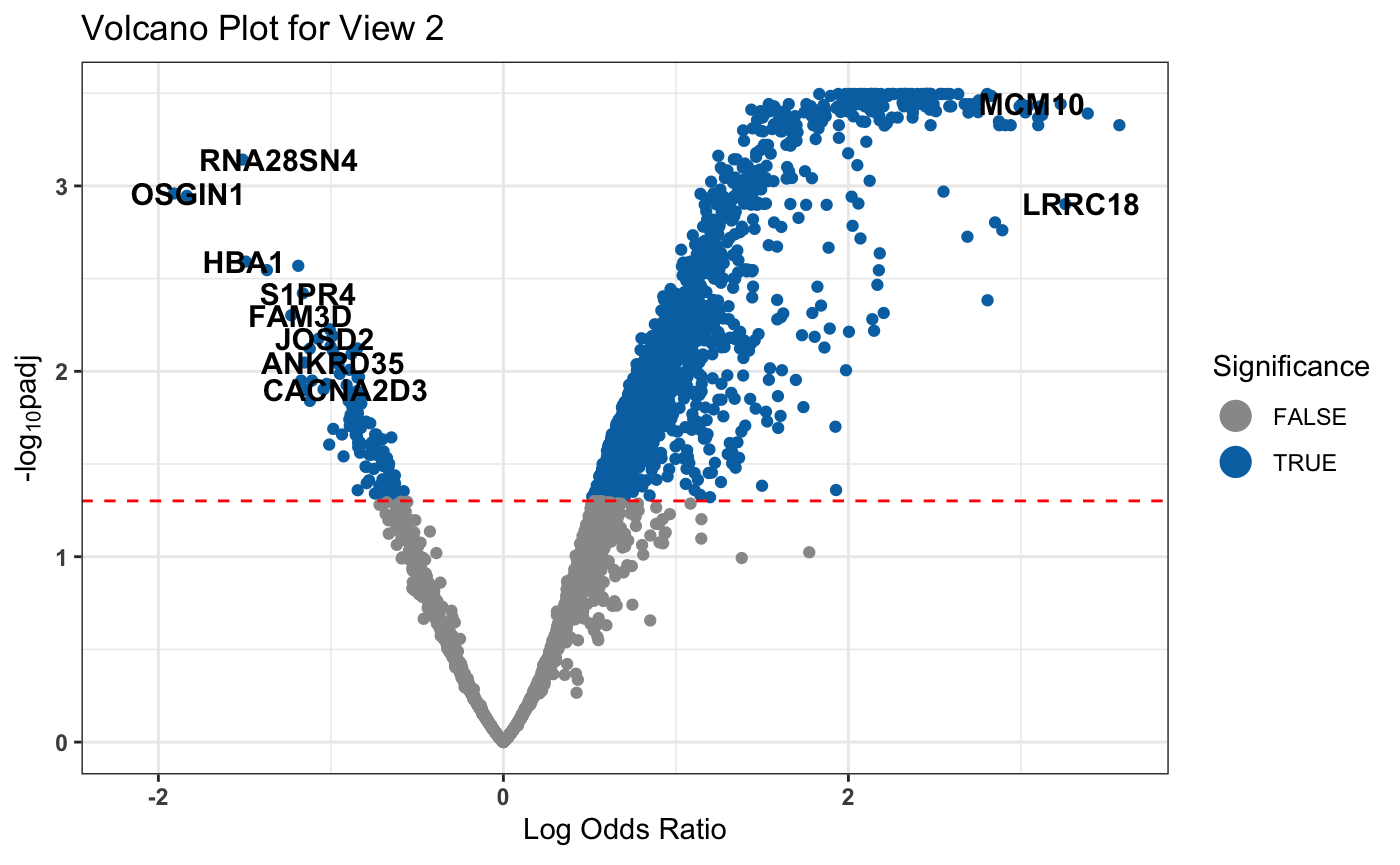}
    \caption{Volcano plots for View 1 (Proteins) and View 2 (Genes).}
    \label{fig:volcanoplots}
\end{figure*}

    \begin{figure*}[!hb]
    \centering
         \includegraphics[width=\textwidth]{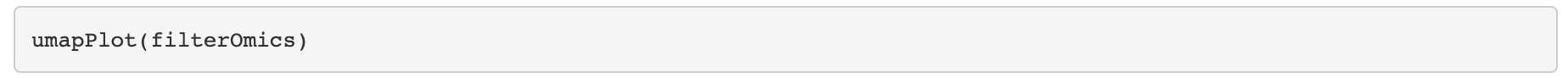} 
    \includegraphics[width=\textwidth]{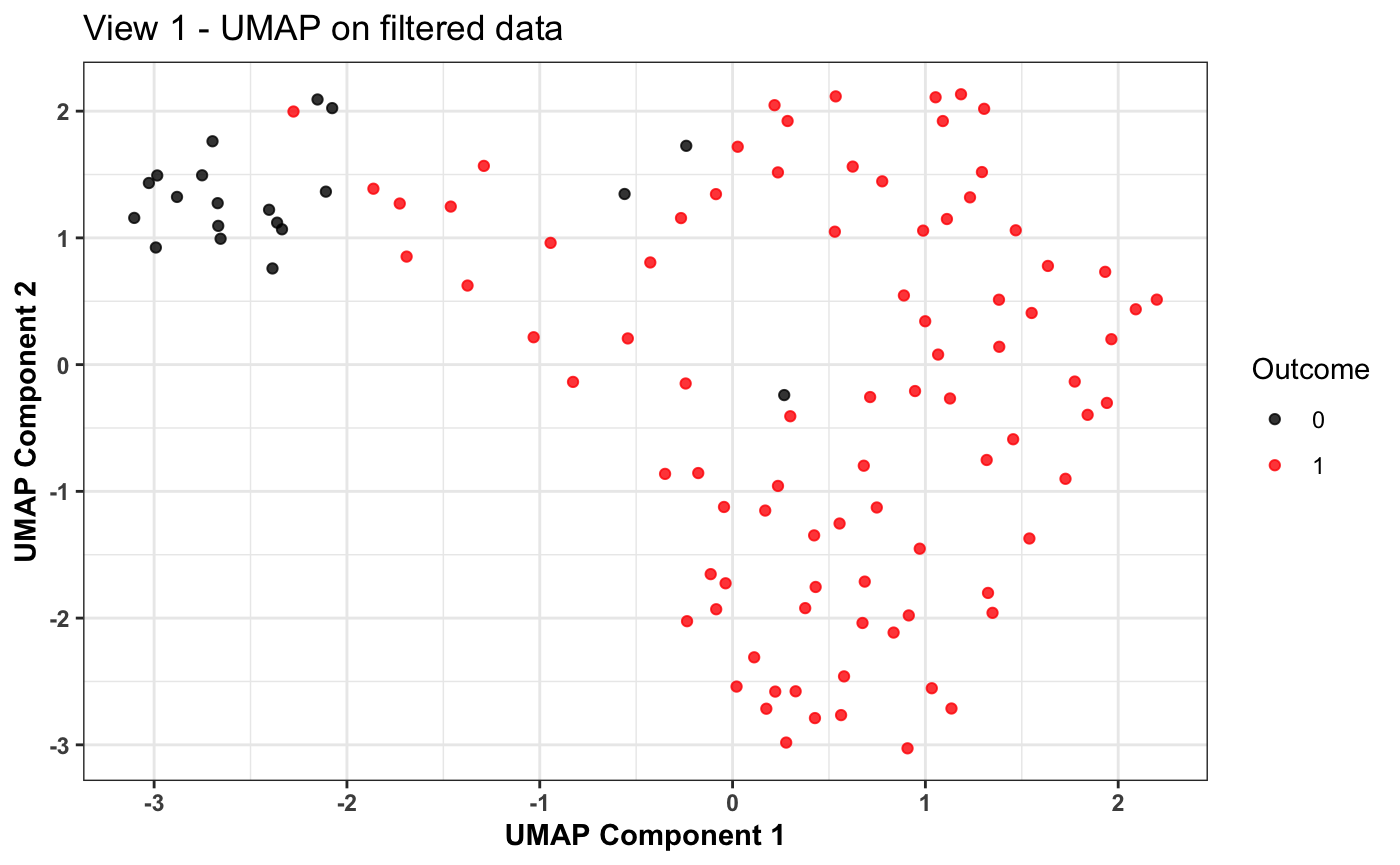}  \includegraphics[width=\textwidth]{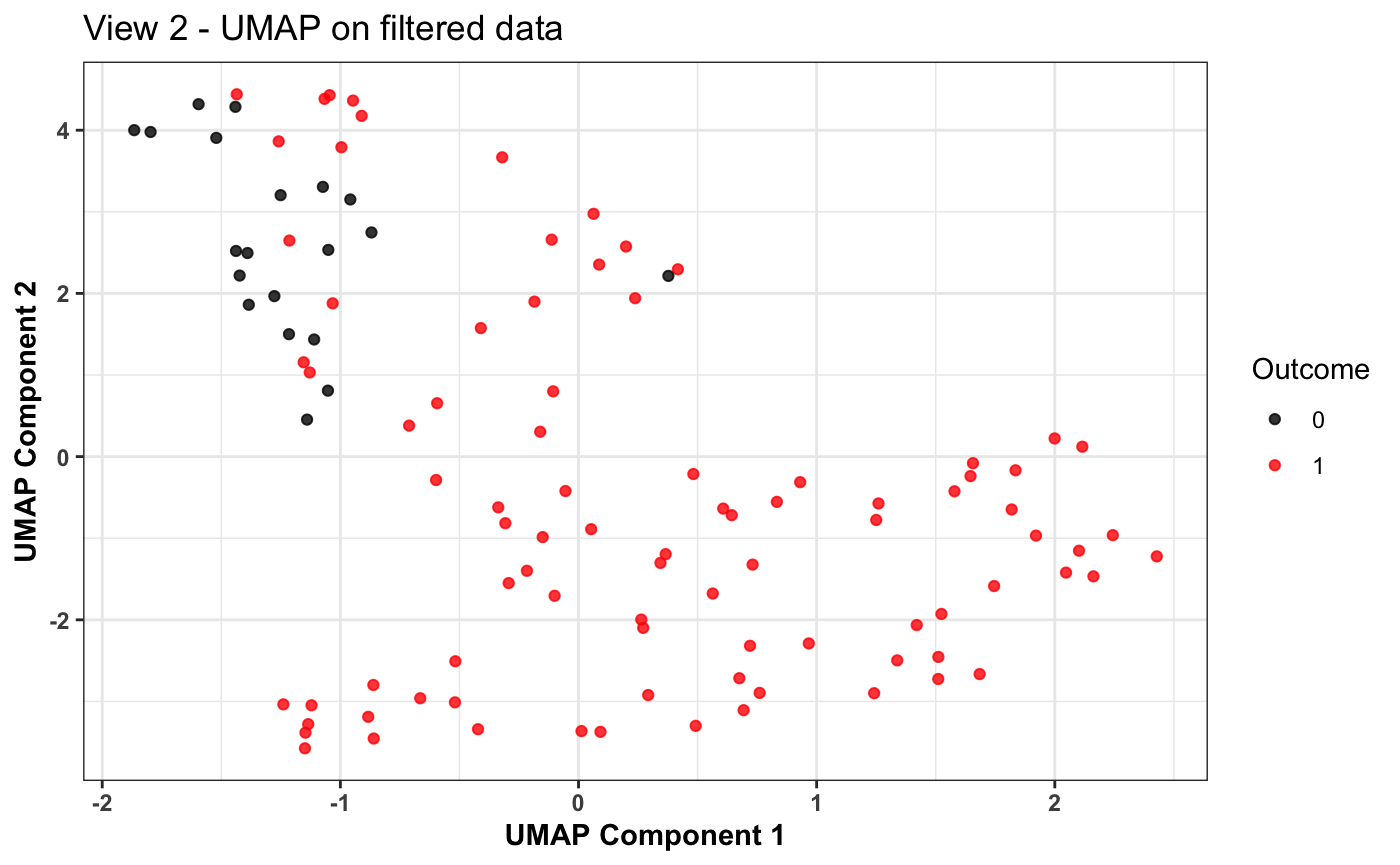}
    \caption{UMAP plots for View 1 (Proteins) and View 2 (Genes) on filtered data. }
    \label{fig:umapplots}
\end{figure*}

\begin{figure*}[!hb]
    \centering
           \includegraphics[width=\textwidth]{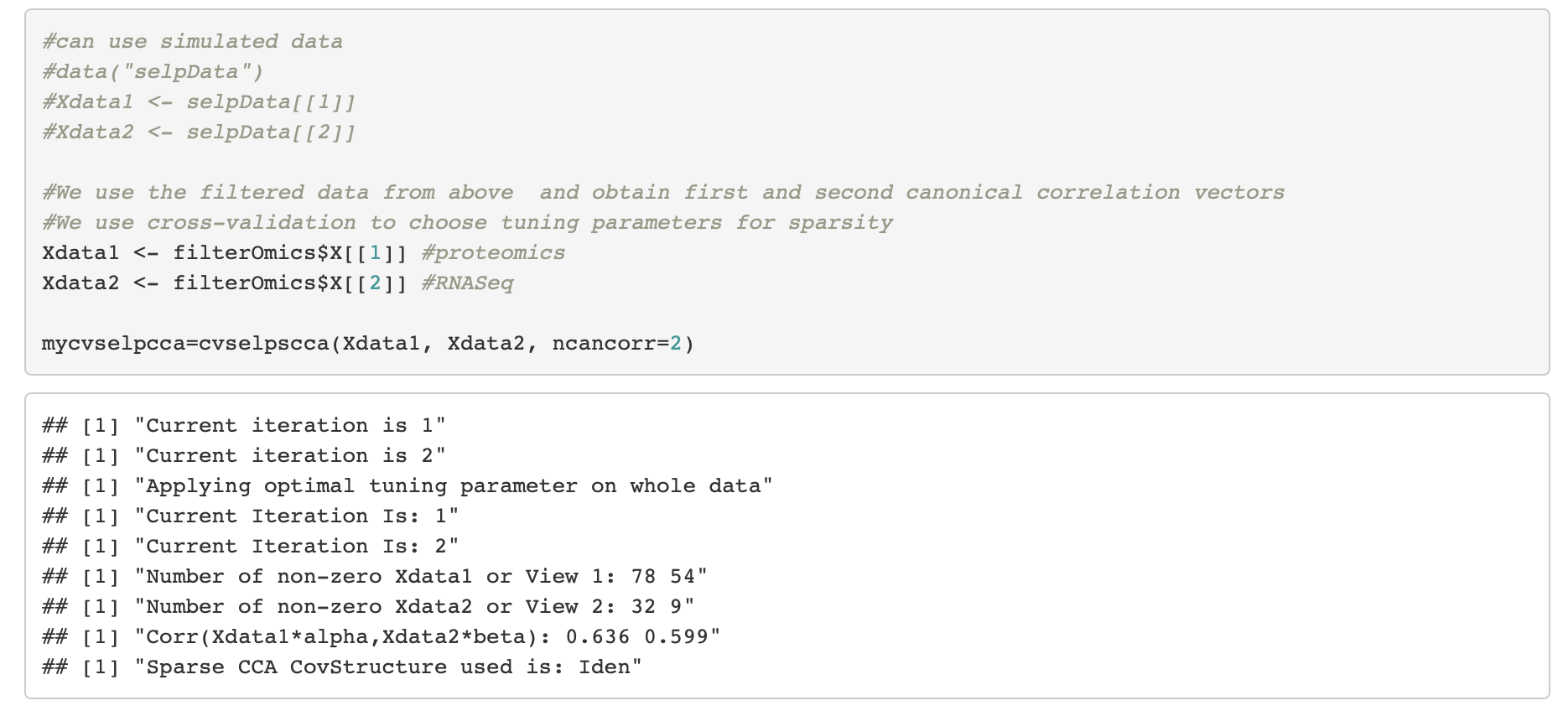}
    \caption{Fitting SELPCCA with cross-validation on training data to choose optimal hyperparameters and to obtain overall canonical correlation vectors on whole data based on the optimal hyperparameter. We use the function \texttt{cvselpscca()}. 
    }
    \label{fig:fitsida}
\end{figure*}

    \begin{figure*}[!hb]
    \centering
       \includegraphics[width=\textwidth]{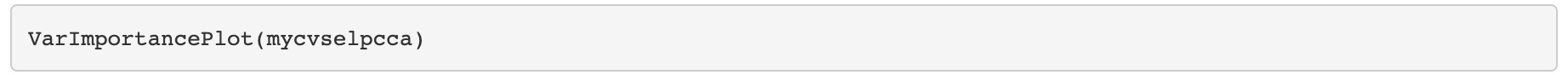} 
    \includegraphics[width=\textwidth]{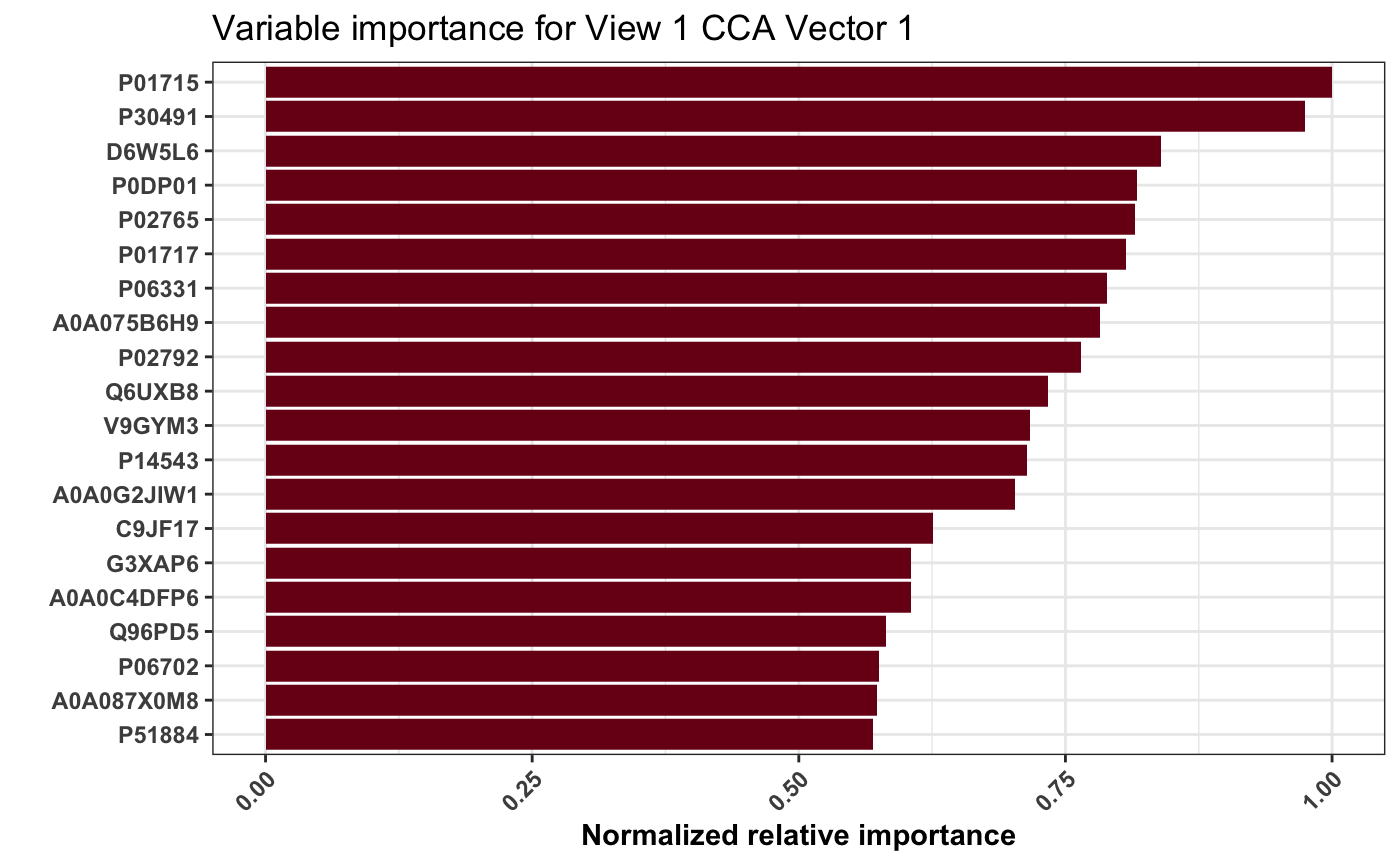}  \includegraphics[width=\textwidth]{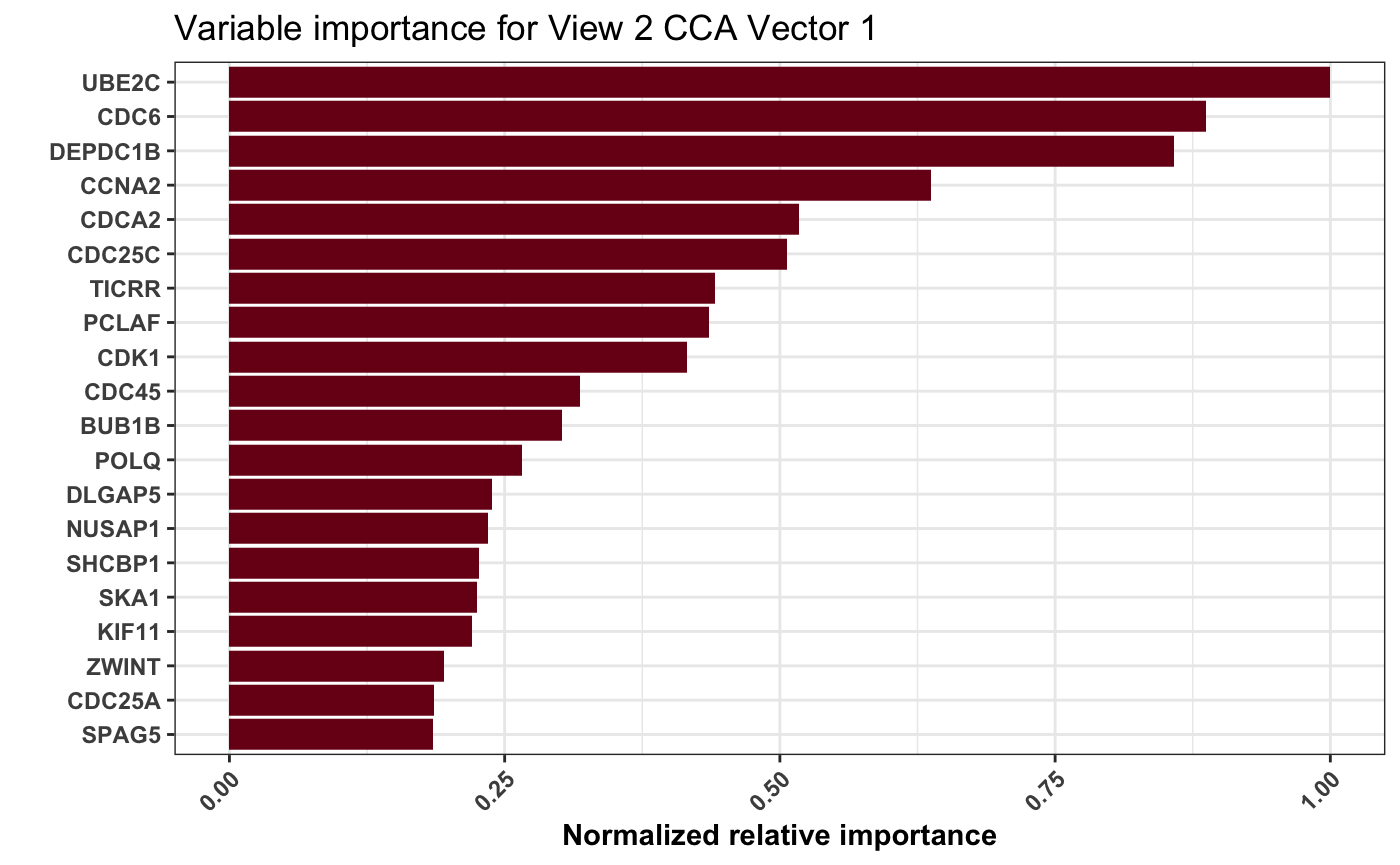}
    \caption{Variable important plots for the first canonical vector for View 1 (Proteins) and View 2 (Genes). Top 20 variables with largest absolute weights for first canonical correlation vector from application of SELPCCA are shown. For proteins, Uniprot IDs are shown on variable importance plot. }
    \label{fig:varimp}
\end{figure*}

\begin{figure*}[!hb]
    \centering
               \includegraphics[width=\textwidth]{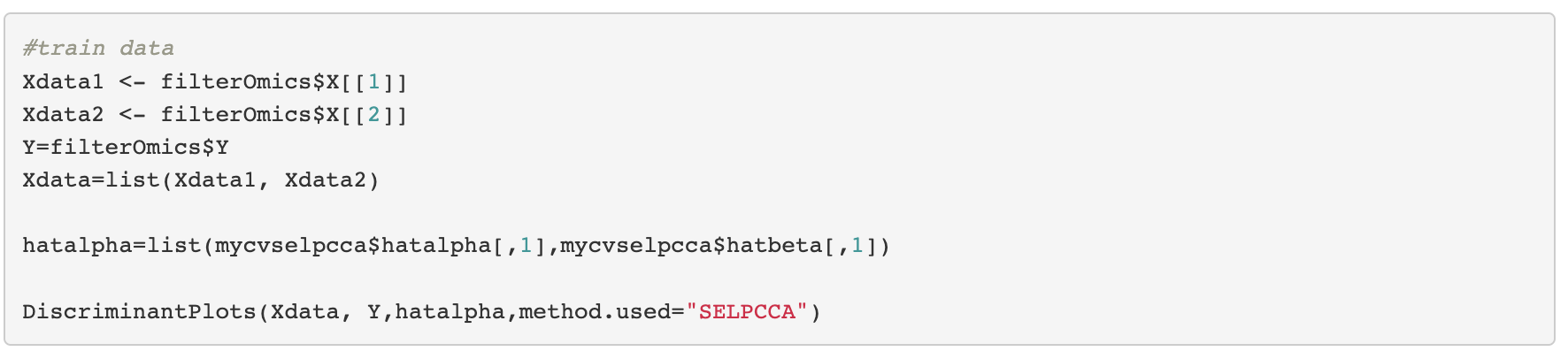}
    \includegraphics[width=5.5in]{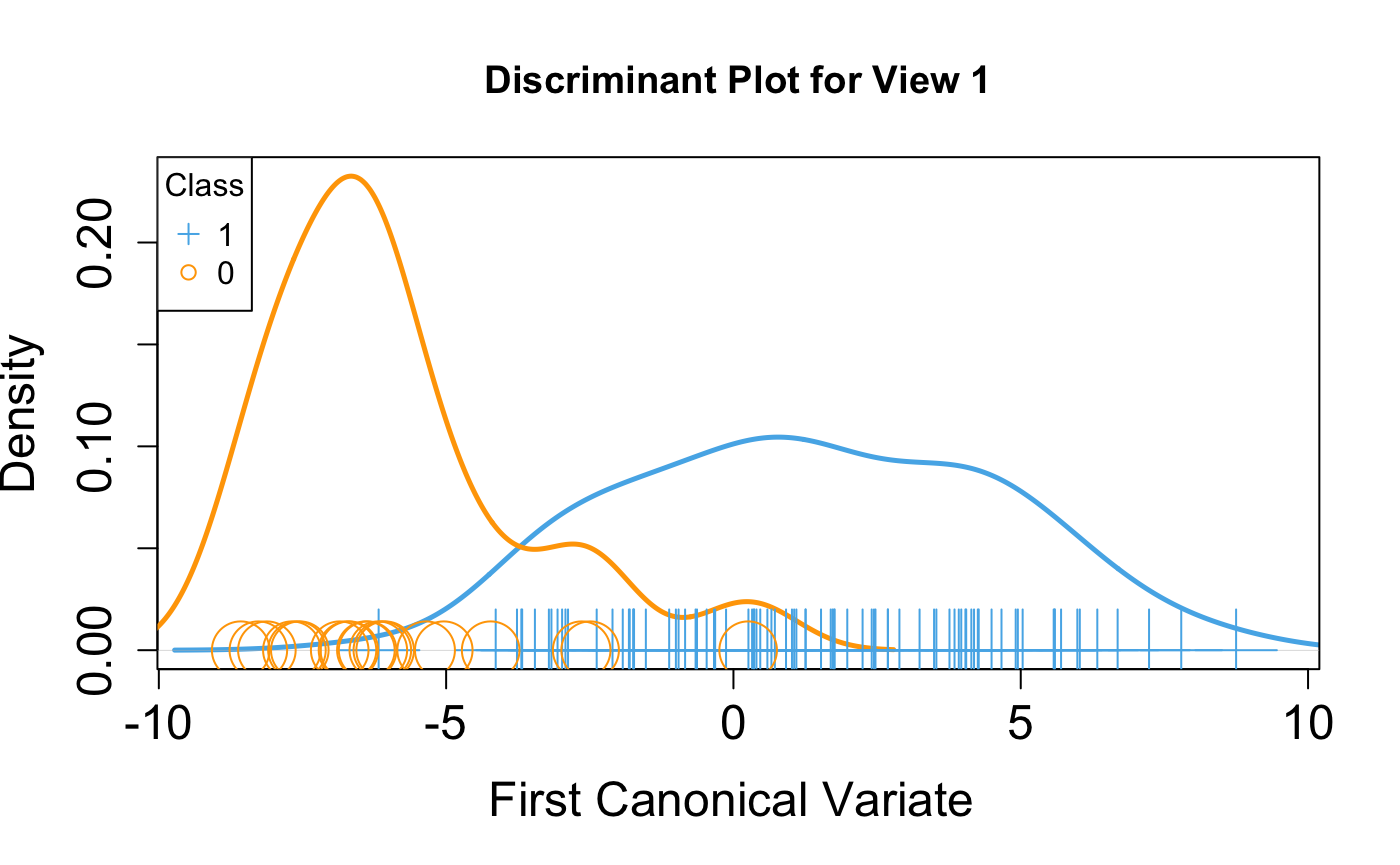}  \includegraphics[width=5.5in]{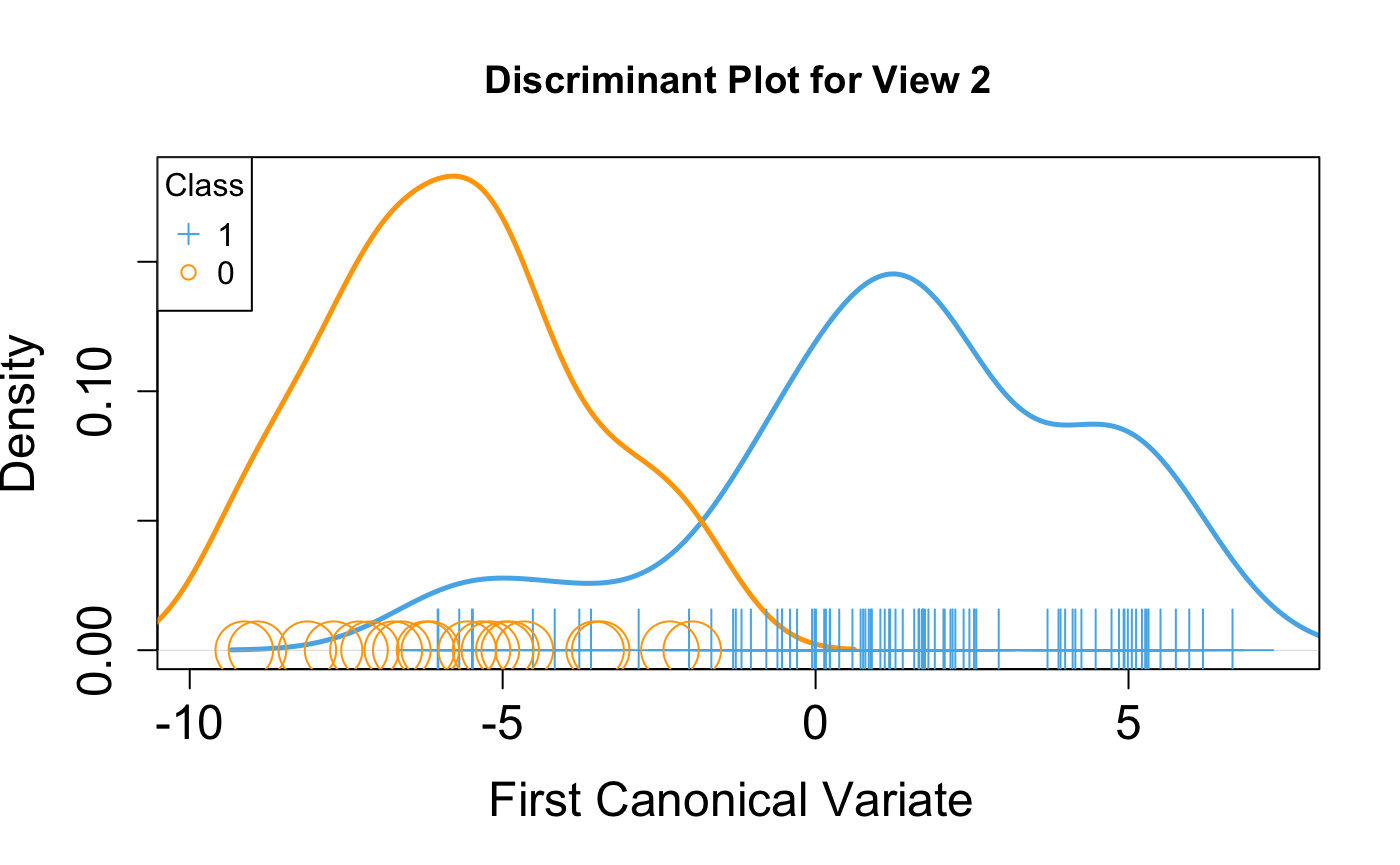}
    \caption{Discriminant plots for SELPCCA based on training data. Top Panel: proteins and Bottom panel: genes. We use the function \texttt{DiscriminantPlots()} to generate these plots. Class 0 is non-COVID-19 samples. Class 1 is COVID-19 samples.}
    \label{fig:discplotselp}
\end{figure*}

\begin{figure*}[!hb]
    \centering
               \includegraphics[width=\textwidth]{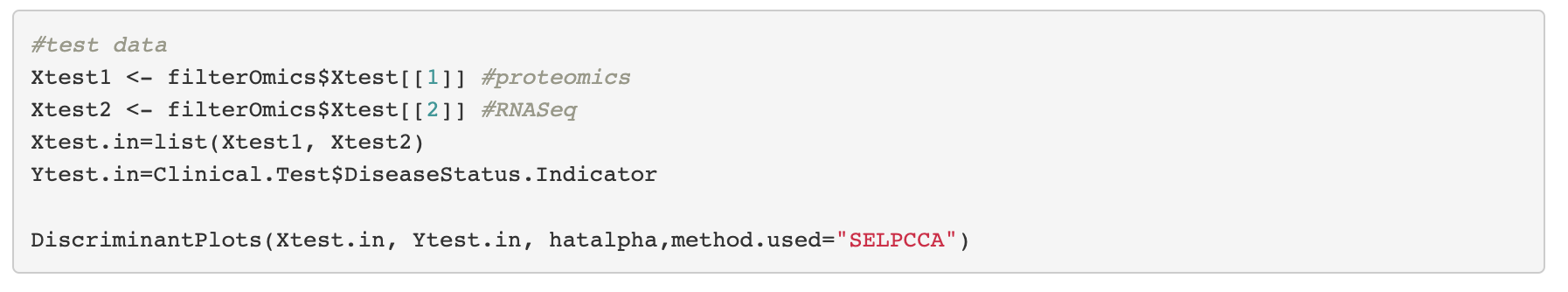}
    \includegraphics[width=5.5in]{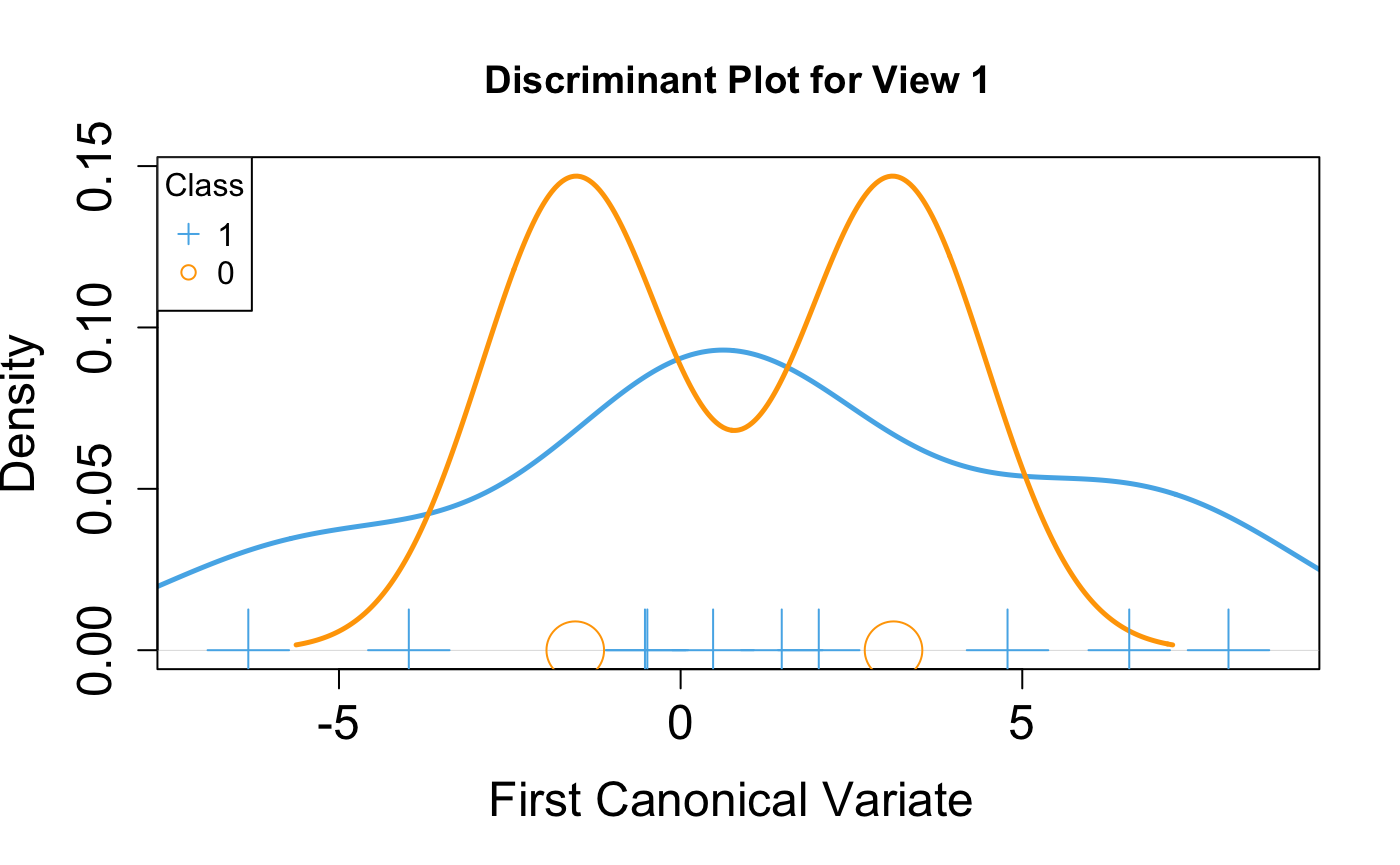}  \includegraphics[width=5.5in]{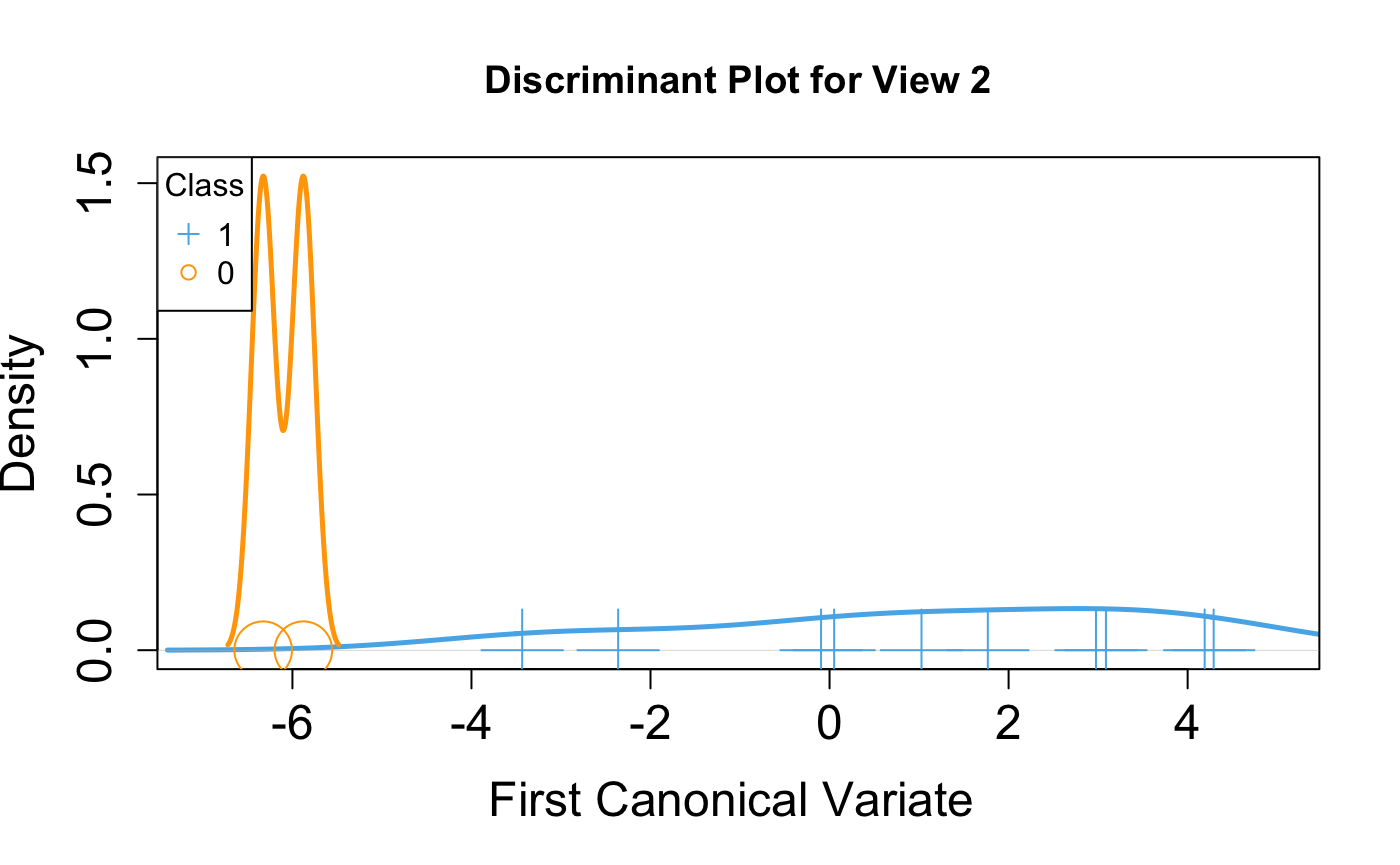}
    \caption{Discriminant plots for SELPCCA based on testing data. Top Panel: proteins and Bottom panel: genes. We use the function \texttt{DiscriminantPlots()} to generate these plots. Class 0 is non-COVID-19 samples. Class 1 is COVID-19 samples.}
    \label{fig:discplottestselp}
\end{figure*}

    \begin{figure*}[!hb]
    \centering
           \includegraphics[width=5.0in]{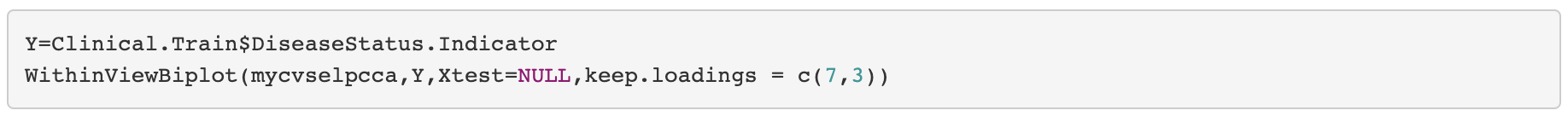} 
    \includegraphics[width=5.0in]{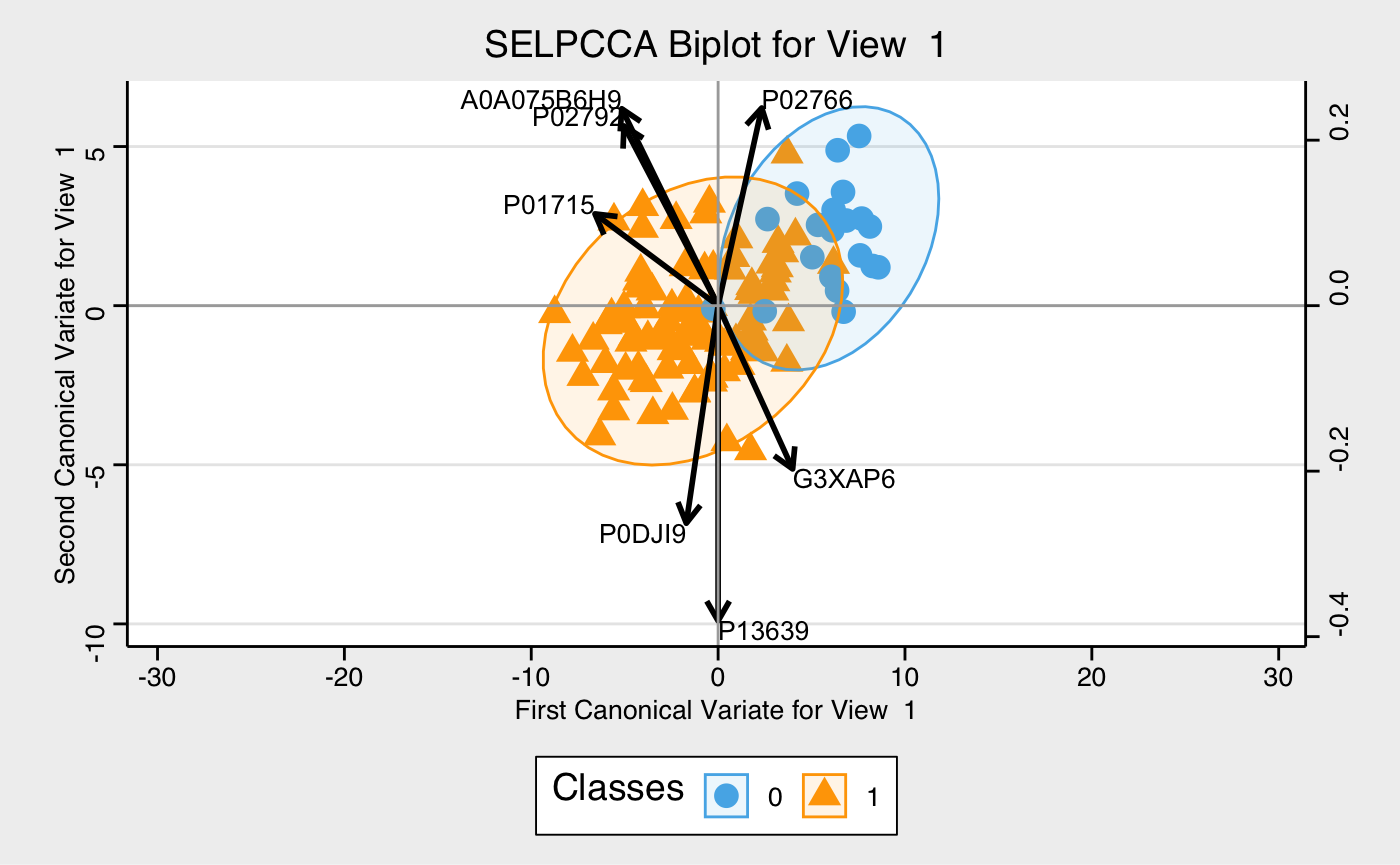}  \includegraphics[width=5.0in]{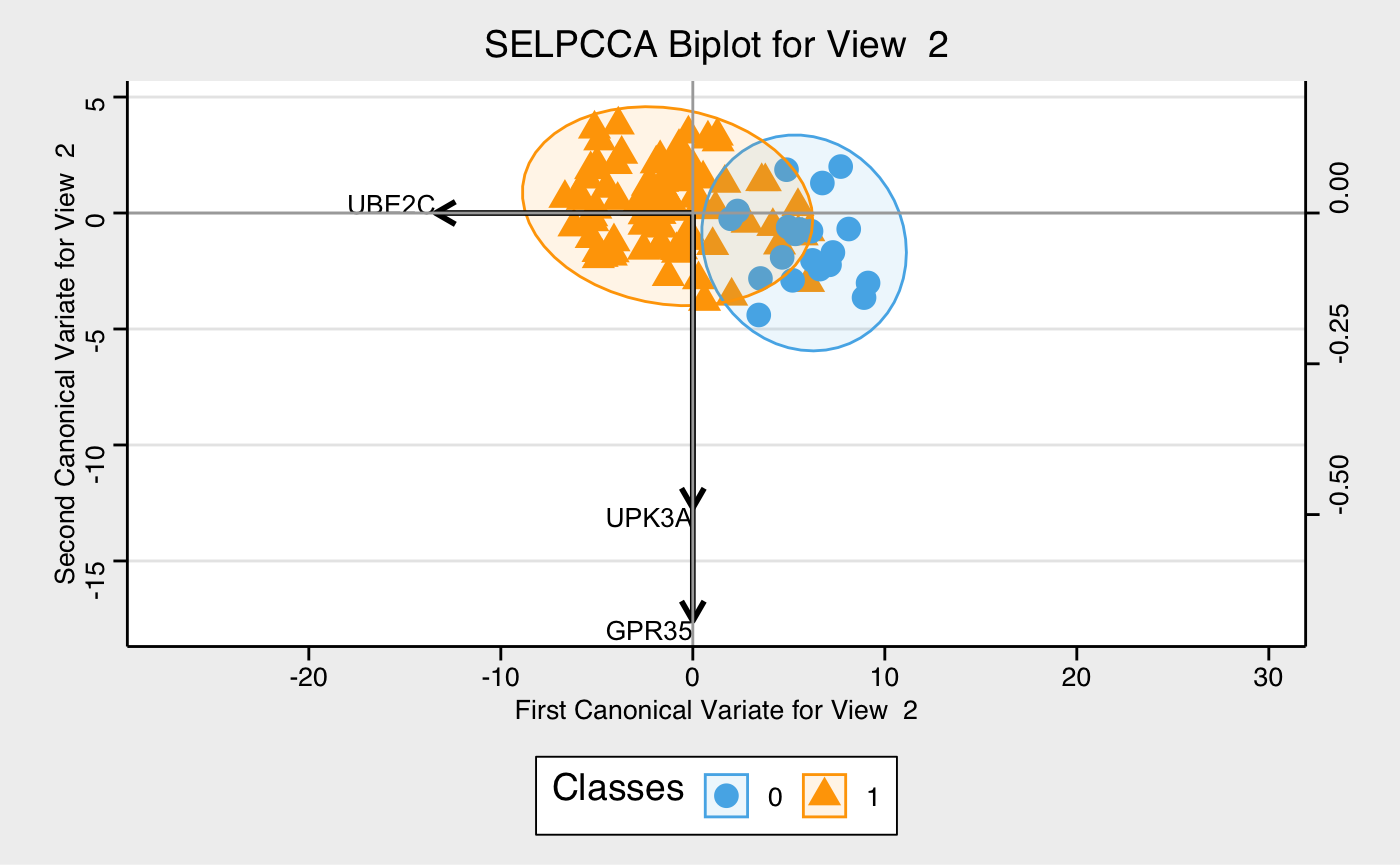}
    \caption{Within-view biplots. Biplots are useful for representing both loadings plot and discriminant scores/canonical correlation variates. We provide the function \texttt{WithinViewBiplot()} to visualize the scores and loadings for each view separately. In this plot, we only show labels for top 7 proteins  and 3 genes with largest absolute weights. Top Panel: Protein IDs shown are loaded on both the first and second canonical vectors. Protein ID A0A075B6H9 appears to be highly correlated with P02792. Bottom Panel: the gene UBE2C is loaded on the first canonical vector; genes UPK3A and GPR35 are loaded on the second canonical vector. Class 0 is non-COVID-19 samples. Class 1 is COVID-19 samples.  }
    \label{fig:wvbselp}
\end{figure*}

   \begin{figure*}[!hb]
    \centering
             \includegraphics[width=6.0in]{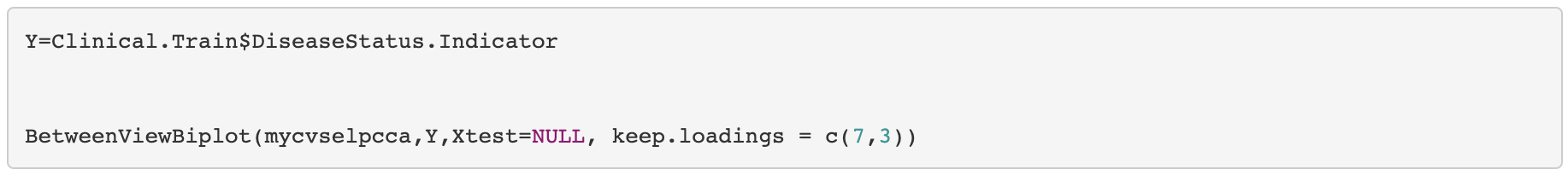} 
    \includegraphics[width=6.0in]{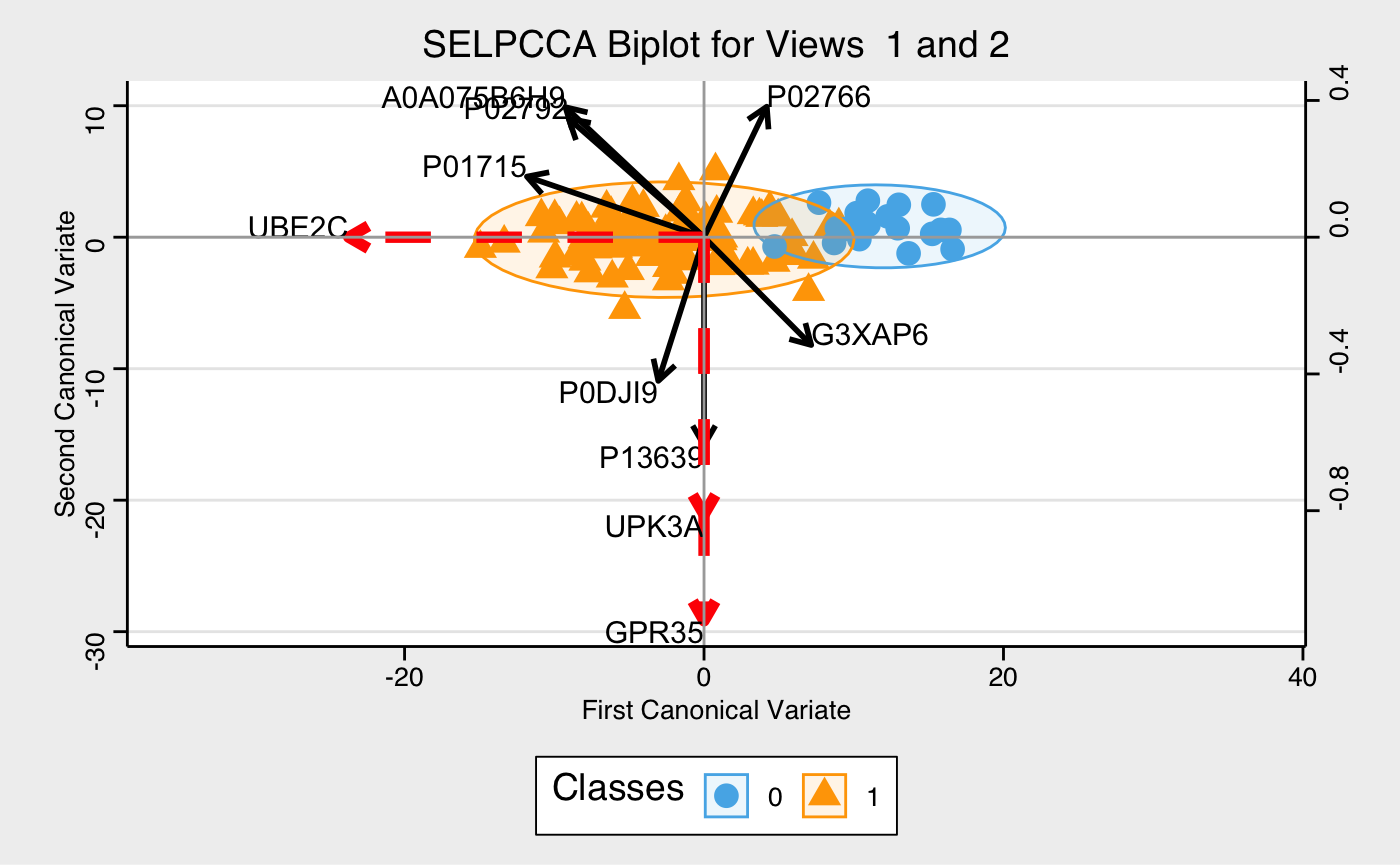} 
    \caption{Between-view biplots are useful for visualizing biplots for both views. This plot allows us to assess how genes and proteins are related.  Solid black vectors represent loading plots for the first view (proteins). Dashed red vectors represent loadings plot for the second view (genes).  We generate this plot with the function \texttt{BetweenViewBiplot()}. In this plot, we only show labels for the top 7 proteins  and 3 genes with largest absolute weights.  The protein  Immunoglobulin lambda variable 3-1 (UID P01715) appears to be positively correlated with the gene UBE2C. The protein P02792 appears to be negatively correlated with the genes UPK3A and GPR35.
    Class 0 is non-COVID-19 samples. Class 1 is COVID-19 samples.}
    \label{fig:bvbselp}
\end{figure*}

  \begin{figure*}[!hb]
    \centering
      \includegraphics[width=\textwidth]{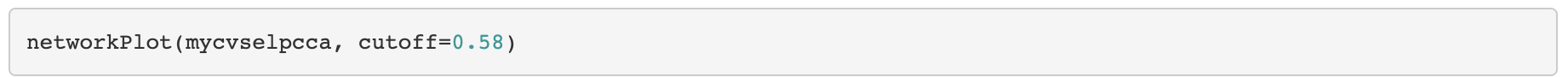} 
    \includegraphics[width=7.5in]{Plots/RNSELPCCA.png} 
    \caption{Relevance network plot. The nodes of the graph represent variables for the pairs of views, and edges represent the correlations between pairs of variables. Dashed and solid lines indicate negative and positive correlations, respectively. Circle nodes are View 1 variables (proteins), and rectangular nodes are View 2 variables (genes). We show edges with correlations at least 0.58.  The plot suggest that the protein P01715 is highly positively correlated with many genes, and the protein P02765 is highly negatively correlated with many genes.  }
    \label{fig:rnselp}
\end{figure*}

   \begin{figure*}[!hb]
    \centering
                 \includegraphics[width=\textwidth]{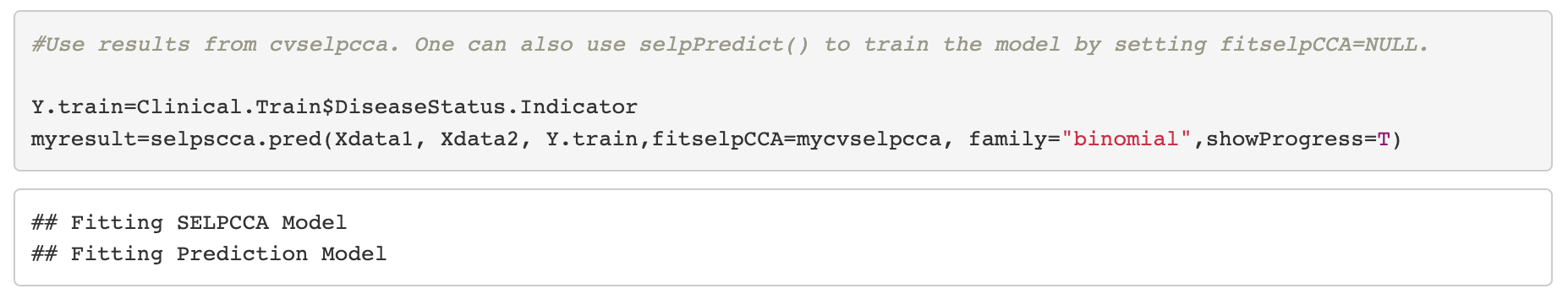}                                 \includegraphics[width=\textwidth]{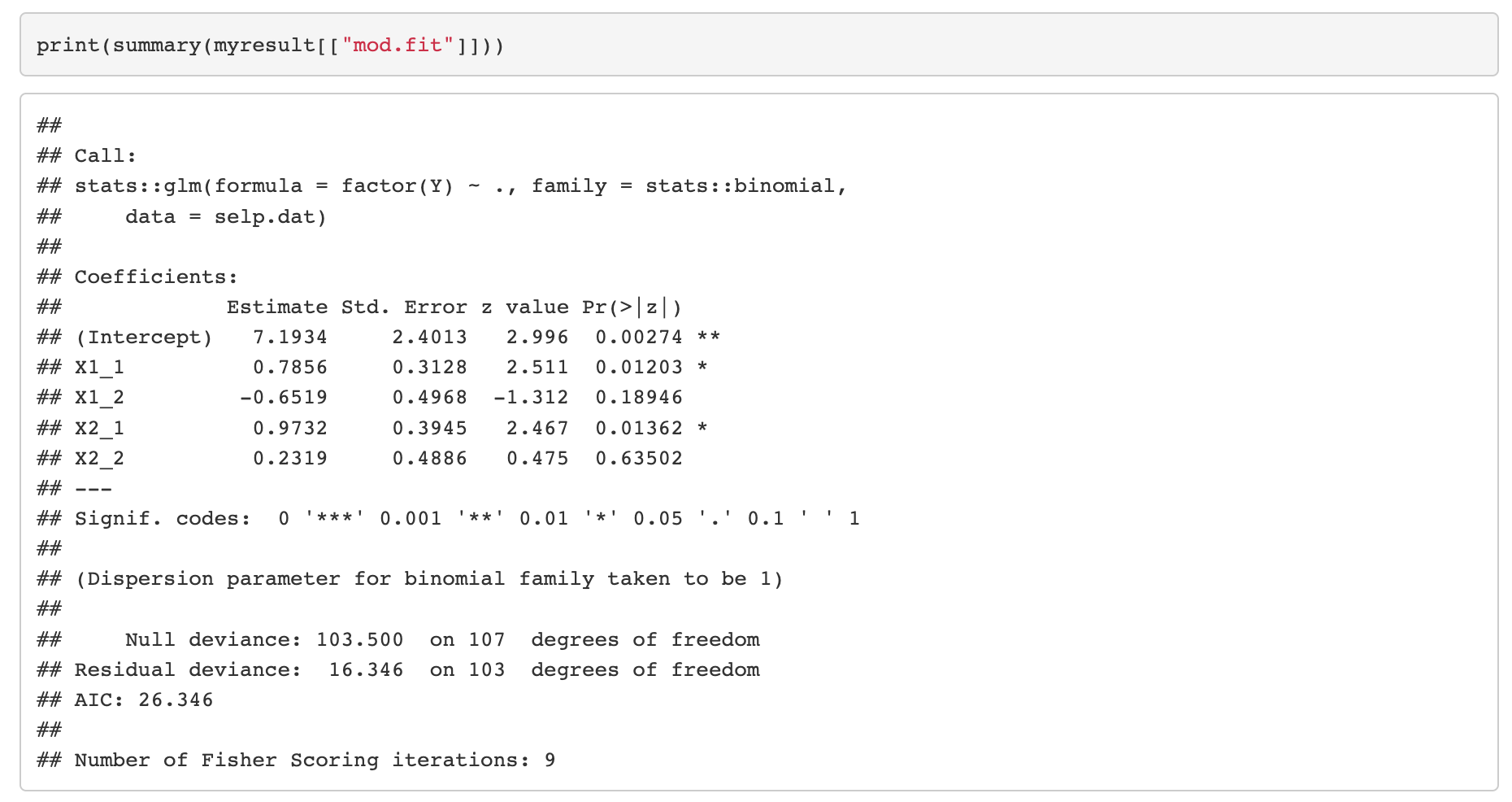} 
    \caption{Prediction with the first two canonical variates from SELPCCA. Results suggest that the first canonical variate for view 1 and view 2 are significantly associated with COVID-19 status (p-value $<$ 0.05). That is, the first canonical variates for views 1 (proteins) and 2 (genes) are able to discriminate between those with and without COVID-19. }
    \label{fig:predictionselp}
\end{figure*}

 \begin{figure*}[!hb]
    \centering
        \includegraphics[width=6.5in]{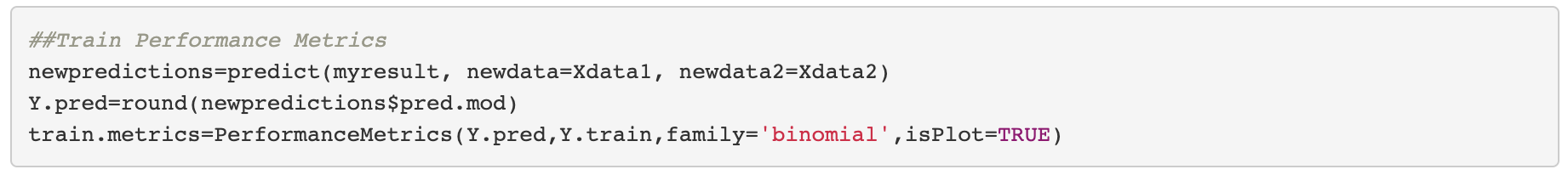}  
    \includegraphics[width=6.5in]{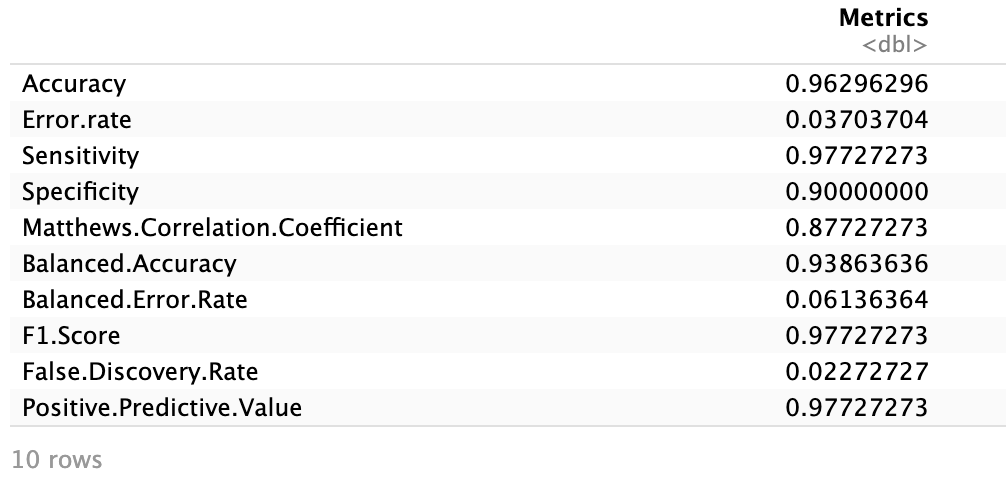} 
    \caption{Train Performance metrics. We use the function \texttt{PerformanceMetrics()} to obtain these metrics. Top panel: Train predictions. }
    \label{fig:performancemetrics}
\end{figure*}

\begin{figure*}[!hb]
    \centering
       \includegraphics[width=6.5in]{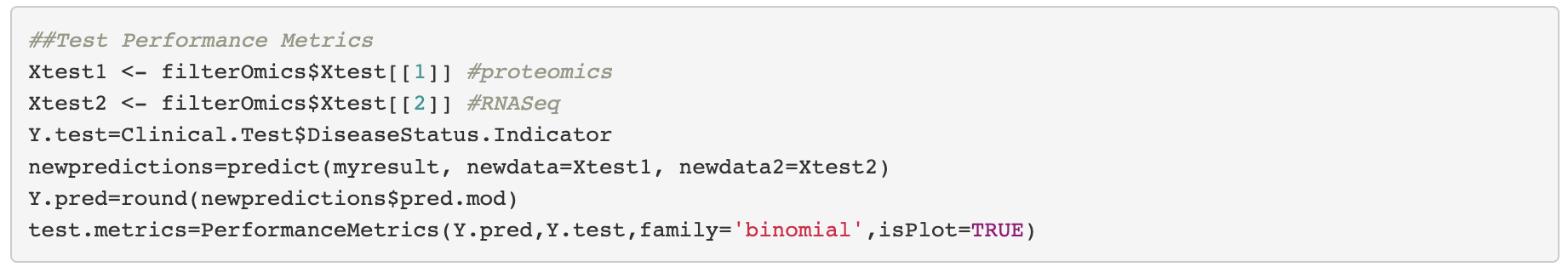}
        \includegraphics[width=6.5in]{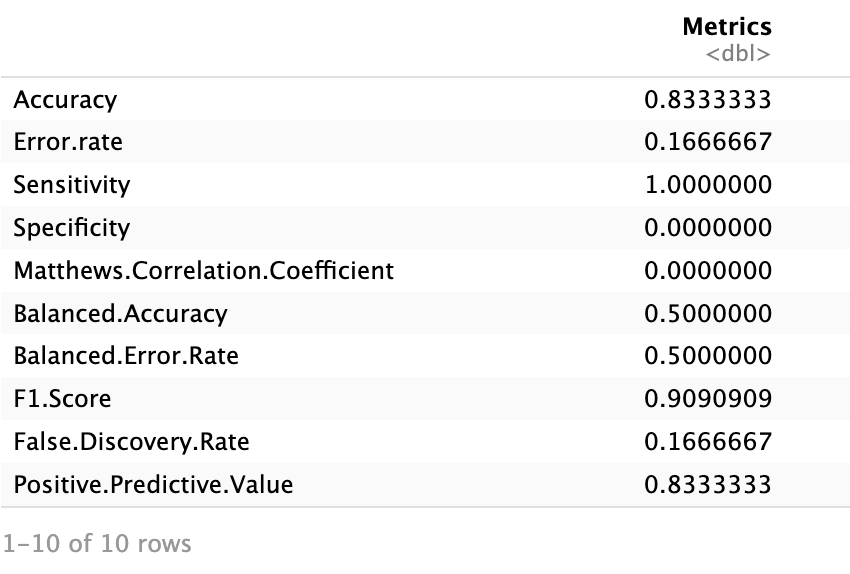}  
    \caption{Test Performance metrics. We use the function \texttt{PerformanceMetrics()} to obtain these metrics. Top panel: Train predictions. }
    \label{fig:performancemetricstest}
\end{figure*}

\clearpage
\section{Demonstration of SIDA on COVID-19 Data}
\begin{figure*}[!hb]
    \centering
           \includegraphics[width=\textwidth]{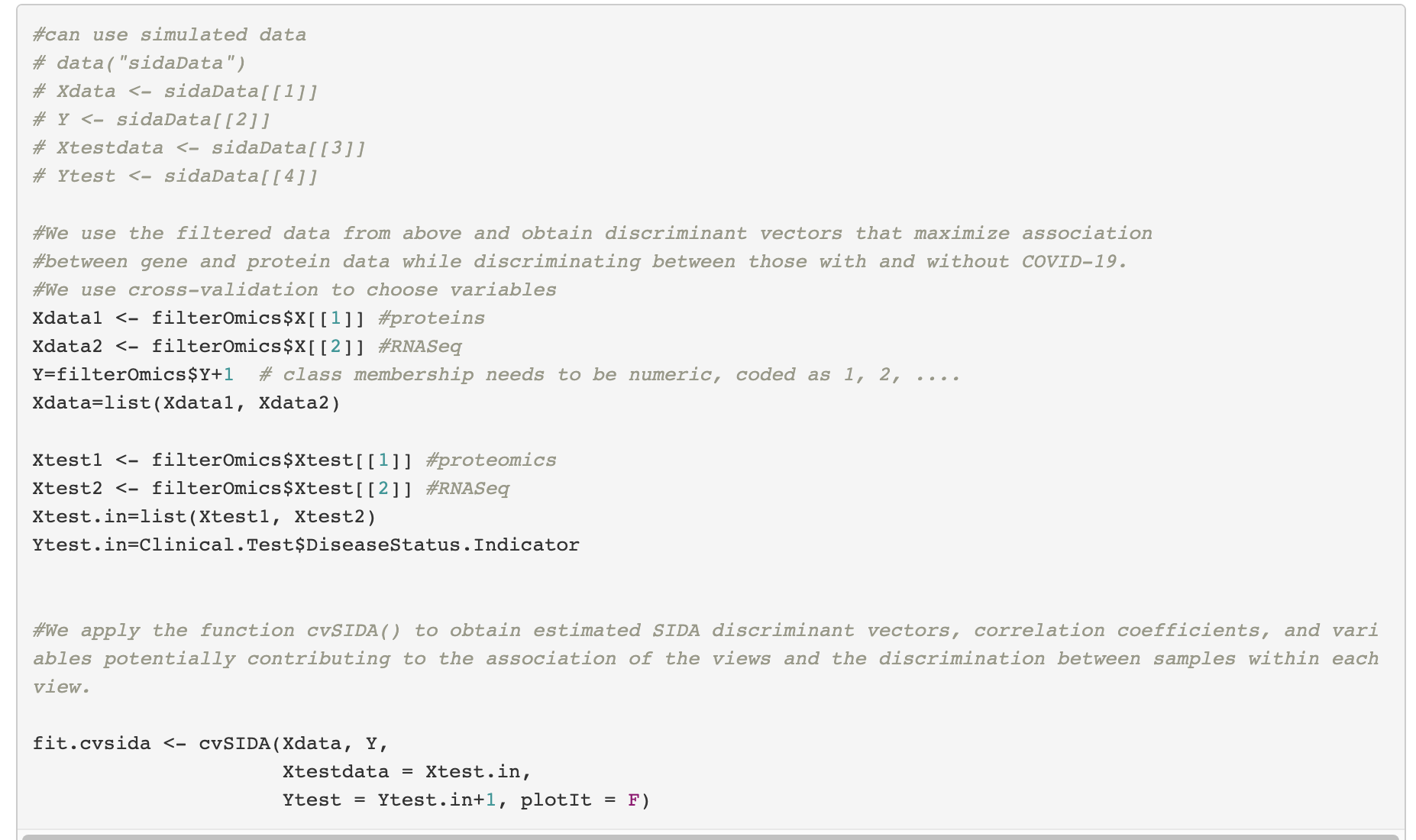}
            \includegraphics[width=\textwidth]{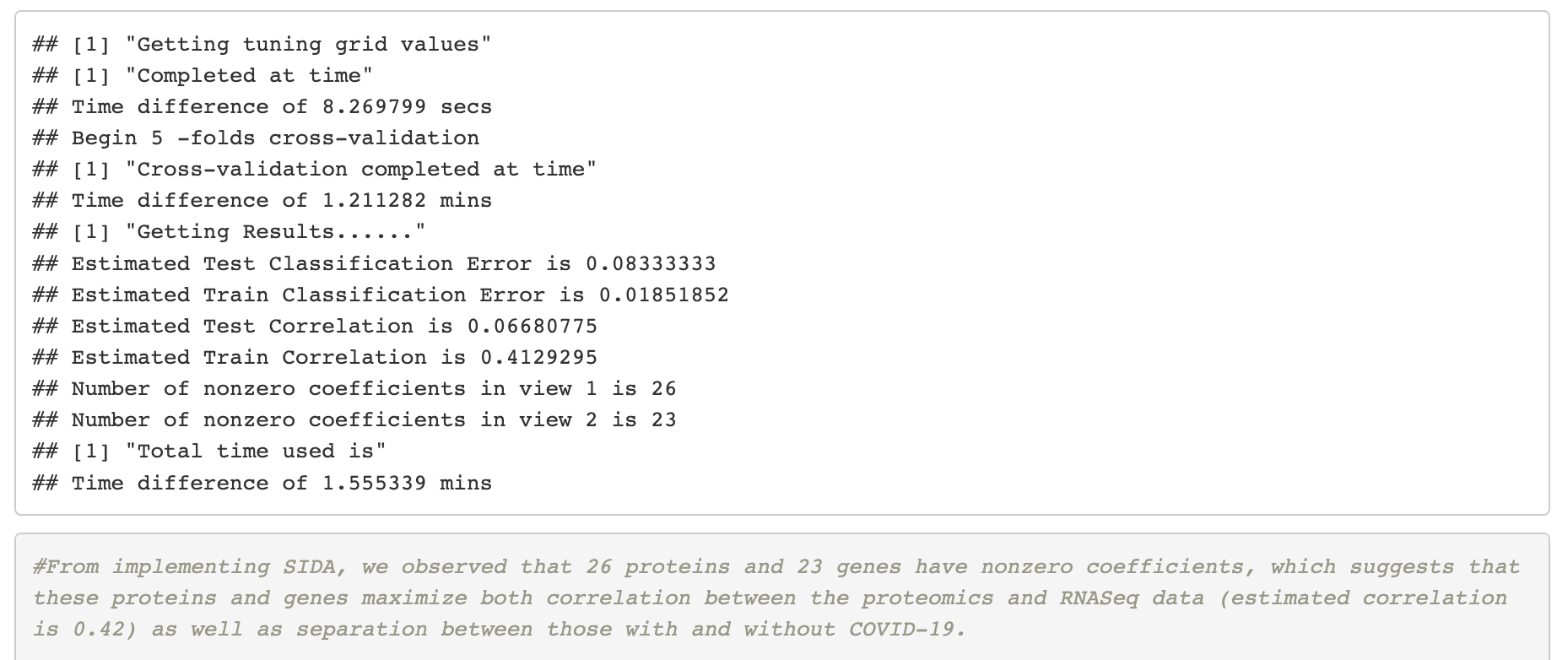}
    \caption{Fitting SIDA with cross-validation on training data to choose optimal hyperparameters and to obtain overall discriminant vectors based on the optimal hyperparameter. We use the function \texttt{cvSIDA()}. 
    }
    \label{fig:fitsida}
\end{figure*}

  \begin{figure*}[!hb]
    \centering
           \includegraphics[width=\textwidth]{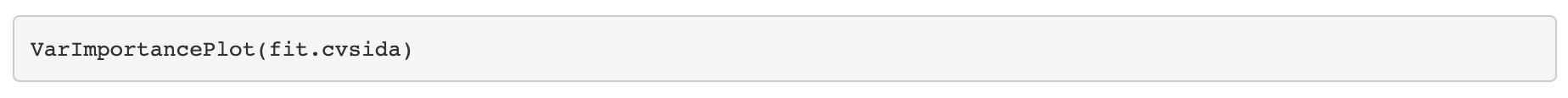}
    \includegraphics[width=\textwidth]{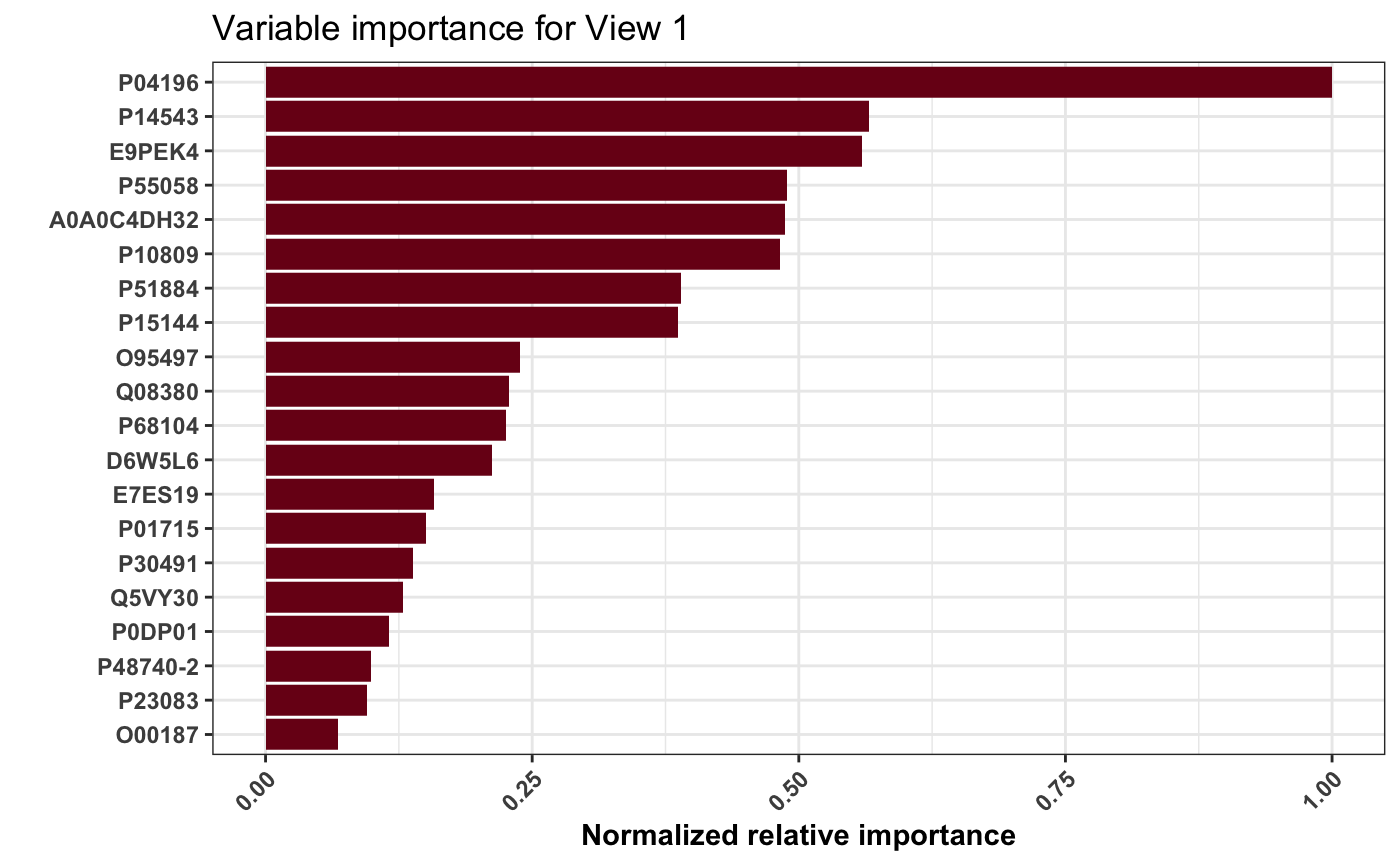}  \includegraphics[width=\textwidth]{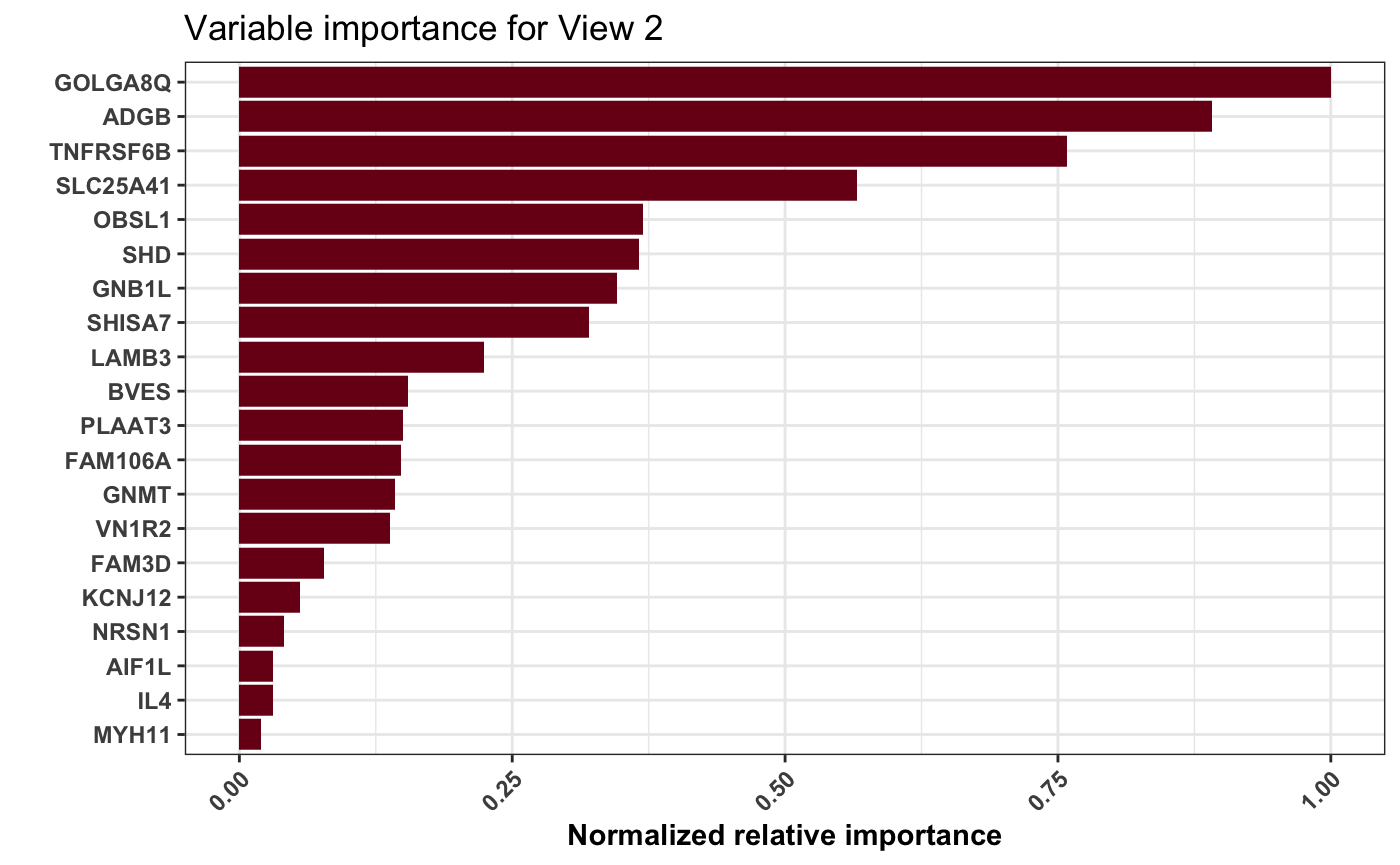}
    \caption{Variable important plots for  View 1 (Proteins) and View 2 (Genes) after implementing SIDA. Top 20 variables with largest absolute weights are shown. For proteins, Uniprot IDs are shown on variable importance plot.}
    \label{fig:varimpsida}
\end{figure*}

 \begin{figure*}[!hb]
    \centering
               \includegraphics[width=\textwidth]{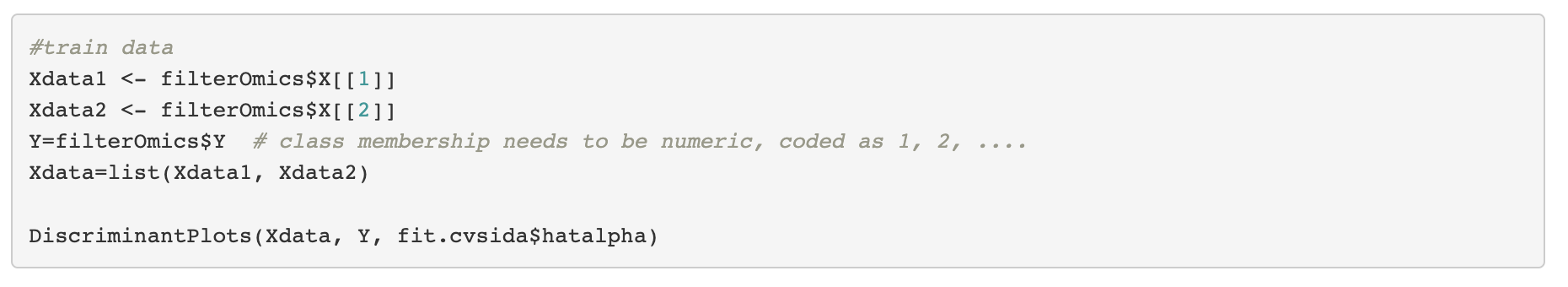}
    \includegraphics[width=5.5in]{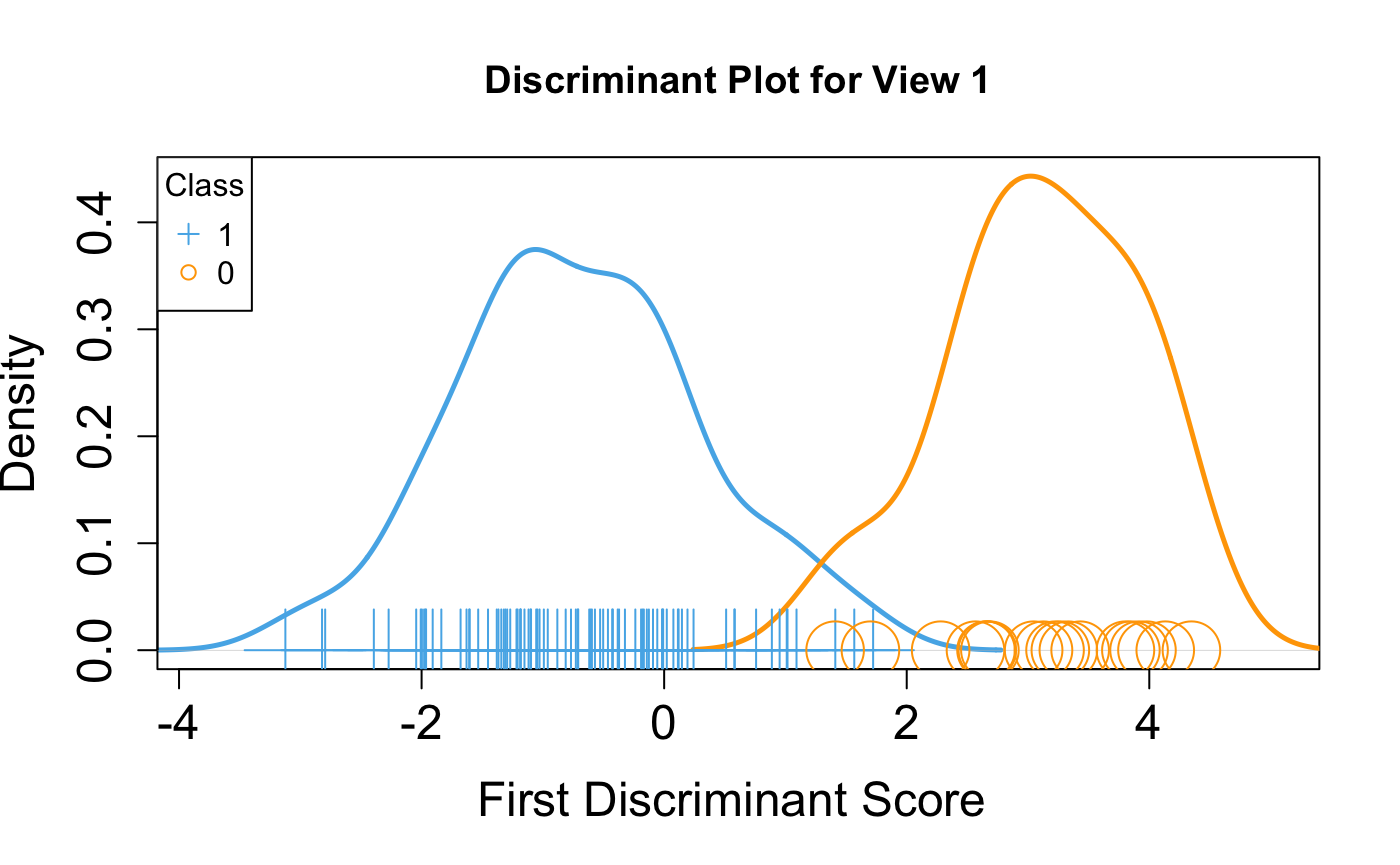}  \includegraphics[width=5.5in]{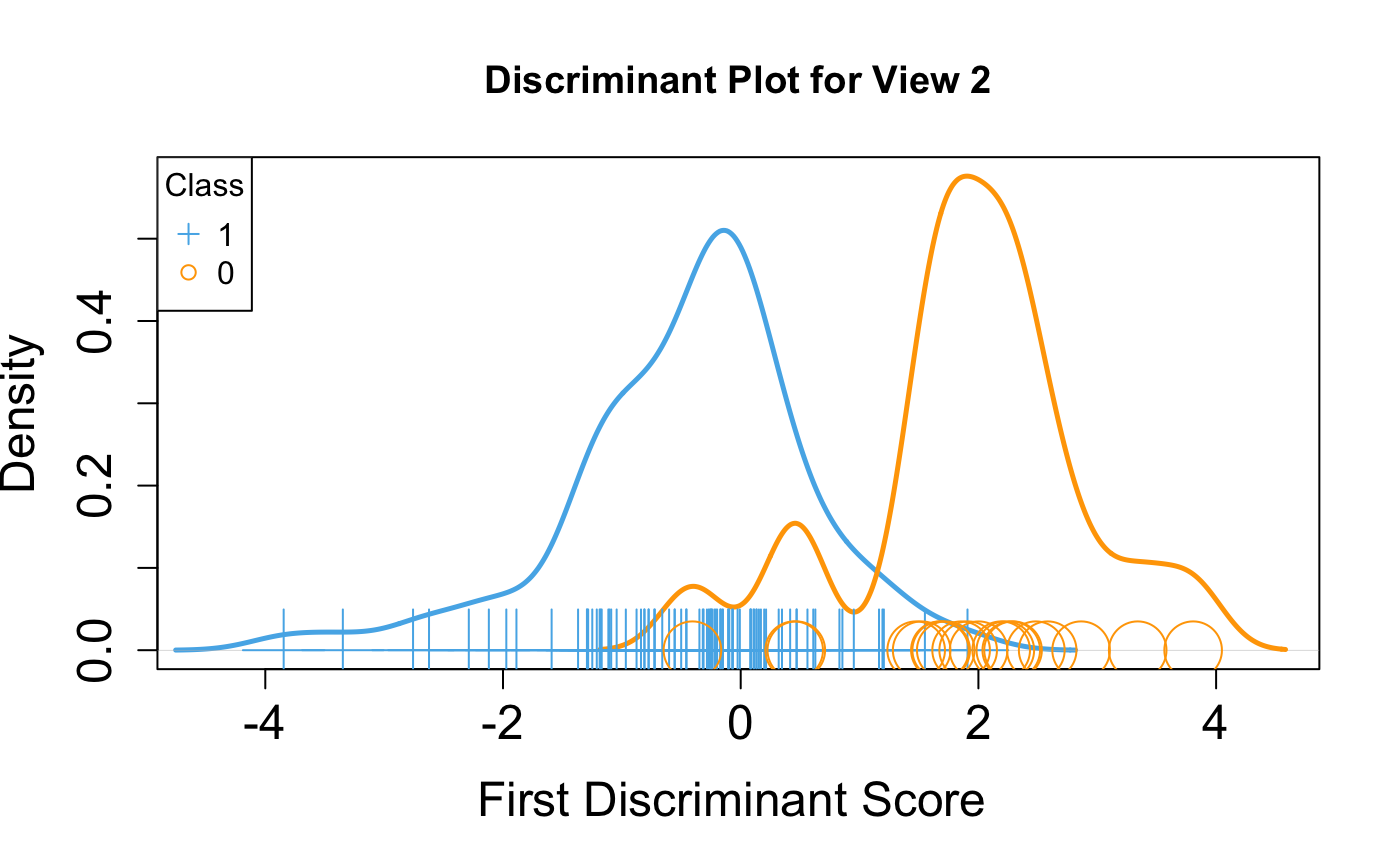}
    \caption{Discriminant plots based on training data. Top Panel: proteins and Bottom panel: genes. We use the function \texttt{DiscriminantPlots()} to generate these plots. Class 0 is non-COVID-19 samples. Class 1 is COVID-19 samples.}
    \label{fig:discplotstrain}
\end{figure*}

 \begin{figure*}[!hb]
    \centering
      \includegraphics[width=\textwidth]{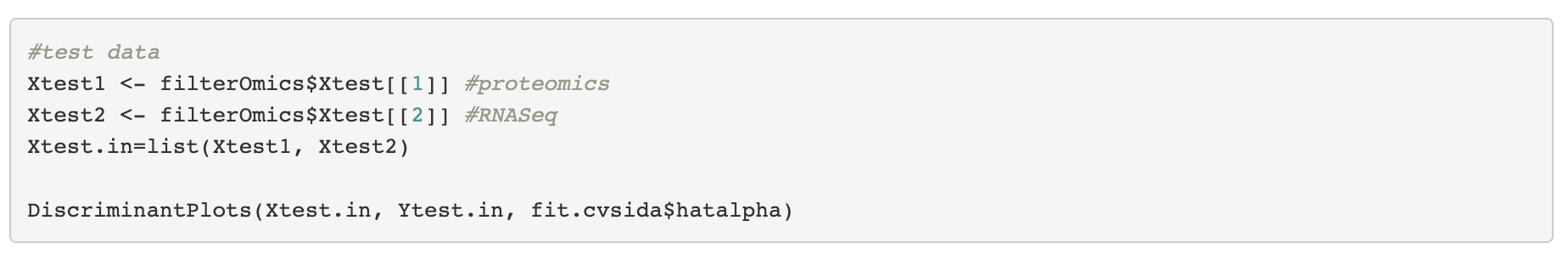}
     \includegraphics[width=5.5in]{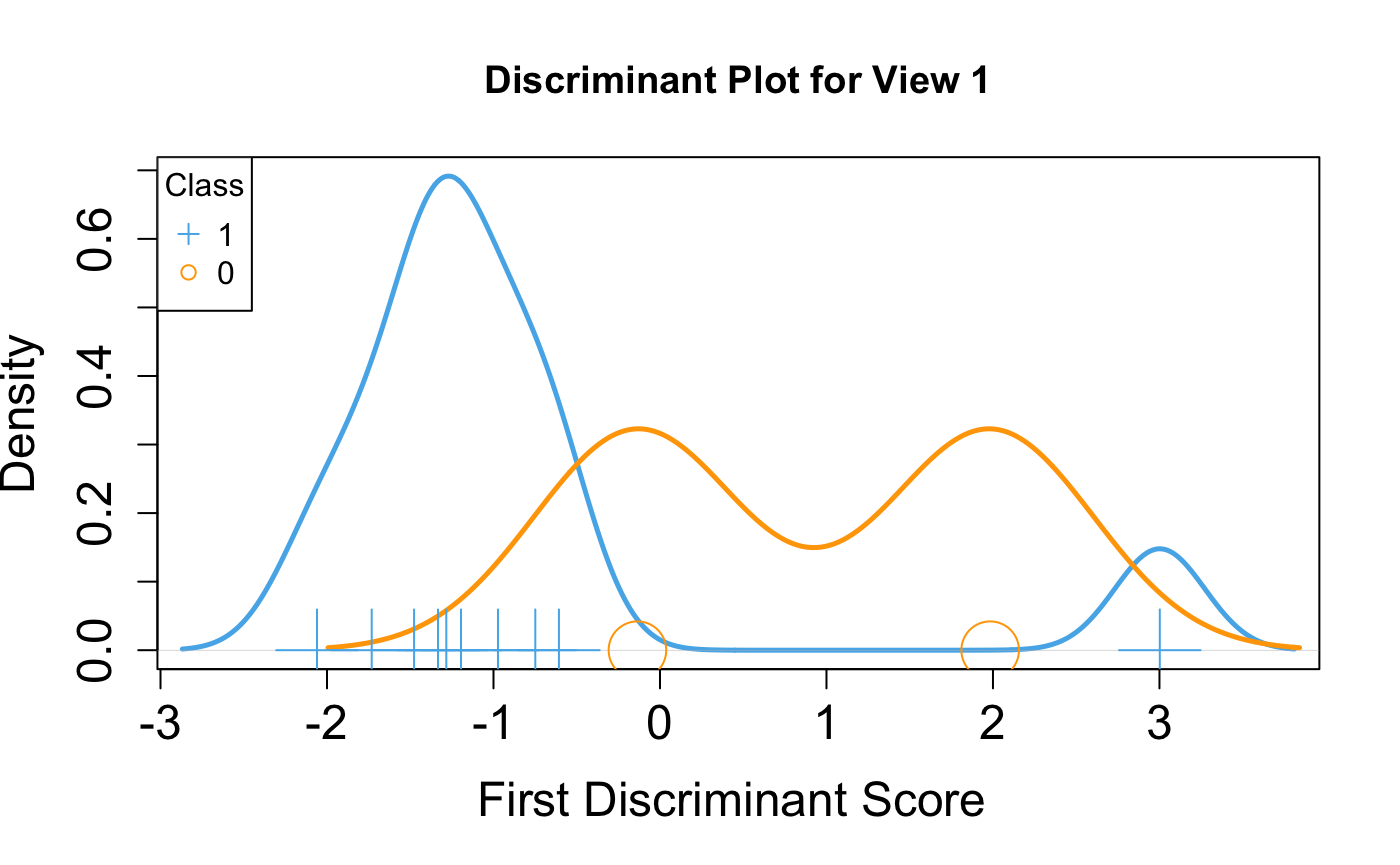} \includegraphics[width=5.5in]{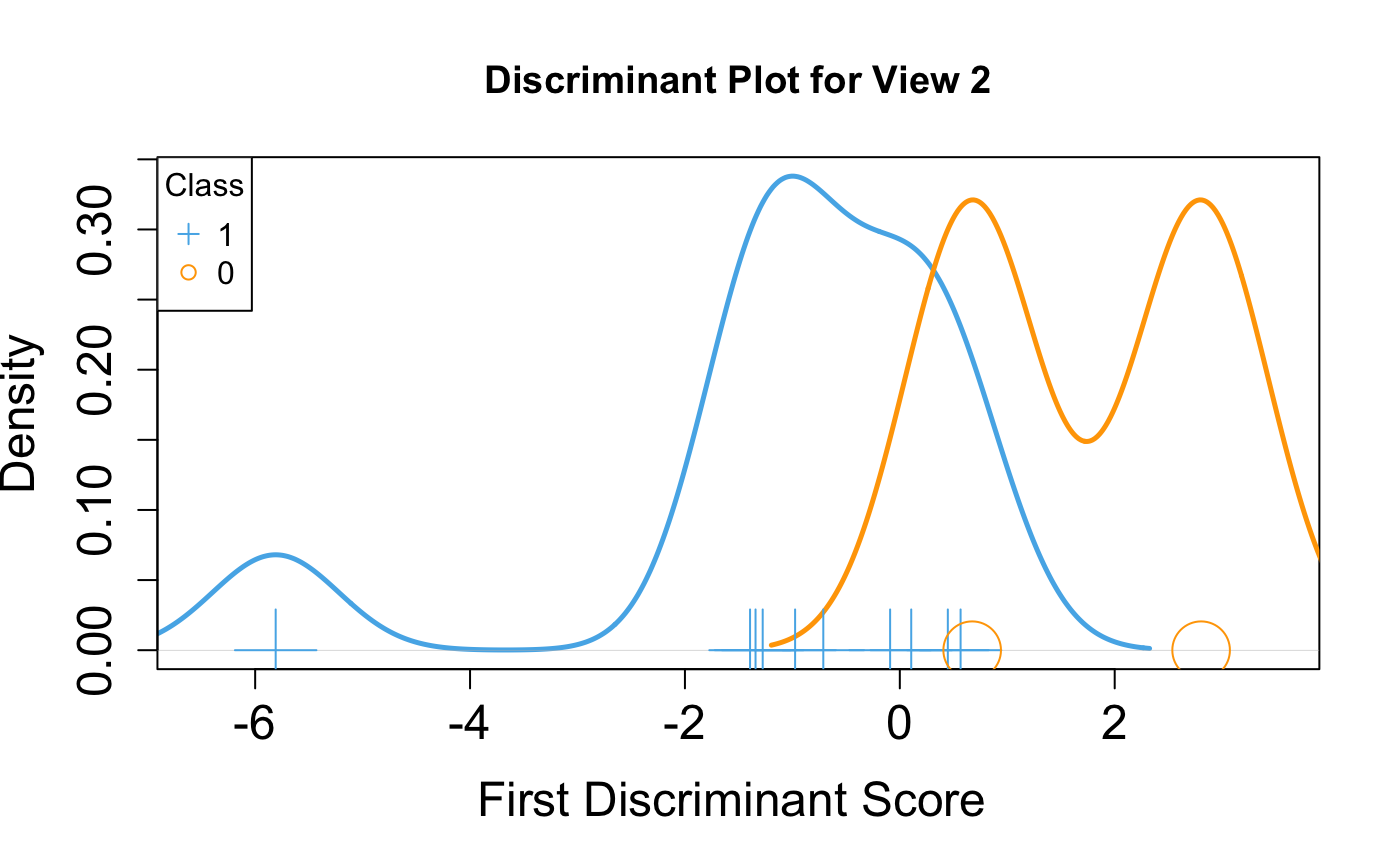}
    \caption{Discriminant plots based on testing data. Top Panel: proteins and Bottom panel: genes. We use the function \texttt{DiscriminantPlots()} to generate these plots. Class 0 is non-COVID-19 samples. Class 1 is COVID-19 samples.}
    \label{fig:discplots}
\end{figure*}

\begin{figure*}[!hb]
    \centering
      \includegraphics[width=\textwidth]{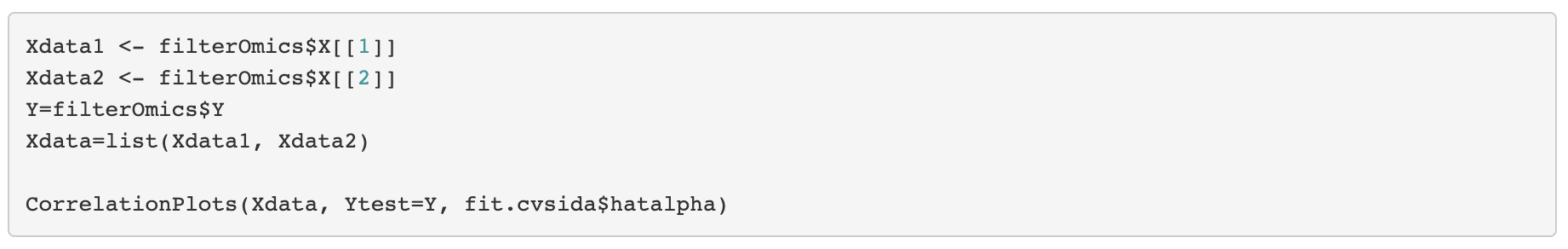}
     \includegraphics[width=5.5in]{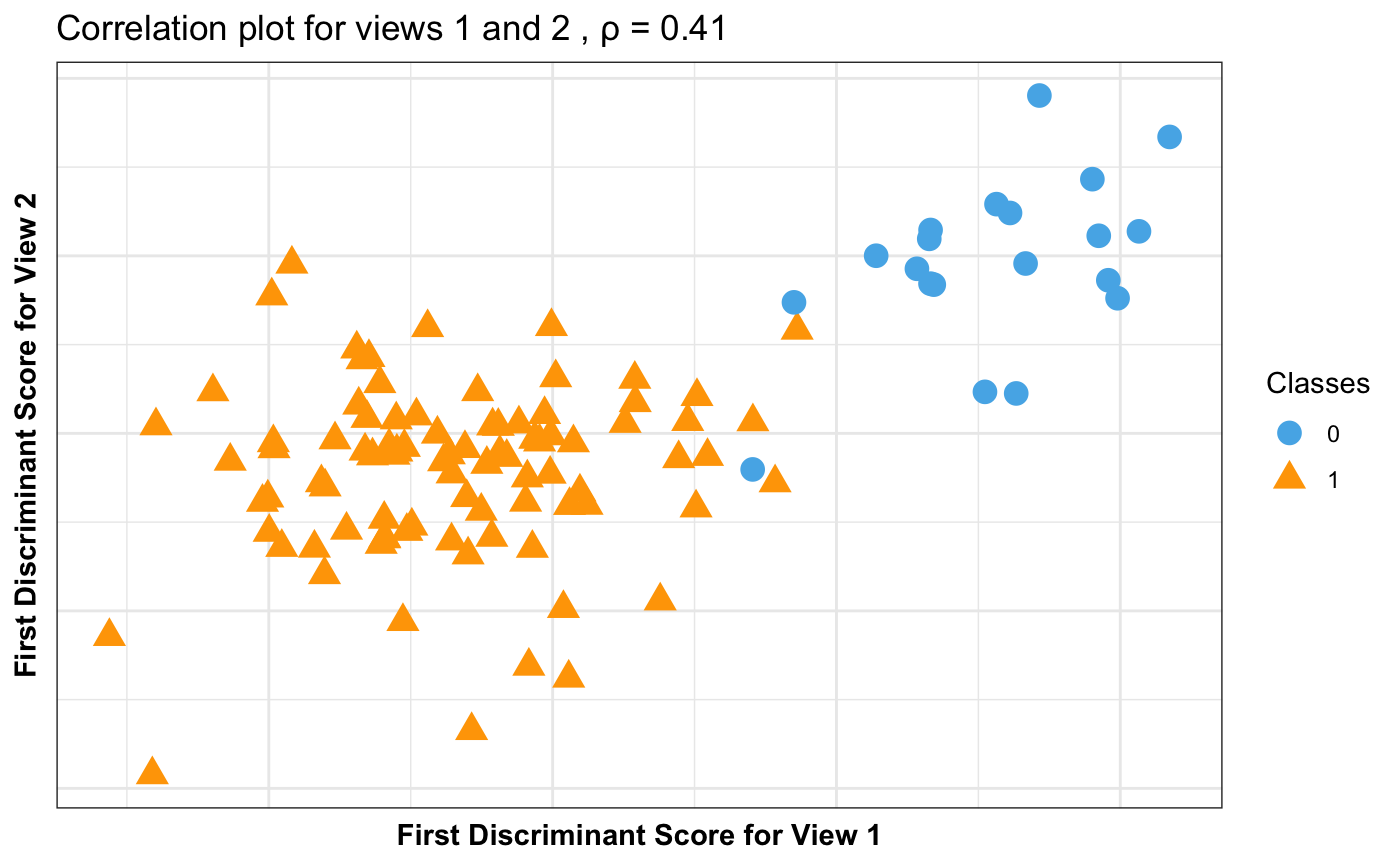} 
    \caption{Correlation plots based on training data. We use the function \texttt{CorrelationPlots()} to generate these plots. Class 0 is non-COVID-19 samples. Class 1 is COVID-19 samples.}
    \label{fig:corrplots}
\end{figure*}

    \begin{figure*}[!hb]
    \centering
     \includegraphics[width=\textwidth]{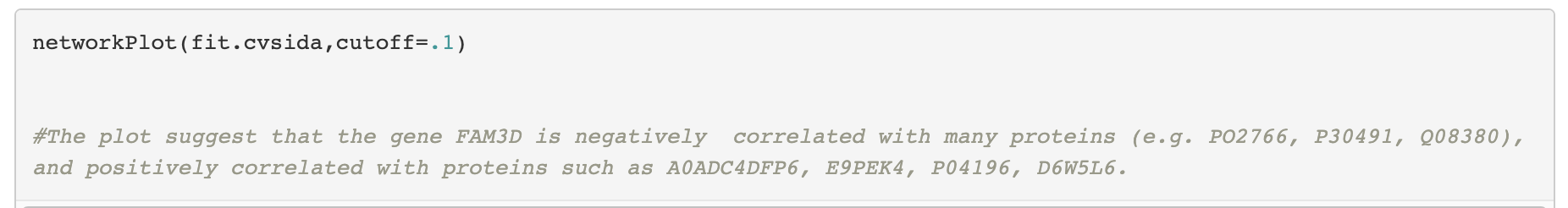}
    \includegraphics[width=7.5in]{Plots/RNSIDA.png} 
    \caption{Relevance network plot for SIDA. The nodes of the graph represent variables for the pairs of views, and edges represent the correlations between pairs of variables. Dashed and solid lines indicate negative and positive correlations, respectively. Circle nodes are View 1 variables (proteins), and rectangular nodes are View 2 variables (genes). We show edges with correlations at least 0.1.  The plot suggest that the gene FAM3D is negatively  correlated with many proteins (e.g. PO2766, P30491, Q08380), and positively correlated with proteins such as A0ADC4DFP6, E9PEK4, P04196, D6W5L6.  }
    \label{fig:rnsida}
\end{figure*}

 \begin{figure*}[!hb]
    \centering
           \includegraphics[width=6.5in]{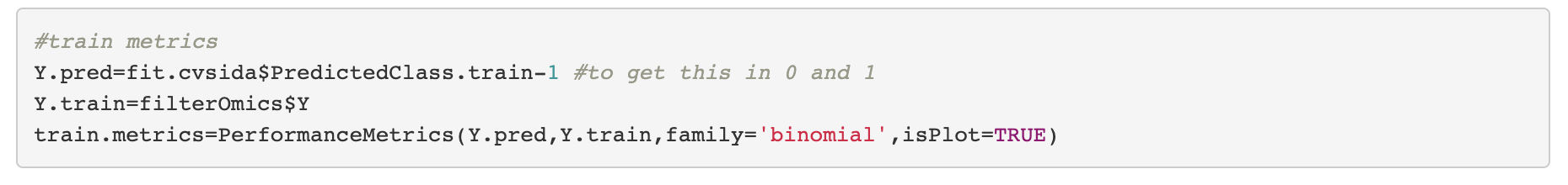}
    \includegraphics[width=6.5in]{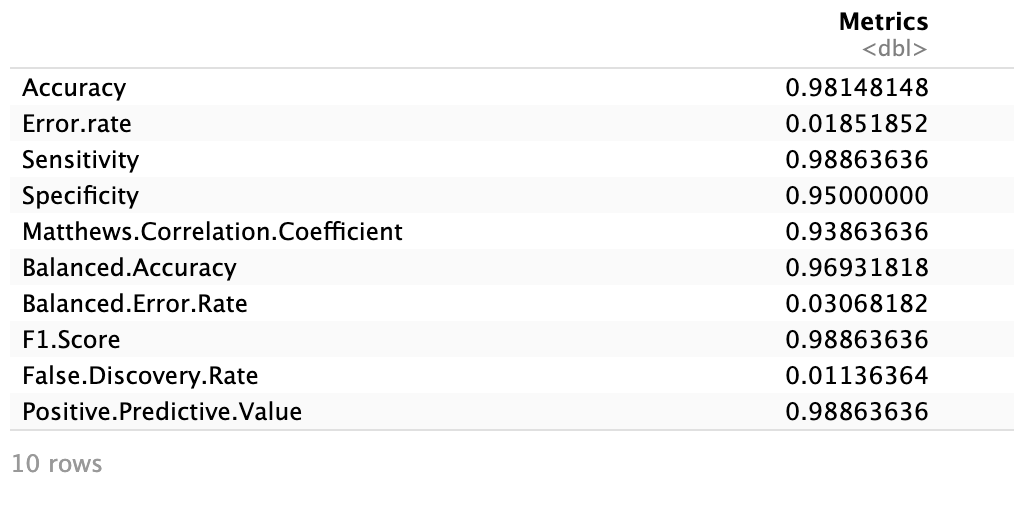}  \includegraphics[width=7.0in]{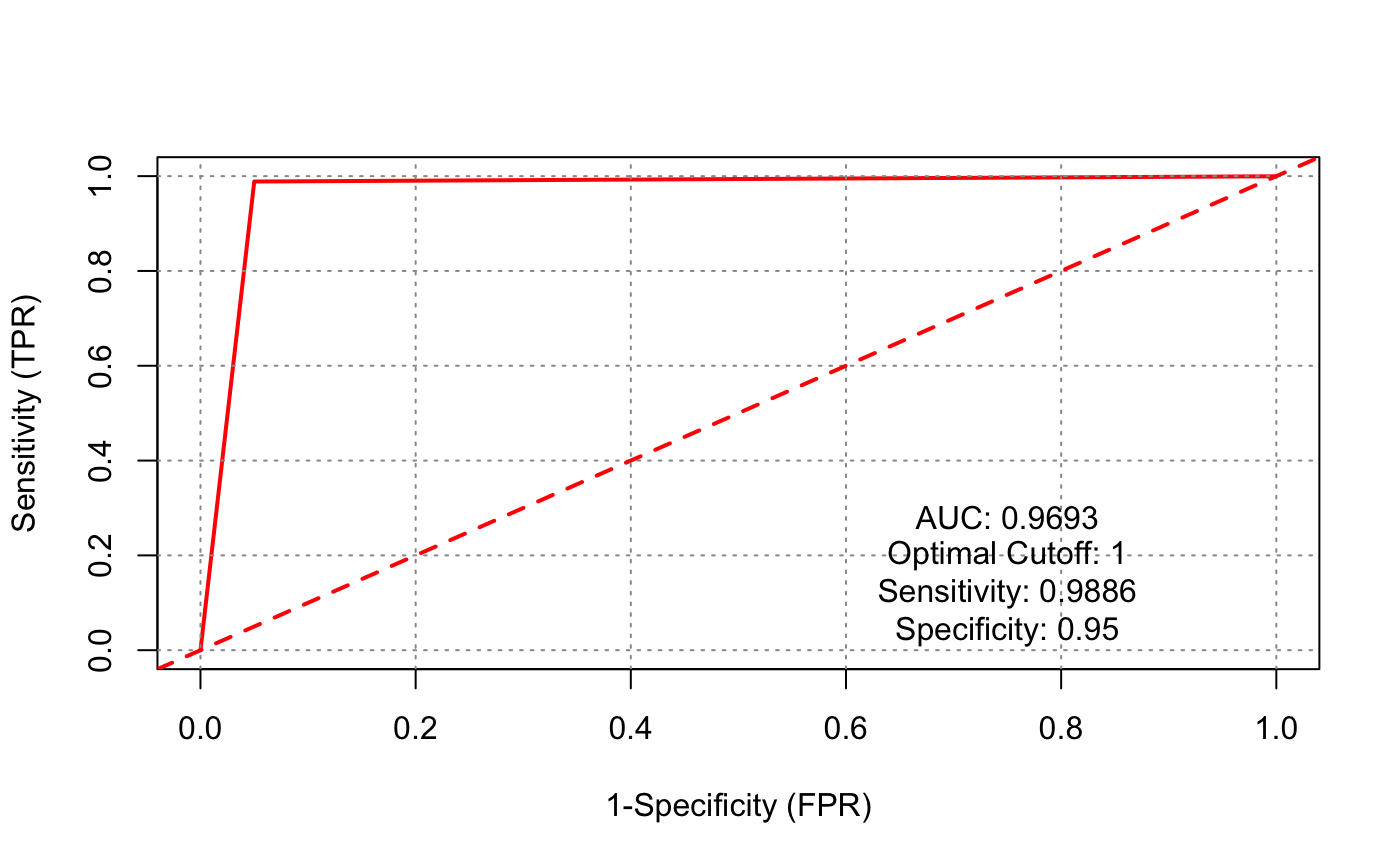}
    \caption{Performance metrics. We use the function \texttt{PerformanceMetrics()} to obtain these metrics. Top panel: Train predictions. Bottom panel: AUC plot.}
    \label{fig:performancemetricssidatrain}
\end{figure*}

 \begin{figure*}[!hb]
    \centering
               \includegraphics[width=6.5in]{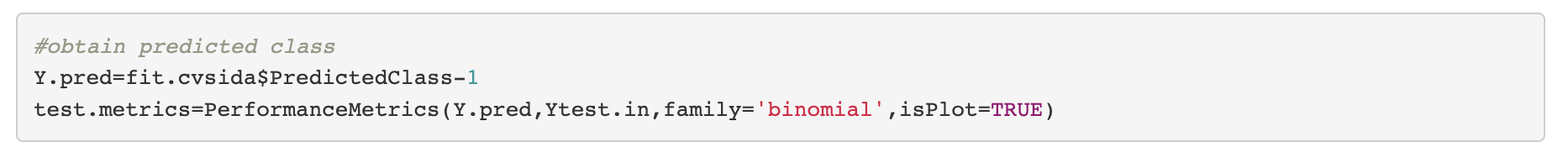}
    \includegraphics[width=6.5in]{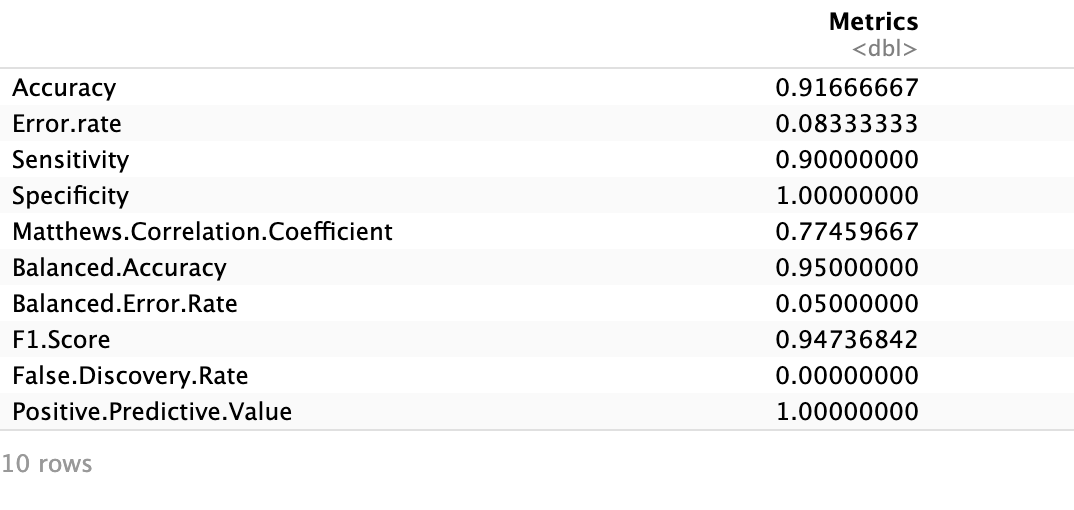}  \includegraphics[width=6.5in]{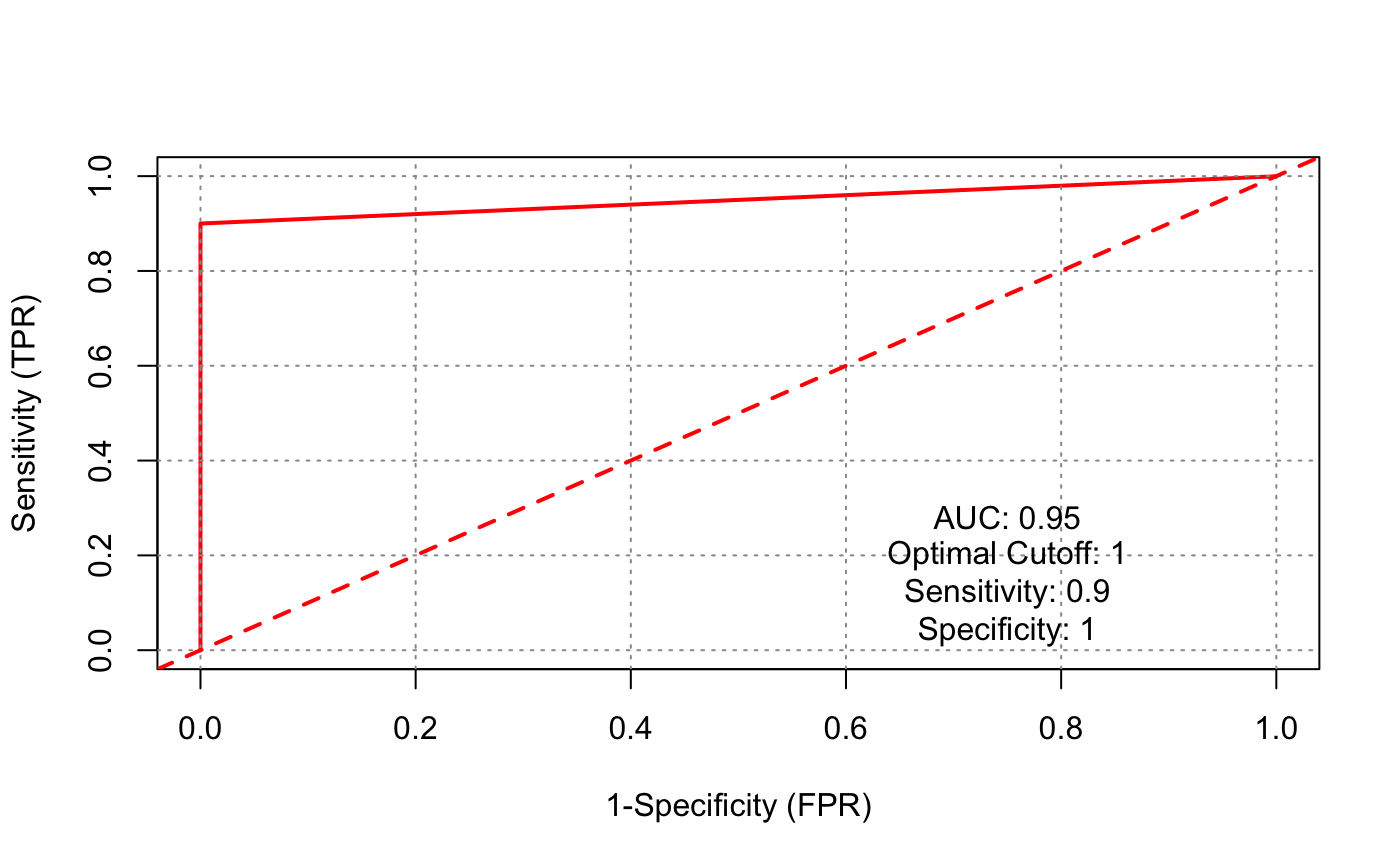}
    \caption{Performance metrics. We use the function \texttt{PerformanceMetrics()} to obtain these metrics. Top panel: Test predictions. Bottom panel: AUC plot.}
    \label{fig:performancemetricssidatest}
\end{figure*}

\begin{sidewaystable}[!hb]
    \centering
    \begin{tabular}{lll}
    \hline
    \hline
Purpose	&Functions	&Description\\
\hline
\hline
Importing Data	&\texttt{data(COVIDData)}	&Multiomics data pertaining to COVID-19\\
	&\texttt{data(selpData)}&	Simulated data example for SELPSCCA\\
	&\texttt{data(sidaData)}	&Simulated data example for SIDA\\
	&\texttt{data(sidanetData)}&	Simulated data example for SIDANet\\
 \midrule
Filtering& \texttt{filter.unsupervised()}&	Unsupervised Filtering. Options are variance and IQR filtering. Default is variance.	\\
&\texttt{filter.supervised()}&	Supervised Filtering. Filtering options are linear, logistic, t-test, and Kruskal-Wallis.\\
 \midrule
Unsupervised data integration	&\texttt{cvselpscca()}	&Performs n-fold cross validation to select optimal 
tuning parameters for SELPCCA \\
&~&based on training data.\\
&	\texttt{cvtunerange()}&	Obtain upper and lower bounds of tuning parameters of SELPCCA for each \\
~& &canonical correlation vector.\\
&	\texttt{multiplescca()}&	Estimates the sparse canonical correlation vectors for fixed tuning parameters.\\
 \midrule
Supervised data integration	&\texttt{selpscca.pred()}	&A two-step approach for integration and prediction based on SELPCCA. \\
~&&Performs n-fold cross validation to select optimal tuning parameters for SELPCCA \\
&&based on training data. Then uses the results to build a GLM or \\
~&& survival model for a pre-specified outcome.\\
&	\texttt{predict.SELPCCA()}&	Prediction for out-of-sample data for SELPCCA predict.\\
&	\texttt{sida()}&	Joint integration and discriminant analysis for fixed tuning parameters. \\
&	\texttt{cvSIDA()}&	Performs nfolds cross validation to select optimal tuning parameters for sida \\
& &based on training data, which are then used with the training  or \\
&& testing data to predict class membership.\\
&	\texttt{sidanet()}&	Joint integration and discriminant analysis with incorporation of prior biological \\
&& information (variable-variable relationships), for fixed tuning parameters.\\
&	\texttt{cvSIDANet()}&	Performs nfolds cross validation to select optimal tuning parameters for sida \\
&&with network information, based on training data, which are then used with \\
&&the training or testing data to predict class membership.\\
&	\texttt{sidatunerange()}&	Function to provide tuning parameter grid values for each view, \\
&&not including covariates, if available, for SIDA.\\
&	\texttt{sidatunerange()}&	Function to provide tuning parameter grid values for each view, \\
&&not including covariates, if available, for SIDANet.\\
&	\texttt{sidaclassify()}&	Classification approach for SIDA and SIDANet\\
 \midrule
Visualizations	&\texttt{volcanoPlot()}&	Wrapper function for volcano plots of the results after supervised filtering.\\
&	\texttt{umapPlot()}&	Wrapper function to plot a UMAP of the results after supervised filtering.\\
&	\texttt{VarImportancePlot()}	&A graph of the absolute loadings for variables selected. \\
&	\texttt{DiscriminantPlots()}&	Plots discriminant scores (for SIDA) and canonical variates (for SELPCCA) \\
&&for visualizing class separation.\\
&	\texttt{CorrelationPlots()}&	Plots for visualizing correlation between estimated discriminant vectors for pairwise data. \\
&	\texttt{LoadingsPlots()}&	Plots discriminant and canonical vectors to visualize how selected variables contribute \\
&&to the first and second discriminant (for SIDA and SIDANet) or canonical correlation \\
&&(for SELPCCA) vectors.\\
&	\texttt{networkPlot()}&	Network visualization of selected variables from integrative analysis methods.\\
&	\texttt{BetweenViewBiplot()}	&Biplots to visualize discriminant scores/ canonical variates between pairs of views.\\
&	\texttt{WithinViewBiplot()}&	Biplots for Discriminant Scores or Canonical Correlation Variates for each View.\\
 \midrule
Performance Estimation	&\texttt{PerformanceMetrics()}&	Estimates performance metrics for a predicted model. Currently works for binary \\
&&and continuous outcomes.\\
\bottomrule
 \end{tabular}
\caption{Description of current functions in \texttt{mvlearnR}}
\label{tab:my_label}
%\end{table}
\end{sidewaystable}

\clearpage
%%%%% References %%%%%
% \bibliographystyle{elsarticle-num} 
% \bibliography{references.bib}

% \begin{thebibliography}{1}
% \expandafter\ifx\csname url\endcsname\relax
%   \def\url#1{\texttt{#1}}\fi
% \expandafter\ifx\csname urlprefix\endcsname\relax\def\urlprefix{URL }\fi
% \expandafter\ifx\csname href\endcsname\relax
%   \def\href#1#2{#2} \def\path#1{#1}\fi

[1] W.~Zhang, C.~Wendt, R.~Bowler, C.~P. Hersh, S.~E. Safo, Robust integrative biclustering for multi-view data, Statistical methods in medical research 31~(11) (2022) 2201--2216.

[2] J.~Wang, S.~E. Safo, Deep {I}da: A deep learning method for integrative
  discriminant analysis of multi-view data with feature ranking--an application to covid-19 severity, ArXiv (2021).

[3] H.~Wang, H.~Lu, J.~Sun, S.~E. Safo, Interpretable deep learning methods for
  multiview learning (2023).

[4] S.~E. Safo, H.~Lu, Scalable randomized kernel methods for multiview data integration and prediction, arXiv preprint arXiv:2304.04692 (2023).

[5] S.~E. Safo, J.~Ahn, Y.~Jeon, S.~Jung,
  \href{https://onlinelibrary.wiley.com/doi/abs/10.1111/biom.12886}{Sparse
  generalized eigenvalue problem with application to canonical correlation
  analysis for integrative analysis of methylation and gene expression data}, Biometrics 74~(4) (2018) 1362--1371.

[6] S.~E. Safo, E.~J. Min, L.~Haine, Sparse linear discriminant analysis for
  multiview structured data, Biometrics 78~(2) (2022) 612--623.

[7] H.~Hotelling, Relations between two sets of variables, Biometrika (1936)
  312--377.

[8] A.~J. Butte, P.~Tamayo, D.~Slonim, T.~R. Golub, I.~S. Kohane, Discovering
  functional relationships between rna expression and chemotherapeutic
  susceptibility using relevance networks, Proceedings of the National Academy
  of Sciences 97~(22) (2000) 12182--12186.

[9] I.~Gonz{\'a}lez, K.-A.~L. Cao, M.~J. Davis, S.~D{\'e}jean, Visualising
  associations between paired ‚Äòomics‚Äô data sets, BioData mining 5 (2012)
  1--23.

%\end{thebibliography}